\documentclass[11pt]{article}
\usepackage[T1]{fontenc}
\usepackage[utf8]{inputenc}
\usepackage{mathptmx}
\usepackage[a4paper,margin=25mm]{geometry}
\usepackage{amsmath,amssymb}
\DeclareSymbolFont{txletters}{OML}{txmi}{m}{it}
\DeclareMathSymbol{\epsilon}{\mathord}{txletters}{"0F}
\DeclareFontFamily{OML}{cmmX}{\skewchar\font=127}
\DeclareFontShape{OML}{cmmX}{m}{it}{<->cmmi10}{}
\DeclareSymbolFont{cmletters}{OML}{cmmX}{m}{it}
\DeclareMathSymbol{g}{\mathalpha}{cmletters}{`g}
\usepackage{xcolor}
\usepackage{graphicx}
\usepackage{array}
\usepackage[super,sort&compress,comma]{natbib}
\usepackage{setspace}
\usepackage[hidelinks]{hyperref}

\newcommand{\ES}{E_{\mathrm{S}}}

\title{Band edges and d levels approaching GW accuracy\\ at semilocal cost}
\author{Ryo Nagai\textsuperscript{1,*}, Shigeru Iwase\textsuperscript{1} \& Chikashi Shinagawa\textsuperscript{1}\\[1.5ex]
\normalsize \textsuperscript{1}Preferred Networks, Inc., 1-6-1 Otemachi, Chiyoda-ku, Tokyo 100-0004, Japan.\\[0.5ex]
\normalsize \textsuperscript{*}Corresponding author. Email: rnagai@preferred.jp}
\date{}

\begin{document}
\maketitle
\onehalfspacing

\begin{abstract}
Band gaps and band-edge positions determine how a material absorbs light and exchanges charge, but accurate methods are prohibitively expensive for large, disordered or finite-temperature systems. We introduce a band correction from a simple physical model with few global parameters, none fitted to measured gaps. A screened spectral term, evaluated once after a self-consistent calculation with a semilocal exchange--correlation functional at integer occupation, shifts only one-electron levels and typically adds 20--40\% to the cost of our meta-generalized-gradient base functional; a shell term corrects localized d levels. On several hundred solids excluded from the fits, corrected gaps are more accurate than those of the HSE06 hybrid, markedly so for wide gaps, and approach GW accuracy; surface ionization potentials are placed correctly on the vacuum scale. Applied directly to molecular-dynamics snapshots, our correction reproduces the measured room-temperature gaps of halide perovskites with a single protocol, extending this accuracy to large, disordered systems.
\end{abstract}

Whether a material can absorb sunlight, drive a photocatalytic reaction or exchange charge across an interface depends on where its electronic levels lie: the band gap, the band edges relative to the vacuum level and the energies of localized d states. Semilocal approximations to the exchange--correlation functional of Kohn--Sham density functional theory (DFT)\cite{hohenberg1964,kohn1965}, such as the generalized gradient approximation of Perdew, Burke and Ernzerhof (PBE)\cite{perdew1996}, are among the least expensive electronic-structure methods and have been used for a wide range of systems, but they systematically underestimate band gaps\cite{perdew1983,sham1983,morisanchez2008}. The screened hybrid functional of Heyd, Scuseria and Ernzerhof (HSE06)\cite{heyd2003,krukau2006} and many-body perturbation theory in the GW approximation to the self-energy\cite{hedin1965,hybertsen1986} remove most of this error, but at a cost several times to orders of magnitude higher.

This cost barrier is highest where accurate levels matter most. A halide perovskite absorber works at room temperature, with its lattice in constant motion and its gap shaped by spin--orbit coupling, and predicting that gap takes supercells of hundreds to thousands of atoms\cite{venkatanarayanan2025}; alloys need many configurations at each composition\cite{kyrtsos2019}, and amorphous solids need large quenched cells. In practice, accurate levels are computed for a small cell, and the effects of temperature, disorder and spin--orbit coupling are added from separate calculations\cite{wang2022}.

Several approaches reduce this cost, each at the expense of other properties. The modified Becke--Johnson (mBJ) potential\cite{tran2009} runs at semilocal cost and is about as accurate as HSE06 on large benchmarks\cite{borlido2019,grossmann2026}, but it yields no total energy and, in its global form, no meaningful absolute band positions\cite{rauch2020}. DFT+U, an on-site Hubbard correction for localized d or f states\cite{anisimov1991,dudarev1998}, needs a $U$ for each material, fitted to experiment or computed by linear response\cite{cococcioni2005,kulik2006}, and a $U$ fitted to one property, such as reaction energies\cite{wang2006,jain2011}, need not reproduce another, such as the gap; other corrections at semilocal cost rely on a parameter chosen for each material or lose accuracy for small gaps\cite{ferreira2008,kuisma2010,tran2018,chan2010}. Dielectric-dependent hybrids\cite{riemelmoser2026,wing2021} and full-frequency and vertex-corrected GW\cite{grossmann2026,varrassi2025} approach experimental accuracy without lowering the cost, and the hybrids need a separate dielectric calculation for each material. Koopmans-compliant functionals\cite{nguyen2018} and orbital-scaling corrections\cite{zheng2011,williams2026}, which restore the piecewise linearity of the energy, localize the orbitals in solids and usually compute the screening of each system by linear response\cite{fan2026}, and a symbolic-regression model of the GW self-energy must be fitted to a GW calculation of each material\cite{pant2026}.

A correction that lowers this barrier is therefore still lacking: one that adds little to the cost of a self-consistent calculation with a numerically stable semilocal functional, does not degrade the total energy and forces of that calculation, places the levels on an absolute scale, approaches GW accuracy and requires no fitted parameter, separate calibration calculation or training for each material.

Here we construct such a correction from the physical origin of the gap error. To leading order, a semilocal band-edge eigenvalue lacks part of the exact exchange experienced by the state; in a solid this exchange is screened, and approximately a fraction $1/\varepsilon_\infty$ of it remains ($\varepsilon_\infty$, electronic dielectric constant), as in the long-range limit of GW and in dielectric-dependent hybrids. We write this screened exchange as a spectral term that is added to a semilocal exchange--correlation functional, the base functional, and acts only on fractionally occupied states. Total energies, densities and forces at integer occupation therefore remain those of the base functional, and one evaluation after the self-consistent calculation shifts only the eigenvalues (Fig.~\ref{fig:method}a): like a one-shot GW calculation, our correction is a quasiparticle correction of the one-electron levels on top of an unchanged ground state. The coefficients of the spectral term are obtained once from calculated data, exact-exchange matrix elements of ten semiconductors and coupled-cluster ionization potentials of 66 molecules; no measured gap is used. For localized d shells, we add to r$^2$SCAN\cite{furness2020}, a regularized form of the strongly constrained and appropriately normed (SCAN) meta-generalized-gradient approximation (meta-GGA)\cite{sun2015}, a shell term that lowers the d levels towards their quasiparticle positions; with the revised Vydrov--van Voorhis (rVV10) dispersion correction\cite{sabatini2013}, we call this base functional LOCUS (LOCalized-shell-Upgraded SCAN; Methods).

On several hundred solids excluded from all coefficient fits, the corrected gaps are more accurate than literature HSE06, mBJ and one-shot plasmon-pole GW results and approach the accuracy of full-frequency and self-consistent GW; applied directly to molecular-dynamics snapshots with spin--orbit coupling, our correction reproduces the measured room-temperature gaps of 11 halide perovskite compositions.

\section*{Results}

\subsection*{A screened spectral correction calibrated on theory}

\begin{figure}[t]
\centering
\includegraphics[width=\textwidth]{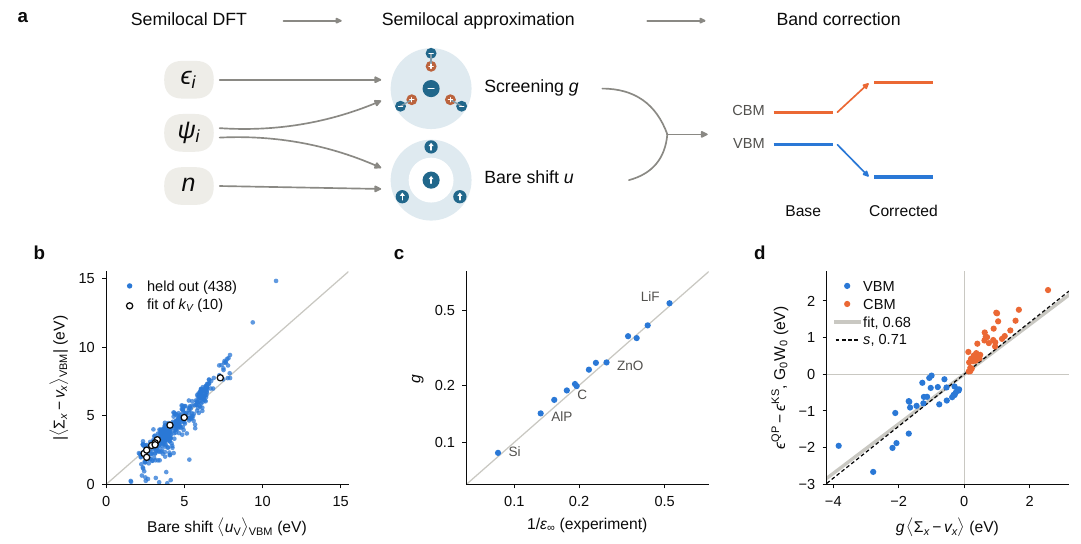}
\caption{\textbf{The spectral correction and its check against theory.}
\textbf{a}, From the eigenvalues $\epsilon_i$, orbitals $\psi_i$ and density $n$ of a semilocal calculation, our correction approximates the screening $g$ and the Fock exchange missing from each state (the bare shift $u$), and shifts the valence-band maximum (VBM) and conduction-band minimum (CBM).
\textbf{b}, Exact-exchange correction of the VBM against its bare shift, on PBE states, for 448 solids of the Borlido benchmark\cite{borlido2019}; $k_V$ was fitted to ten of them.
\textbf{c}, $g$ (LOCUS) against $1/\varepsilon_\infty$ from experiment\cite{skone2014} for 13 solids.
\textbf{d}, G$_0$W$_0$@PBE corrections of the band edges of 34 solids against the screened exchange; the fitted slope agrees with $s$ fitted to 66 molecules for the PBE base; the VBM of Ne ($-9.3$, $-5.7$~eV) lies outside the plotted range. Grey diagonals in \textbf{b} and \textbf{c}, $y = x$.}
\label{fig:method}
\end{figure}

Our correction is a spectral term added to the base functional,
\begin{equation}
  \ES = \sum_i f_i (1 - f_i) \int W_i(\mathbf{r})\, |\psi_i(\mathbf{r})|^2 \, \mathrm{d}\mathbf{r},
  \qquad W_i(\mathbf{r}) = s\, g\, u_i(\mathbf{r}),
  \label{eq:es}
\end{equation}
where $f_i$ are occupation numbers and $\psi_i$ Kohn--Sham orbitals. The weight $W_i$ is the product of a factor $s$ fixed for each base functional, the screening $g$ of the material and the bare exchange shift $u_i$ (bare shift for short), with one form below the gap and another above it. Because $f(1-f)$ vanishes at integer occupation, a single evaluation on the converged orbitals of the base shifts each eigenvalue by $(1-2f_n)\langle\psi_n|W_n|\psi_n\rangle$ and leaves the total energy and forces unchanged (Methods). Each of the three factors can be compared with the many-body quantity that it represents.

The bare shift $u(\mathbf{r})$ estimates from the local density the exact exchange that a state lacks in the semilocal potential, that is, the exact-exchange correction $|\langle\Sigma_x - v_x\rangle|$, with $\Sigma_x$ the Fock exchange operator and $v_x$ the semilocal exchange potential (Methods). With the coefficient of its occupied-state form fitted to ten prototypical semiconductors, it reproduces this correction for the valence-band maximum (VBM) in 438 other solids with a correlation coefficient of 0.93 and a mean absolute error (MAE) of 0.54~eV, or 13\% of the mean exact-exchange correction (Fig.~\ref{fig:method}b). The screening $g$ is the inverse of the independent-particle electronic dielectric constant computed from the Kohn--Sham eigenvalues and momentum matrix elements of each material, with no fitted parameter. It agrees with the experimental $1/\varepsilon_\infty$\cite{skone2014} over almost an order of magnitude, from silicon to lithium fluoride, with a root-mean-square deviation of 6\% (Fig.~\ref{fig:method}c).

The factor $s$ is fitted for each base functional, with $g = 1$ as for any isolated molecule, to coupled-cluster (CCSD(T)) ionization potentials of 66 small molecules of the GW100 benchmark\cite{vansetten2015,krause2015} (Methods). Its transfer to solids can be tested against GW: for the band edges of 34 solids, the one-shot GW (G$_0$W$_0$) correction on PBE is proportional to the screened exchange $g\langle\Sigma_x - v_x\rangle$ with a slope of 0.68, close to $s = 0.71$ obtained from molecules alone for the same PBE base (Fig.~\ref{fig:method}d; Supplementary Section~4 relates $s$ to the quasiparticle renormalization). Fitted separately, the slopes are 0.62 for the VBM and 1.07 for the conduction-band minimum (CBM); a single factor can therefore only be a compromise: $s$ is closer to the VBM slope, and our correction underestimates the shift of the CBM (Discussion).

\begin{figure}[t]
\centering
\includegraphics[width=\textwidth]{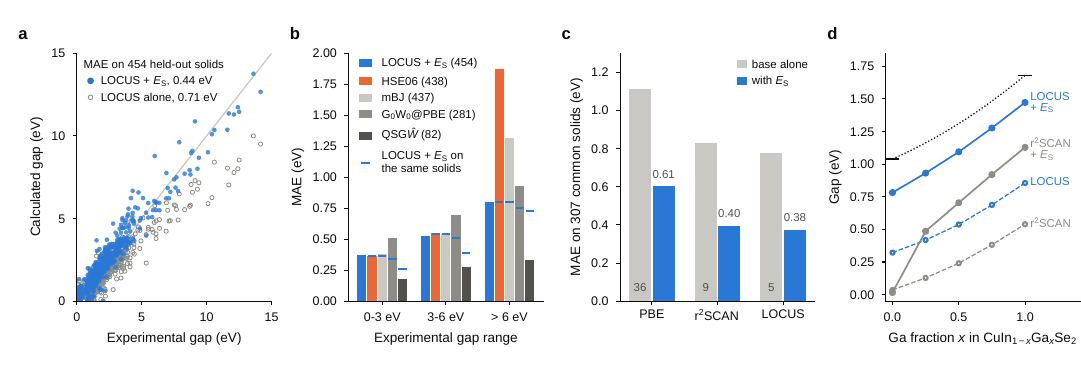}
\caption{\textbf{Band gaps of solids not used in the calibration.} \textbf{a}, Calculated against experimental gaps for 454 ordered crystals (438 of the Borlido benchmark\cite{borlido2019} and 16 added), with our correction on LOCUS (blue) and for LOCUS alone without spin--orbit coupling (open grey). \textbf{b}, MAE by range of the experimental gap for our correction on LOCUS and for HSE06, mBJ, G$_0$W$_0$@PBE and QSG$\hat{W}$ from ref.~\citenum{grossmann2026}; numbers in parentheses, solids with values for each method; blue ticks, our correction on the same solids. The GW values exist only for subsets of the solids (281 for G$_0$W$_0$@PBE and 82 for QSG$\hat{W}$). Only QSG$\hat{W}$ includes spin--orbit coupling; Supplementary Section~6 compares the other literature values shifted by the spin--orbit change of the PBE gap. \textbf{c}, MAE of the base alone (grey) and with our correction (blue) on 307 solids computed with three bases; numbers in the grey bars, solids whose corrected gap stays below half of the experimental one. \textbf{d}, Gaps of CuIn$_{1-x}$Ga$_x$Se$_2$ (special quasirandom structures of 64 atoms) on r$^2$SCAN (grey) and LOCUS (blue), for the base alone (dashed, open symbols) and with our correction (solid); black bars, experimental end members\cite{rincon2001}; dotted line, interpolation with the experimental bowing parameter.}
\label{fig:bench}
\end{figure}

\subsection*{Band gaps of 454 solids}

On 454 ordered crystals excluded from all coefficient fits, 438 from the benchmark of Borlido et al.\cite{borlido2019} and 16 added for this work, the corrected gaps, which include spin--orbit coupling, have an MAE of 0.44~eV against experiment (0.43~eV on the 438 crystals of the benchmark), compared with 0.71~eV for the base functional alone and 0.77~eV for the base functional with spin--orbit coupling (Fig.~\ref{fig:bench}a; Supplementary Table~18). Without the 36 solids of the benchmark whose experimental gap lies well below independent hybrid and GW results (low-reference flag; Methods), the MAE on the remaining 402 is 0.36~eV. No experimental gap enters any coefficient. We note, however, that the rules marked as chosen in Table~\ref{tab:coeff} were selected by inspecting this benchmark; Methods gives their effect and that of the choice of calibration solids. Without spin--orbit coupling, the MAE of the corrected gaps rises by 0.06~eV. On 50 compounds drawn at random from a compilation of experimental gaps\cite{zhuo2018} and absent from all sets used here (49 computed), the corrected gaps have an MAE of 0.28~eV, against 0.68~eV for the base functional alone, on the 34 whose reference is not below the gap of the base functional. The other 15 references are not reliable as fundamental gaps: because semilocal functionals underestimate gaps, a measured value below the base gap most likely belongs to another polymorph or composition, to defect absorption or to a compilation error (Supplementary Section~13).

The comparison with other methods depends on the magnitude of the gap. Fig.~\ref{fig:bench}b compares our correction, on the same solids, with literature values for HSE06, mBJ, one-shot G$_0$W$_0$ on PBE in the plasmon-pole approximation (G$_0$W$_0$@PBE) and vertex-corrected quasiparticle self-consistent GW with spin--orbit coupling (QSG$\hat{W}$)\cite{grossmann2026}. Below 3~eV our correction matches HSE06 and mBJ (0.36~eV, compared with 0.36 and 0.38~eV), and between 3 and 6~eV all three have MAEs of about 0.54~eV. Above 6~eV, HSE06 and mBJ underestimate the gaps, with MAEs of 1.9 and 1.3~eV, whereas our correction has an MAE of 0.80~eV; on the systems with QSG$\hat{W}$ data its MAE is more than twice that of QSG$\hat{W}$ (0.72 against 0.33~eV). Over the 281, 132 and 82 solids with values of G$_0$W$_0$@PBE, full-frequency quasiparticle G$_0$W$_0$ with spin--orbit coupling (QPG$_0$W$_0$) and QSG$\hat{W}$, the MAE of our correction is 0.42~eV in each case, against 0.60, 0.31 and 0.25~eV. Our correction is therefore more accurate than G$_0$W$_0$@PBE but 0.1--0.2~eV less accurate than full-frequency and self-consistent GW; Supplementary Table~15 compares all methods on the 81 solids with values of each.

The choice of base functional matters most through whether its own Kohn--Sham gap opens: our correction cannot recover a gap that the base closes, because a closed gap makes the screening metallic and $g$ small. Of the 307 solids computed with three bases, the corrected gap remains below half of the experimental one for 36 with PBE, 9 with r$^2$SCAN and 5 with LOCUS (Fig.~\ref{fig:bench}c), most of them copper and silver d$^{10}$ compounds and narrow-gap pnictides (Supplementary Section~9); on all 307 solids the MAEs are 0.61, 0.40 and 0.38~eV. LOCUS was constructed to remove such qualitative failures of r$^2$SCAN for closed d shells, and its average error differs little from that of r$^2$SCAN. In CuInSe$_2$ (Fig.~\ref{fig:bench}d), the Kohn--Sham gap on r$^2$SCAN is only 0.04~eV and the corrected gap 0.02~eV, against 1.04~eV in experiment; the shell term of LOCUS places the copper 3d band deeper and opens the Kohn--Sham gap to 0.32~eV, after which our correction gives 0.78~eV, and over the series CuIn$_{1-x}$Ga$_x$Se$_2$ LOCUS with our correction follows the experimental composition dependence 0.2~eV below it, with a bowing of 0.13~eV, compared with 0.20~eV in experiment\cite{rincon2001}.

\begin{figure}[t]
\centering
\includegraphics[width=\textwidth]{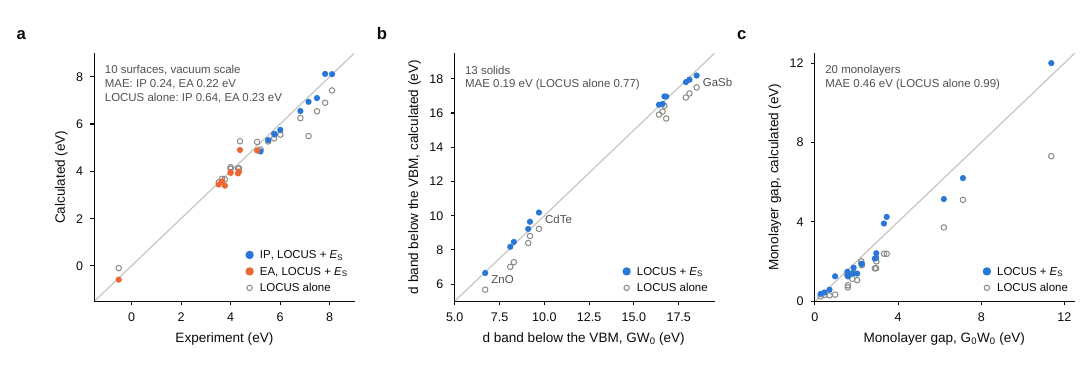}
\caption{\textbf{Absolute band edges, d bands and monolayers.} \textbf{a}, Ionization potentials (IP, blue) and electron affinities (EA, orange) of ten nonpolar surfaces on the vacuum scale against experiment. \textbf{b}, Depth of the d-band centre below the VBM against GW$_0$@PBE\cite{shishkin2007}. \textbf{c}, Gaps of 20 monolayers against G$_0$W$_0$ from C2DB\cite{haastrup2018}. Open symbols, LOCUS alone without spin--orbit coupling; grey diagonals, $y = x$.}
\label{fig:levels}
\end{figure}

\subsection*{Absolute band edges and d levels}

Because our correction leaves the density and the electrostatic potential of the base functional unchanged, the corrected bulk band edges can be placed on the vacuum scale with the bulk--slab alignment of the base (Methods). For ten nonpolar surfaces, the ionization potentials (IP) have an MAE of 0.24~eV, compared with 0.64~eV for the base functional alone (Fig.~\ref{fig:levels}a); on the nine surfaces with GW values, GW with vertex corrections has an MAE of 0.17~eV\cite{hinuma2014,chen2014}, against 0.27~eV for our correction. The electron affinities (EA) have an MAE of 0.22~eV, close to that of the base functional (0.23~eV), because on these surfaces the gap error of the base lies mostly in its VBM.

The same weight, with a separate coefficient for localized d states, shifts the d bands. For 13 semiconductors, the shell term of LOCUS already lowers the closed d shells by 0.8--2.2~eV relative to r$^2$SCAN + rVV10 and reduces the error against partially self-consistent GW (GW$_0$)\cite{shishkin2007} from 2.1 to 0.8~eV; with our correction, their d-band positions have an MAE of 0.19~eV (Fig.~\ref{fig:levels}b; Supplementary Table~21). Six of the ten surfaces and eight of these 13 solids are calibration solids; Methods gives the IP and d-band statistics without them. For 20 monolayers of the Computational 2D Materials Database (C2DB)\cite{haastrup2018}, the gaps have an MAE of 0.46~eV against G$_0$W$_0$, compared with 0.99~eV for the base (Fig.~\ref{fig:levels}c). Isolated molecules are the limit $g = 1$ of our correction: for 34 molecules of GW100 that were not used to fit $s$, the corrected ionization potentials have an MAE of 0.40~eV against CCSD(T)\cite{krause2015}, compared with 0.30~eV for G$_0$W$_0$@PBE\cite{vansetten2015} and 3.1~eV for the base alone (Supplementary Section~5). For the transport gaps of four aromatic hydrocarbon crystals and solid C$_{60}$\cite{rangel2016,lof1992}, the MAE is 0.33~eV (Supplementary Table~27).

\begin{figure}[t]
\centering
\includegraphics[width=\textwidth]{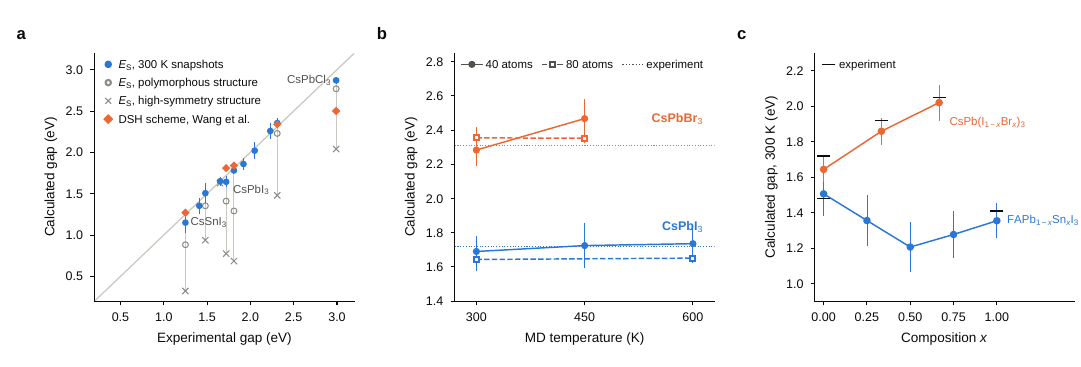}
\caption{\textbf{Halide perovskites at finite temperature, with spin--orbit coupling.} Calculated gaps are LOCUS + $\ES$ with spin--orbit coupling. \textbf{a}, Gaps averaged over molecular-dynamics snapshots at 300~K (bars, standard deviation over snapshots) in the largest cell of up to 96 atoms, gaps of the high-symmetry crystal structures (crosses) and of polymorphous structures relaxed at fixed cell from the last 300~K snapshot (open circles), and gaps from the scheme of ref.~\citenum{wang2022} based on the doubly screened hybrid (DSH) functional, a dielectric-dependent range-separated hybrid, for the five compositions it covers, against experiment\cite{wang2022,sutton2016,lau2016,eperon2014,koh2015,dar2016}. \textbf{b}, Gap against the temperature of the molecular dynamics for CsPbI$_3$ and CsPbBr$_3$ in cells of 40 and 80 atoms. \textbf{c}, Composition series of FAPb$_{1-x}$Sn$_x$I$_3$ and CsPb(I$_{1-x}$Br$_x$)$_3$ at 300~K\cite{wang2022,sutton2016,lau2016,eperon2014,koh2015}.}
\label{fig:pvk}
\end{figure}

\subsection*{Halide perovskites at finite temperature}

Halide perovskites illustrate why the cost barrier matters. With spin--orbit and thermal corrections added, PBE and HSE06 have MAEs of 1.5 and 0.8~eV on the inorganic metal halides of ref.~\citenum{wang2022}, hybrid functionals underestimate the gaps even in symmetry-broken structures\cite{sabino2024}, and on molecular-dynamics structures a hybrid reproduces the room-temperature gap of CsPbBr$_3$ only with a fraction of exact exchange fitted to that gap\cite{zhao2025}. Because dielectric-dependent hybrids and GW are too expensive for the many large snapshots that finite temperatures require, a recent predictive scheme applies a dielectric-dependent hybrid\cite{cui2018} only to the experimental crystal structure and adds spin--orbit, zero-point and thermal corrections from PBE, reaching an MAE of 0.17~eV; its harmonic thermal correction excludes the cubic phases\cite{wang2022}. Our correction, which adds only one step to a semilocal calculation, can instead be applied directly to snapshots of molecular dynamics with spin--orbit coupling included, in cubic phases at 300--600~K and at mixed compositions alike, with no parameter fitted to any gap.

For 11 compositions of lead and tin halides with caesium, methylammonium and formamidinium, including mixed halides, the gaps averaged over snapshots at 300~K have an MAE of 0.05~eV against experiment\cite{wang2022,sutton2016,lau2016,eperon2014,koh2015,dar2016} (Fig.~\ref{fig:pvk}a), within the uncertainty of the protocol, 0.1--0.3~eV, which comes mainly from the cell size (Fig.~\ref{fig:pvk}b; Supplementary Section~10). The base functional without spin--orbit coupling reaches an MAE of 0.31~eV on the same snapshots only by a cancellation of errors: with spin--orbit coupling it underestimates every gap, by 0.87~eV on average. Applied to the high-symmetry cubic structures instead, our correction gives gaps 0.5--1.1~eV below the measured room-temperature values (Fig.~\ref{fig:pvk}a): the gap at room temperature is a property of the thermally disordered structure. The same protocol follows the composition dependence of mixed-halide and lead--tin perovskites, including the bowing of FAPb$_{1-x}$Sn$_x$I$_3$ with a minimum near $x = 0.5$ (Fig.~\ref{fig:pvk}c), as observed\cite{prasanna2017}.

Hybrid functionals are far more expensive for such cells. For a 40-atom snapshot of CsPbI$_3$, the LOCUS ground state and our correction together took 59~min on one node, whereas HSE06 had not completed its first exchange-operator build after 63~min (Supplementary Section~12).

\begin{figure}[t]
\centering
\includegraphics[width=\textwidth]{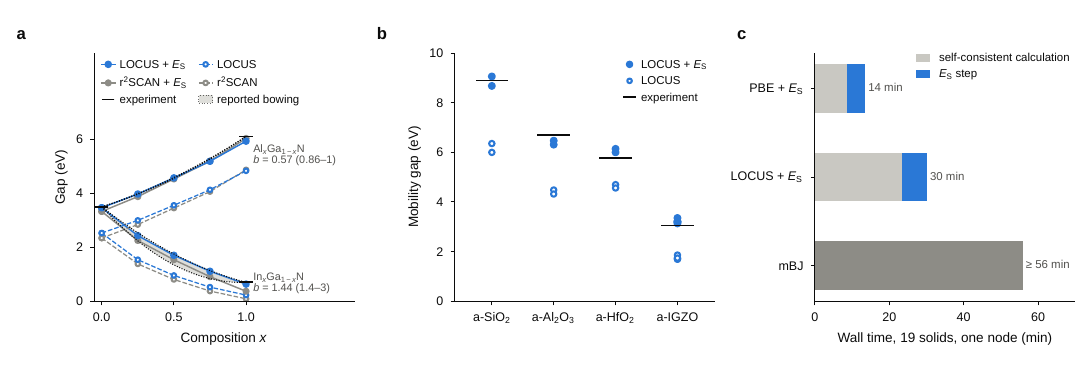}
\caption{\textbf{Alloys, amorphous oxides and cost.} \textbf{a}, Gaps of In$_x$Ga$_{1-x}$N and Al$_x$Ga$_{1-x}$N modelled by special quasirandom structures\cite{zunger1990} and averaged over structures at each composition, on LOCUS (blue) and r$^2$SCAN (grey), with $\ES$ (filled symbols, solid lines) and for the base alone (open symbols, dashed lines); horizontal bars mark the experimental end members, and the shaded band between dotted curves spans the reported bowing parameters\cite{wu2002,alam2020}, a band narrower than the line width for Al$_x$Ga$_{1-x}$N. Labels give the bowing parameter $b$ fitted to the LOCUS~+~$\ES$ series and the reported range (eV). \textbf{b}, Mobility gaps of melt-quenched amorphous oxides with and without $\ES$ against experiment\cite{distefano1971,ding2014,fung2008}. \textbf{c}, Wall time summed over 19 solids of up to 14 atoms (Supplementary Table~33), computed with GPAW (plane waves of 600~eV, the $k$-point meshes of the benchmark) with every step of a solid on one node (16 cores of an Intel Xeon Gold 6254, 16 Message Passing Interface (MPI) processes): the self-consistent calculation of each base, including its PBE start, and the $\ES$ step with empty bands up to twice the occupied ones; mBJ for comparison is a lower bound, as five of its calculations stopped unconverged after 300 iterations.}
\label{fig:alloys}
\end{figure}

\subsection*{Alloys, disordered systems and cost}

The approach applies equally to systems that need large cells or many configurations. For special quasirandom structures\cite{zunger1990} of In$_x$Ga$_{1-x}$N and Al$_x$Ga$_{1-x}$N, the corrected gaps of the end members InN, GaN and AlN lie within 0.17~eV of experiment, whereas LOCUS and r$^2$SCAN alone place them 0.47--1.27~eV below it, and the bowing parameters fitted to the calculated series, 1.44~eV for InGaN and 0.57~eV for AlGaN, compare with reported ranges of 1.4--3~eV and 0.86--1.00~eV\cite{wu2002,alam2020}; the low value for AlGaN comes mainly from the calculated AlN gap, 0.17~eV below experiment (Fig.~\ref{fig:alloys}a; Supplementary Table~23). For amorphous SiO$_2$, Al$_2$O$_3$, HfO$_2$ and InGaZnO$_4$, the mobility gaps, averaged over melt-quenched cells, lie within 0.32~eV of experiment\cite{distefano1971,ding2014,fung2008}, whereas those of LOCUS alone lie 1.1--2.7~eV below it (Fig.~\ref{fig:alloys}b; Methods; Supplementary Tables~24, 26 and 25).

The cost of our correction is that of one step after the self-consistent calculation, spent mostly on the additional empty bands needed for the screening. With LOCUS, this step adds a median of 21\% to the time of the self-consistent calculation over the benchmark calculations (59\% with the faster PBE base; Supplementary Table~34) and 39\% for 19 solids of up to 14 atoms timed on one node (Fig.~\ref{fig:alloys}c). For these 19 solids, with our correction, PBE, r$^2$SCAN + rVV10 and LOCUS take 1.6, 2.4 and 3.1 times the time of PBE alone (medians over the solids; 14, 26 and 30~min in total), against at least 56~min for mBJ alone (Supplementary Table~33).

\section*{Discussion}

A single evaluation after a semilocal self-consistent calculation, with global coefficients fitted to theoretical data, gives band gaps more accurate than those of HSE06 (Supplementary Table~15), with an MAE 0.1--0.2~eV above that of full-frequency and self-consistent GW, and places band edges on the vacuum scale and d bands close to GW, also in structures for which hybrid functionals and GW become impractical.

The accuracy does not result from fitting to gaps. Each factor follows the many-body quantity that it represents (Fig.~\ref{fig:method} and Methods). Replacing the screening or the bare shifts of each material by their means over the calibration solids shows what each contributes (Supplementary Table~4): below 3~eV even a rigid shift of each edge (both factors replaced) is about as accurate as the full form (0.37~eV), whereas above 6~eV the MAE rises from 0.80~eV to 2.19~eV with a constant screening and to 1.30~eV with constant bare shifts. The base functional contributes as well: the CBM shift of our correction is about half of that of G$_0$W$_0$, and on LOCUS the opening of the generalized Kohn--Sham gap of the meta-GGA base\cite{perdew2017} supplies most of the difference; on PBE it does not, and the corrected gaps are underestimated on average (mean error $-0.45$~eV). This explains why our correction approaches GW accuracy on LOCUS but only HSE06 accuracy on PBE (MAE 0.53~eV where the PBE gap opens; Supplementary Section~9).

Our correction shares with orbital-scaling corrections\cite{zheng2011,williams2026} and Koopmans-compliant functionals\cite{dabo2010,nguyen2018} the occupation form $f(1-f)$, but its weight is local, so that it acts on Bloch states without localizing orbitals, and unlike dielectric-dependent hybrids it needs no separate dielectric calculation, because its screening comes from the Kohn--Sham spectrum. The shell term addresses, for filled d shells, the error that DFT+U corrects, with one rule and one strength for all elements and no projectors on chosen sites; on 542 experimental formation enthalpies it lowers the error of the 3d oxides from 0.089 to 0.066~eV per atom (Supplementary Table~9), but, acting on the density alone, it cannot open the gaps of Mott--Hubbard insulators such as CoO and FeO.

In practice, PBE is the least expensive base where its gap opens, and LOCUS is preferable for copper and silver d$^{10}$ compounds, for narrow-gap semiconductors and wherever absolute levels or d bands are needed.

An important limitation concerns wide-gap insulators, where the MAE is more than twice that of QSG$\hat{W}$. The largest errors are in the rare-gas solids and the alkali halides, where the bare shift underestimates the exchange correction of the diffuse CBM by about 30\%, the same deficit that overestimates the electron affinities of bound anions by 1.0~eV on average (Supplementary Section~5). A second limitation is the d$^0$ oxides such as TiO$_2$ and SrTiO$_3$, whose corrected gaps are overestimated by 0.46~eV on average (13 oxides without the low-reference flag), compared with 0.17~eV for HSE06; G$_0$W$_0$ (Supplementary Section~4) shows that this error comes from the VBM, an oxygen 2p state that, next to empty cation d states, is screened much more strongly than the global $g$ implies. Both errors call for GW-based bare shifts and screening of diffuse conduction bands and of such anion p states. Open-shell transition-metal oxides are only partly described (Supplementary Section~11). The shell term does not act on the 5d and 4f shells, whose levels are left to our correction alone; extending it to these shells is left to future work. Defect formation energies and charge-transition levels, which are differences of total energies, remain those of the base functional. Finally, the screening is a single number for a calculation and so misplaces band offsets at interfaces between materials of different dielectric constants; a local screening $g(\mathbf{r})$ that gives each material the $g$ of its bulk, without a new parameter, keeps the offsets within 0.03~eV of the alignment of the corrected bulk edges (Supplementary Section~7).

Our correction thus extends accurate band-edge calculations from small static cells to the structures in which materials are used: combined with machine-learned interatomic potentials, which generated the structures here, it can follow how band edges change with temperature, composition and disorder, and with no parameter for each material it can screen materials databases by gaps and band-edge positions. Wherever the base functional opens the gap, band edges approaching GW accuracy come within reach of the calculations already run with semilocal functionals.

\section*{Methods}

\subsection*{The spectral term}
The bare shift $u$ of equation~\eqref{eq:es} is
\begin{equation}
  u_\mathrm{V}(\mathbf{r}) = k_V\, k_F\, e^{-\alpha}\, G(s_r), \qquad
  u_\mathrm{C}(\mathbf{r}) = k_C\, k_F\, G(s_r)
  \label{eq:bare}
\end{equation}
for occupied (V) and empty (C) states, where $k_F = (3\pi^2 n)^{1/3}$ is the local Fermi wavevector, $\alpha$ is the iso-orbital indicator of meta-GGAs and $G(s_r) = 1 + c_s s_r^2/(1 + s_r^2/s_c^2)$ is a saturating function of the reduced gradient $s_r$ ($c_s = 3$, $s_c = 2$); occupied states with localized d character use $k_d u_\mathrm{V}$. The bare shift has the dimension of an energy, and its expectation value over a state approximates the magnitude of the exact-exchange correction $|\langle\Sigma_x - v_x\rangle|$ of that state (Fig.~\ref{fig:method}b). $\ES$ is evaluated once on the converged calculation of the base functional, which it leaves unchanged at integer occupation, and shifts each eigenvalue by exactly $(1 - 2f_n)\langle\psi_n|W_n|\psi_n\rangle$: because $f(1-f)$ vanishes at integer occupation, $\ES$ and its derivatives with respect to the orbitals and the density vanish for a calculation with integer occupations, and only the derivative with respect to the occupation remains. The occupation form $f(1-f)$ is that of corrections that restore the piecewise linearity of the energy\cite{zheng2011,dabo2010}; here it defines level shifts that leave the energies at integer occupation unchanged, and we do not claim that the corrected energy is piecewise linear between integers. For the band edges of a calculation at integer occupation, including each molecular-dynamics snapshot, the eigenvalue shift gives
\begin{equation}
  \epsilon'_n = \epsilon_n + s\,g\,\hat{D}_n + \Delta^\mathrm{SOC}_n, \qquad
  \hat{D}_\mathrm{VBM} = -\langle u_\mathrm{V} \rangle_n, \qquad
  \hat{D}_\mathrm{CBM} = +\langle u_\mathrm{C} \rangle_n,
\end{equation}
where $\langle\cdot\rangle_n$ is the expectation value over $|\psi_n|^2$ including the projector augmented-wave (PAW) spheres and $\Delta^\mathrm{SOC}_n$ is the shift from spin--orbit coupling (see below). For calculations with fractional occupations (metals, Fermi smearing), $\ES$ is not added to the energy, and only the level shifts are evaluated, with $\hat{D}_n = -f_n \langle u_\mathrm{V} \rangle_n + (1 - f_n) \langle u_\mathrm{C} \rangle_n$. The part of the gap opening that the base functional itself already provides is not subtracted from these shifts. For an s or p band edge localized on one atom, identified from the ratio of its on-site interaction to its bandwidth (above 2) and from its effective mass (above 2.5), the global $g$ is replaced by the screening averaged over the atomic-shell form factor of the state (Supplementary Section~2 gives the definitions and thresholds).

The forms of the bare shifts follow the physics of each edge. The bare shift of an occupied state is motivated by the removal of its own self-interaction, which is largest where one orbital dominates the density; empirically, the factor $e^{-\alpha}$, which emphasizes such regions, lowers the error of the VBM bare shift against the exact-exchange data from 20\% to 13\%. An empty state has no self-interaction, and including $e^{-\alpha}$, a property of the occupied density, in the CBM bare shift worsens its agreement with the exact-exchange data. The exact-exchange correction of the CBM varies little between bulk solids; for half of them it lies between 1.7 and 2.6~eV. Accordingly, a constant CBM bare shift gives the same accuracy as the local one for bulk solids and monolayers (Supplementary Table~3). The local form is nevertheless retained because it makes our correction local: it then vanishes in vacuum and follows the state at surfaces, at interfaces and in molecules, which a constant cannot do.

\subsection*{Screening}
The screening is $g = 1/(\varepsilon_\mathrm{KS} + \varepsilon_\mathrm{TF})$. Here $\varepsilon_\mathrm{KS}$ is the independent-particle electronic dielectric constant without local fields, a sum over Kohn--Sham transitions with momentum matrix elements (Supplementary Section~2):
\begin{equation}
  \varepsilon_\mathrm{KS} = 1 + \frac{8\pi}{\Omega} \sum_{\mathbf{k}} w_\mathbf{k} \sum_{n<m} f_{n\mathbf{k}}(1 - f_{m\mathbf{k}})\, \frac{2\,|\langle\psi_{m\mathbf{k}}|\hat{\mathbf{p}}|\psi_{n\mathbf{k}}\rangle|^2}{3\,\omega_{nm\mathbf{k}}^3}\, R_{nm\mathbf{k}},
\end{equation}
in atomic units, with the cell volume $\Omega$, $k$-point weights $w_\mathbf{k}$, transition energies $\omega_{nm\mathbf{k}}$ and a factor $R$ that averages the integrand over the $k$-point cell around direct-gap minima. The sum is extrapolated in the number of empty bands, and transition energies below 0.1~eV are set to 0.1~eV. For partial occupations the factor $f_n(1 - f_m)$ differs from the $f_n - f_m$ of linear response; it keeps $\varepsilon_\mathrm{KS}$ continuous through band crossings, and the intraband screening of metals is carried by $\varepsilon_\mathrm{TF} = 4\pi D_F L^2$, a Thomas--Fermi term with the density of states $D_F$ at the Fermi level (0.05~eV broadening) and $L = 6$~bohr, which dominates $g$ in metals. For isolated systems, a smooth switch based on the variation of the transition energies over the $k$-points sets $g$ to 1; a cell sampled at a single $k$-point is treated as isolated only when it has a vacuum gap of at least 5~\AA{} along every periodic axis. For monolayers, the cell-averaged dielectric constant depends on the vacuum thickness, and an effective dielectric constant of the layer, for the adopted thickness $d$, is obtained as $\varepsilon_s = 1 + (\varepsilon_\mathrm{cell} - 1)L_c/d$, with $L_c$ the cell length normal to the layer and $d$ the atomic thickness plus 3.1~\AA, the interlayer spacing of layered parent compounds. None of these quantities is fitted.

\subsection*{Spin--orbit coupling}
Spin--orbit coupling is added in second variation within a window of bands on each side of the gap: 8 bands in the cells of the benchmark and, in supercells (molecular-dynamics snapshots, amorphous cells and special quasirandom structures), as many bands as the cell has atoms, which keeps the energy range of the window when the bands fold. A window is used because the spin--orbit shifts of lead halides do not converge with the number of empty bands in second variation (Supplementary Section~10). Spin--orbit coupling is omitted for elements whose setups carry f projectors, such as La and Ce. The screening and the bare shifts are computed from the scalar-relativistic states, and the spin--orbit shifts $\Delta^\mathrm{SOC}_n$ are obtained independently and added; diagonalizing the spin--orbit coupling on the corrected levels instead changes the gap of CsPbI$_3$ by less than 0.01~eV (Supplementary Section~10).

\subsection*{Calibration}
The three bare-shift coefficients were fitted by least squares through the origin to the exact-exchange corrections $\langle\Sigma_x - v_x^\mathrm{PBE}\rangle$ of PBE states of ten semiconductors (Si, Ge, GaAs, GaP, InP, InSb, GaN, ZnO, ZnSe and CdTe). These were chosen as prototypical group-IV, III--V and II--VI semiconductors that span experimental gaps from 0.24 to 3.5~eV and include the semicore d states of Ga, In, Zn and Cd needed for $k_d$; the choice was not optimized, and calibrations on randomly drawn solids give nearly the same accuracy (Supplementary Section~14). The fit gives $k_V = 2.897$~eV\,bohr from the VBM, $k_C = 0.651$~eV\,bohr from the CBM and the dimensionless $k_d = 1.493$ from the semicore d states. The factor $s$ was fitted for each base, with $g = 1$, to the CCSD(T)/def2-TZVPP ionization potentials of the GW100 molecules that contain only elements from H to Ar and at most 12 atoms (66 molecules)\cite{vansetten2015,krause2015}; the other 34 molecules, which contain heavier elements (K to Xe, including Ti, Cu and Ag) or more atoms, form the molecular test set. The molecular levels of the base functionals were computed with PySCF\cite{sun2020}. This gives $s = 0.658$ for LOCUS, 0.708 for PBE and 0.652 for r$^2$SCAN. No experimental gap was used. Removing any of the rules chosen by inspecting the benchmark (Table~\ref{tab:coeff}) changes the benchmark MAE by at most 0.02~eV, and when the ten calibration solids are drawn at random, the 95th percentile of the MAE lies 0.04~eV above the value of the adopted calibration (Supplementary Sections~1 and 14). The ten calibration solids and the 66 molecules are excluded from the statistics of gaps and molecular levels. Six of the ten surfaces of Fig.~\ref{fig:levels}a (CdTe, GaAs, GaP, InP, ZnO and ZnSe) are of calibration solids, whose bulk band edges entered the fits of $k_V$ and $k_C$; the IP MAE is 0.24~eV on these six and 0.25~eV on the other four. Eight of the 13 solids of the d-band comparison (Fig.~\ref{fig:levels}b) are calibration solids, whose d states entered the fit of $k_d$ through their exact-exchange corrections; no GW or measured d-band energy entered any fit. On the other five (CdS, CdSe, GaSb, InAs and ZnS) the d-band MAE is 0.21~eV, against 0.18~eV on the eight. The strength of the shell term was set from calculations on d$^{10}$ atoms and ions (below).

\subsection*{Shell term}
LOCUS is r$^2$SCAN\cite{furness2020} with the rVV10 nonlocal dispersion correction\cite{sabatini2013} and a shell term
\begin{equation}
  E_\mathrm{sh} = c\int \Theta(\mathbf{r})\, e_x^\mathrm{LDA}\big(n(\mathbf{r})\big)\,\mathrm{d}\mathbf{r}, \qquad c = 0.2,
  \label{eq:shell}
\end{equation}
where $e_x^\mathrm{LDA}$ is the exchange energy density of the local-density approximation (LDA) and $\Theta$ is a gate described below; the term is evaluated on the total density and therefore introduces no spin-dependent potential (Supplementary Section~3). The gate is a product of smooth factors in the density, the reduced gradient and a shell descriptor, the ratio of the density averaged over a Gaussian of width $4/k_F$ to the local density. This is a nonlocal density descriptor of the kind introduced for machine-learned functionals\cite{nagai2020}, in the form used for the CIDER features\cite{bystrom2022}. The gate depends on the density alone, not on an occupation or an oxidation state. It covers dense multi-orbital d shells, fully for the filled 3d shells of Cu to Ge and 4d shells of In to Sb, and partly for the 3d shells of Fe to Ni, closed or open, and the 4d shells of Ag and Cd; the open 3d shells of Ti to Mn lie below its density floor in the free atoms but are partly covered in oxides, where the cation shells contract. It excludes the second-row L shells and bonding regions. The filled 5d shells of Au to Bi and the 4f shells are not covered: the 5d shells have no flat maximum that the gate can enclose, and the density cap set to keep the term off the empty 4f orbitals of La to Tm, which would otherwise narrow the gaps of their oxides, also excludes the filled 4f$^{14}$ shells (Supplementary Section~3). The boundaries of the gate in the shell descriptor lie between successive shell maxima of all-electron atoms from Li to Bi; only the gradient threshold and a density cap for the N shells (principal quantum number 4) were chosen, and their sensitivity is given in Supplementary Table~7. The upper cut of the gate is a $C^2$ polynomial that makes the term and its first two derivatives vanish in the uniform electron gas. With $c = 0.2$ the term removes 15--20\% of the Koopmans error of the d shells it covers in d$^{10}$ atoms and ions\cite{dabo2010}, a screened fraction comparable to $1/\varepsilon_\infty$ of d compounds. The same $c$ is used for every shell the gate covers. On oxides of several oxidation states the term changes the energies of 15 oxidation steps between oxides of the same metal by at most 0.27~eV per metal atom (MAE 0.23~eV with the term and 0.24~eV without it on the 14 with grade A or B references) and the site moments of magnetic oxides by less than 0.08~$\mu_\mathrm{B}$ (Supplementary Section~3). The forces of LOCUS include the explicit dependence of the shell term's one-centre evaluation on the atomic positions and agree with finite differences of the energy ($4 \times 10^{-5}$~eV~\AA$^{-1}$ in NiO); in NVE molecular dynamics of NiO the total energy is conserved as well as without the term (Supplementary Section~3). The gate was checked on 27 solids with d bands, which include the 13 solids of the d-band comparison; that comparison is therefore not an independent test of the d-band positions. On 13 solids the MAE of the lattice constants against experiment corrected for zero-point anharmonic expansion\cite{hao2012} is 0.66\% with the shell term and 0.62\% without it (Supplementary Table~8), and on 542 experimental formation enthalpies\cite{wang2021} the MAE is 0.081~eV per atom, against 0.077~eV for r$^2$SCAN + rVV10 (Supplementary Table~9). The rVV10 term is included for structures and energies and has little effect on the band gaps (on 338 solids, by 0.03~eV on average for the Kohn--Sham gaps and 0.04~eV after our correction).

\subsection*{Calculations}
Solids were computed with GPAW\cite{mortensen2024} in the PAW method\cite{blochl1994} with plane waves of 600~eV and libxc\cite{lehtola2018}; the corrected gaps of solids, except the 320-atom CsPbI$_3$ cells, were evaluated on an earlier variant of LOCUS that agrees with the final form within 0.02~eV on 40 semiconductors and insulators (Supplementary Section~3). For 39 elements we used PAW setups generated with the GPAW generator during the development of the base functional; these setups are distributed with the code, and their PBE lattice constants were validated against all-electron values. The standard setups 24.11.0 were used for the other elements. For Ca and Ga, the generated setups include a 3d projector (Ca) and the 3d shell in the valence (Ga), because the Ca 3d states form the conduction band of Ca compounds and the Ga 3d band is among the d levels compared with GW. Without them, the PBE lattice constants of Ca, CaO and CaF$_2$ were underestimated by 2--4\% and that of GaN was overestimated by 0.8\%; with them, the errors are 0.3\% for Ca and 0.4\% for GaN. The meta-GGA functionals were converged from a PBE starting point; $\ES$ uses empty bands up to twice the number of occupied ones. Open-shell oxides were computed in high-spin antiferromagnetic order (type II for the rock-salt monoxides). Surfaces were symmetric relaxed slabs; IP $= [V_\mathrm{vac} - \langle V\rangle]_\mathrm{slab} + [\langle V\rangle - \epsilon_\mathrm{VBM}]_\mathrm{bulk}$ with the slab dipole of each functional\cite{hinuma2014}, and EA $=$ IP $-$ gap. Monolayers were taken from C2DB\cite{haastrup2018}. Alloys used special quasirandom structures\cite{zunger1990}: $2\times2\times2$ and $3\times3\times2$ wurtzite supercells (32 and 72 atoms) for the nitrides and $2\times2\times2$ supercells of the primitive chalcopyrite cell (64 atoms) for CuIn$_{1-x}$Ga$_x$Se$_2$. Brillouin zones were sampled with $\Gamma$-centred meshes of $n_i \approx 30$~\AA$/a_i$ points along each lattice vector $a_i$, replaced by meshes that contain the zone-boundary points for solids whose band edges lie there. Molecular dynamics and structure relaxations were carried out with the universal neural-network potential PFP (PreFerred Potential)\cite{takamoto2022}, version 9.0.0, on Matlantis\cite{matlantis}, in its r$^2$SCAN calculation mode\cite{shinagawa2026}, which matches the meta-GGA level of LOCUS. Perovskite snapshots were taken from Langevin molecular dynamics at fixed cell, in $2\times2\times2$ cubic supercells (40 atoms for the caesium and 96 for the formamidinium compounds), in 80-atom cells, for MAPbI$_3$ in the 48-atom orthorhombic cell and, for three snapshots of CsPbI$_3$, in a $4\times4\times4$ supercell (320 atoms). Time steps were 2~fs for the caesium compounds and 1~fs with a hydrogen mass of 4~u for those with methylammonium or formamidinium. Snapshots were taken every 2~ps after 6~ps of equilibration, 3--9 per composition, temperature and cell. Amorphous cells of 96--112 atoms were melt-quenched with the same potential at experimental densities: liquid at 3,000--4,000~K, quenched within 13.5~ps and relaxed to residual forces below 0.02~eV\,\AA$^{-1}$. For these cells every state near the gap was corrected, and mobility gaps are the gaps between extended states, those whose edge-uniformity index is at least 0.4.

\subsection*{References and statistics}
Experimental gaps of solids are those of Borlido et al.\cite{borlido2019}; zero-point renormalization is not removed\cite{miglio2020}. Two flags mark references that may not represent the fundamental gap. The low-reference flag (36 solids) is set when at least three of HSE06, the hybrid PBE0, mBJ and GW (Supplementary Section~6) agree to within 35\% of their median and all exceed the experiment, which lies more than 0.5~eV below their median; the exciton flag (58 solids) is set when a Wannier--Mott estimate from the band-edge masses and $\varepsilon = 1/g$ gives a binding energy of at least 0.5~eV or a radius of at most 4~\AA. The flags do not establish that an experiment is wrong; all statistics are reported for the full set, and the unflagged subsets serve as a sensitivity analysis. The references mix fundamental and optical gaps; amorphous oxides are compared through mobility gaps with optical or photoconductivity onsets, and molecular crystals through transport gaps. HSE06, mBJ, G$_0$W$_0$@PBE, QPG$_0$W$_0$ and QSG$\hat{W}$ values are from Gro{\ss}mann et al.\cite{grossmann2026}; d-band references from Shishkin and Kresse (GW$_0$)\cite{shishkin2007} and photoemission\cite{fuchs2007} (Supplementary Table~21); surface IPs from refs~\citenum{hinuma2014,chen2014}; monolayer G$_0$W$_0$ gaps from C2DB\cite{haastrup2018} as tabulated in ref.~\citenum{rauch2020}; molecular crystals from refs~\citenum{rangel2016,lof1992}; experimental $\varepsilon_\infty$ as compiled in ref.~\citenum{skone2014}; perovskite gaps from refs~\citenum{wang2022,sutton2016,lau2016,eperon2014,koh2015,dar2016}; alloy end members and bowing parameters from refs~\citenum{wu2002,alam2020,rincon2001}; amorphous oxides from refs~\citenum{distefano1971,ding2014,fung2008}; open-shell oxides from refs~\citenum{sawatzky1984,vanelp1991mno,vanelp1991coo,bowen1975,parkinson2016,kasikov2019} (Supplementary Table~32). For alloys, experimental values are used for the end members only, and intermediate compositions are compared through bowing parameters. Times in Fig.~\ref{fig:alloys}c are wall times with all steps of a solid run in sequence on the same node (16 cores of an Intel Xeon Gold 6254, 16 MPI processes), for held-out Borlido solids chosen to span sp semiconductors, d$^{10}$ compounds and oxides; mBJ was run with GPAW's implementation on the PBE setups. The time shares of the $\ES$ step over the benchmark calculations come from runs whose steps may have run on different nodes.

\begin{table}[t]
\centering
\caption{\textbf{Parameters of the spectral correction.} No coefficient was fitted to an experimental gap. The parameters of the shell term of LOCUS are listed in Supplementary Table~6.}
\label{tab:coeff}
\small
\begin{tabular}{>{\raggedright\arraybackslash}p{0.24\textwidth}>{\raggedright\arraybackslash}p{0.20\textwidth}>{\raggedright\arraybackslash}p{0.48\textwidth}}
\hline
Quantity & Value & How it was set \\
\hline
\multicolumn{3}{l}{\emph{Fitted to theory (least squares through the origin)}}\\
VBM bare-shift coefficient $k_V$ & 2.897 eV\,bohr & exact-exchange corrections of the VBM, PBE states, ten semiconductors \\
CBM bare-shift coefficient $k_C$ & 0.651 eV\,bohr & the same for the CBM \\
d-state factor $k_d$ & 1.493 & the same for the semicore d states (8 solids) \\
factor $s$ & 0.658 / 0.708 / 0.652 & CCSD(T) ionization potentials of 66 molecules, $g = 1$; LOCUS / PBE / r$^2$SCAN \\
\multicolumn{3}{l}{\emph{Fixed by construction}}\\
gradient factor & $c_s = 3$, $s_c = 2$ & saturation of $G$ in density tails \\
metallic screening & $L = 6$ bohr & Thomas--Fermi length \\
empty bands & $2N_\mathrm{occ}$ & numerical convergence \\
spin--orbit window & $\pm 8$ bands ($\pm N_\mathrm{atoms}$ in supercells) & fixed energy range; the spin--orbit shifts do not converge with the number of bands (Methods) \\
transition-energy floor; broadening of $D_F$ & 0.1~eV; 0.05~eV & numerical stability \\
isolated-system switch & dispersion-weighted $\varepsilon_\mathrm{disp}$ 1.15, width 0.03; at one $k$-point, vacuum $\geq 5$~\AA{} & between molecules (1) and solids ($\geq 1.33$; Supplementary Section~2) \\
monolayer thickness & atomic thickness + 3.1~\AA & interlayer spacing of layered compounds \\
\multicolumn{3}{l}{\emph{Rules chosen by inspecting the benchmark}}\\
form of the CBM bare shift & no $e^{-\alpha}$ & selected on exact-exchange rows, checked on the benchmark \\
localized-edge screening & on-site interaction / bandwidth $> 2$, effective mass $> 2.5$ & band edges localized on one atom; selected on the benchmark and molecular crystals \\
mobility edge (amorphous cells) & uniformity $\geq 0.4$ & sensitivity in Supplementary Table~26 \\
\hline
\end{tabular}
\end{table}

\section*{Acknowledgements}
We thank Akihide Hayashi, Yu Miyazaki, Tomoya Shiota and So Takamoto for discussions, and Masahiro Takenaga for help with the release of the code, all of Preferred Networks, Inc.

\section*{Author contributions}
R.N. conceived the study, developed the method and the code, performed the calculations, analysed the results and wrote the manuscript. S.I. and C.S. took part in the discussion of the results and in writing the manuscript.

\section*{Competing interests}
R.N., S.I. and C.S. are employees of Preferred Networks, Inc. Preferred Networks, Inc. has filed a US provisional patent application covering the spectral correction $\ES$ and the shell term of LOCUS described in this work (application no.~64/197,285, filed 12 November 2026; inventor R.N.); the application is pending.

\section*{Data availability}
The structures, $k$-point meshes and calculation settings of every system are available in the \texttt{data} directory of \url{https://github.com/pfnet-research/locus-es}; systems taken unchanged from databases are listed there by their identifiers. The per-system results are given in the Supplementary Tables.

\section*{Code availability}
The code is available at \url{https://github.com/pfnet-research/locus-es}. The $\ES$ code is released as a standalone package that works on converged GPAW calculations, including the local screening $g(\mathbf{r})$ for interfaces (Supplementary Section~7), and the shell term as modules for GPAW and PySCF. A prototype exporter applies $\ES$ to PAW calculations with Quantum ESPRESSO\cite{giannozzi2017} on the PBE base.


\begin{thebibliography}{86}
\providecommand{\natexlab}[1]{#1}
\providecommand{\url}[1]{\texttt{#1}}
\expandafter\ifx\csname urlstyle\endcsname\relax
  \providecommand{\doi}[1]{doi: #1}\else
  \providecommand{\doi}{doi: \begingroup \urlstyle{rm}\Url}\fi

\bibitem[Hohenberg and Kohn(1964)]{hohenberg1964}
Hohenberg, P. \& Kohn, W.
\newblock Inhomogeneous electron gas.
\newblock \emph{Phys. Rev.} {\bf 136}, B864--B871 (1964).

\bibitem[Kohn and Sham(1965)]{kohn1965}
Kohn, W. \& Sham, L.~J.
\newblock Self-consistent equations including exchange and correlation effects.
\newblock \emph{Phys. Rev.} {\bf 140}, A1133--A1138 (1965).

\bibitem[Perdew et~al.(1996)Perdew, Burke, and Ernzerhof]{perdew1996}
Perdew, J.~P., Burke, K. \& Ernzerhof, M.
\newblock Generalized gradient approximation made simple.
\newblock \emph{Phys. Rev. Lett.} {\bf 77}, 3865--3868 (1996).

\bibitem[Perdew and Levy(1983)]{perdew1983}
Perdew, J.~P. \& Levy, M.
\newblock Physical content of the exact {Kohn}--{Sham} orbital energies: Band
  gaps and derivative discontinuities.
\newblock \emph{Phys. Rev. Lett.} {\bf 51}, 1884--1887 (1983).

\bibitem[Sham and Schlüter(1983)]{sham1983}
Sham, L.~J. \& Schlüter, M.
\newblock Density-functional theory of the energy gap.
\newblock \emph{Phys. Rev. Lett.} {\bf 51}, 1888--1891 (1983).

\bibitem[Mori-Sánchez et~al.(2008)Mori-Sánchez, Cohen, and
  Yang]{morisanchez2008}
Mori-Sánchez, P., Cohen, A.~J. \& Yang, W.
\newblock Localization and delocalization errors in density functional theory
  and implications for band-gap prediction.
\newblock \emph{Phys. Rev. Lett.} {\bf 100}, 146401 (2008).

\bibitem[Heyd et~al.(2003)Heyd, Scuseria, and Ernzerhof]{heyd2003}
Heyd, J., Scuseria, G.~E. \& Ernzerhof, M.
\newblock Hybrid functionals based on a screened {Coulomb} potential.
\newblock \emph{J. Chem. Phys.} {\bf 118}, 8207--8215 (2003).

\bibitem[Krukau et~al.(2006)Krukau, Vydrov, Izmaylov, and Scuseria]{krukau2006}
Krukau, A.~V., Vydrov, O.~A., Izmaylov, A.~F. \& Scuseria, G.~E.
\newblock Influence of the exchange screening parameter on the performance of
  screened hybrid functionals.
\newblock \emph{J. Chem. Phys.} {\bf 125}, 224106 (2006).

\bibitem[Hedin(1965)]{hedin1965}
Hedin, L.
\newblock New method for calculating the one-particle {Green}'s function with
  application to the electron-gas problem.
\newblock \emph{Phys. Rev.} {\bf 139}, A796--A823 (1965).

\bibitem[Hybertsen and Louie(1986)]{hybertsen1986}
Hybertsen, M.~S. \& Louie, S.~G.
\newblock Electron correlation in semiconductors and insulators: Band gaps and
  quasiparticle energies.
\newblock \emph{Phys. Rev. B} {\bf 34}, 5390--5413 (1986).

\bibitem[Venkatanarayanan et~al.(2025)Venkatanarayanan, Slama, Mukherjee,
  Vezzosi, Rothlisberger, and Carnevali]{venkatanarayanan2025}
Venkatanarayanan, M. et~al.
\newblock Coupled structural and electronic requirements in
  {$\alpha$-FASnI$_3$} imposed by the {Sn(II)} lone pair.
\newblock Preprint at https://arxiv.org/abs/2511.21254 (2025).

\bibitem[Kyrtsos et~al.(2019)Kyrtsos, Matsubara, and Bellotti]{kyrtsos2019}
Kyrtsos, A., Matsubara, M. \& Bellotti, E.
\newblock First-principles study of the impact of the atomic configuration on
  the electronic properties of {Al$_x$Ga$_{1-x}$N} alloys.
\newblock \emph{Phys. Rev. B} {\bf 99}, 035201 (2019).

\bibitem[Wang et~al.(2022)Wang, Tal, Bischoff, Gono, and Pasquarello]{wang2022}
Wang, H., Tal, A., Bischoff, T., Gono, P. \& Pasquarello, A.
\newblock Accurate and efficient band-gap predictions for metal halide
  perovskites at finite temperature.
\newblock \emph{npj Comput. Mater.} {\bf 8}, 237 (2022).

\bibitem[Tran and Blaha(2009)]{tran2009}
Tran, F. \& Blaha, P.
\newblock Accurate band gaps of semiconductors and insulators with a semilocal
  exchange-correlation potential.
\newblock \emph{Phys. Rev. Lett.} {\bf 102}, 226401 (2009).

\bibitem[Borlido et~al.(2019)Borlido, Aull, Huran, Tran, Marques, and
  Botti]{borlido2019}
Borlido, P. et~al.
\newblock Large-scale benchmark of exchange--correlation functionals for the
  determination of electronic band gaps of solids.
\newblock \emph{J. Chem. Theory Comput.} {\bf 15}, 5069--5079 (2019).

\bibitem[Großmann et~al.(2026)Großmann, Thieme, Grunert, and
  Runge]{grossmann2026}
Großmann, M., Thieme, M., Grunert, M. \& Runge, E.
\newblock Many-body perturbation theory vs. density functional theory: a
  systematic benchmark for band gaps of solids.
\newblock \emph{npj Comput. Mater.} {\bf 12}, 25 (2026).

\bibitem[Rauch et~al.(2020)Rauch, Marques, and Botti]{rauch2020}
Rauch, T., Marques, M. A.~L. \& Botti, S.
\newblock Local modified {Becke}-{Johnson} exchange-correlation potential for
  interfaces, surfaces, and two-dimensional materials.
\newblock \emph{J. Chem. Theory Comput.} {\bf 16}, 2654--2660 (2020).

\bibitem[Anisimov et~al.(1991)Anisimov, Zaanen, and Andersen]{anisimov1991}
Anisimov, V.~I., Zaanen, J. \& Andersen, O.~K.
\newblock Band theory and {M}ott insulators: {H}ubbard {U} instead of {S}toner
  {I}.
\newblock \emph{Phys. Rev. B} {\bf 44}, 943--954 (1991).

\bibitem[Dudarev et~al.(1998)Dudarev, Botton, Savrasov, Humphreys, and
  Sutton]{dudarev1998}
Dudarev, S.~L., Botton, G.~A., Savrasov, S.~Y., Humphreys, C.~J. \& Sutton,
  A.~P.
\newblock Electron-energy-loss spectra and the structural stability of nickel
  oxide: {A}n {LSDA+U} study.
\newblock \emph{Phys. Rev. B} {\bf 57}, 1505--1509 (1998).

\bibitem[Cococcioni and de~Gironcoli(2005)]{cococcioni2005}
Cococcioni, M. \& de~Gironcoli, S.
\newblock Linear response approach to the calculation of the effective
  interaction parameters in the {LDA+U} method.
\newblock \emph{Phys. Rev. B} {\bf 71}, 035105 (2005).

\bibitem[Kulik et~al.(2006)Kulik, Cococcioni, Scherlis, and Marzari]{kulik2006}
Kulik, H.~J., Cococcioni, M., Scherlis, D.~A. \& Marzari, N.
\newblock Density functional theory in transition-metal chemistry: {A}
  self-consistent {H}ubbard {U} approach.
\newblock \emph{Phys. Rev. Lett.} {\bf 97}, 103001 (2006).

\bibitem[Wang et~al.(2006)Wang, Maxisch, and Ceder]{wang2006}
Wang, L., Maxisch, T. \& Ceder, G.
\newblock Oxidation energies of transition metal oxides within the {GGA+U}
  framework.
\newblock \emph{Phys. Rev. B} {\bf 73}, 195107 (2006).

\bibitem[Jain et~al.(2011)Jain, Hautier, Ong, Moore, Fischer, Persson, and
  Ceder]{jain2011}
Jain, A. et~al.
\newblock Formation enthalpies by mixing {GGA} and {GGA+U} calculations.
\newblock \emph{Phys. Rev. B} {\bf 84}, 045115 (2011).

\bibitem[Ferreira et~al.(2008)Ferreira, Marques, and Teles]{ferreira2008}
Ferreira, L.~G., Marques, M. \& Teles, L.~K.
\newblock Approximation to density functional theory for the calculation of
  band gaps of semiconductors.
\newblock \emph{Phys. Rev. B} {\bf 78}, 125116 (2008).

\bibitem[Kuisma et~al.(2010)Kuisma, Ojanen, Enkovaara, and Rantala]{kuisma2010}
Kuisma, M., Ojanen, J., Enkovaara, J. \& Rantala, T.~T.
\newblock {Kohn}--{Sham} potential with discontinuity for band gap materials.
\newblock \emph{Phys. Rev. B} {\bf 82}, 115106 (2010).

\bibitem[Tran et~al.(2018)Tran, Ehsan, and Blaha]{tran2018}
Tran, F., Ehsan, S. \& Blaha, P.
\newblock Assessment of the {GLLB-SC} potential for solid-state properties and
  attempts for improvement.
\newblock \emph{Phys. Rev. Mater.} {\bf 2}, 023802 (2018).

\bibitem[Chan and Ceder(2010)]{chan2010}
Chan, M. K.~Y. \& Ceder, G.
\newblock Efficient band gap prediction for solids.
\newblock \emph{Phys. Rev. Lett.} {\bf 105}, 196403 (2010).

\bibitem[Riemelmoser et~al.(2026)Riemelmoser, Xu, and
  Pasquarello]{riemelmoser2026}
Riemelmoser, S., Xu, X. \& Pasquarello, A.
\newblock Dielectric-dependent hybrid functional based on meta-{GGA}.
\newblock \emph{Nat. Commun.} {\bf 17}, 8323 (2026).

\bibitem[Wing et~al.(2021)Wing, Ohad, Haber, Filip, Gant, Neaton, and
  Kronik]{wing2021}
Wing, D. et~al.
\newblock Band gaps of crystalline solids from {Wannier}-localization--based
  optimal tuning of a screened range-separated hybrid functional.
\newblock \emph{Proc. Natl Acad. Sci. USA} {\bf 118}, e2104556118 (2021).

\bibitem[Varrassi et~al.(2025)Varrassi, Ellinger, Flage-Larsen, Wolloch,
  Kresse, Marzari, and Franchini]{varrassi2025}
Varrassi, L. et~al.
\newblock Automated workflow for accurate high-throughput {GW} calculations
  using plane waves.
\newblock \emph{npj Comput. Mater.} {\bf 11}, 351 (2025).

\bibitem[Nguyen et~al.(2018)Nguyen, Colonna, Ferretti, and Marzari]{nguyen2018}
Nguyen, N.~L., Colonna, N., Ferretti, A. \& Marzari, N.
\newblock {Koopmans}-compliant spectral functionals for extended systems.
\newblock \emph{Phys. Rev. X} {\bf 8}, 021051 (2018).

\bibitem[Zheng et~al.(2011)Zheng, Cohen, Mori-Sánchez, Hu, and
  Yang]{zheng2011}
Zheng, X., Cohen, A.~J., Mori-Sánchez, P., Hu, X. \& Yang, W.
\newblock Improving band gap prediction in density functional theory from
  molecules to solids.
\newblock \emph{Phys. Rev. Lett.} {\bf 107}, 026403 (2011).

\bibitem[Williams and Yang(2026)]{williams2026}
Williams, J.~Z. \& Yang, W.
\newblock Correcting delocalization error in materials with localized orbitals
  and linear-response screening.
\newblock \emph{Phys. Rev. B} {\bf 113}, 165147 (2026).

\bibitem[Fan et~al.(2026)Fan, Williams, and Yang]{fan2026}
Fan, Y., Williams, J.~Z. \& Yang, W.
\newblock {olLOSC}: Unified and efficient density functional approximation to
  correct delocalization error in molecules and periodic materials.
\newblock Preprint at https://arxiv.org/abs/2603.21906 (2026).

\bibitem[Pant et~al.(2026)Pant, Ha, Kim, and Giustino]{pant2026}
Pant, N., Ha, V.-A., Kim, D. \& Giustino, F.
\newblock Electron correlation in semiconductors and insulators via symbolic
  regression.
\newblock \emph{Phys. Rev. B} {\bf 114}, 105112 (2026).

\bibitem[Furness et~al.(2020)Furness, Kaplan, Ning, Perdew, and
  Sun]{furness2020}
Furness, J.~W., Kaplan, A.~D., Ning, J., Perdew, J.~P. \& Sun, J.
\newblock Accurate and numerically efficient {r2SCAN} meta-generalized gradient
  approximation.
\newblock \emph{J. Phys. Chem. Lett.} {\bf 11}, 8208--8215 (2020).

\bibitem[Sun et~al.(2015)Sun, Ruzsinszky, and Perdew]{sun2015}
Sun, J., Ruzsinszky, A. \& Perdew, J.~P.
\newblock Strongly constrained and appropriately normed semilocal density
  functional.
\newblock \emph{Phys. Rev. Lett.} {\bf 115}, 036402 (2015).

\bibitem[Sabatini et~al.(2013)Sabatini, Gorni, and de~Gironcoli]{sabatini2013}
Sabatini, R., Gorni, T. \& de~Gironcoli, S.
\newblock Nonlocal van der {Waals} density functional made simple and
  efficient.
\newblock \emph{Phys. Rev. B} {\bf 87}, 041108 (2013).

\bibitem[Skone et~al.(2014)Skone, Govoni, and Galli]{skone2014}
Skone, J.~H., Govoni, M. \& Galli, G.
\newblock Self-consistent hybrid functional for condensed systems.
\newblock \emph{Phys. Rev. B} {\bf 89}, 195112 (2014).

\bibitem[van Setten et~al.(2015)van Setten, Caruso, Sharifzadeh, Ren,
  Scheffler, Liu, Lischner, Lin, Deslippe, Louie, Yang, Weigend, Neaton, Evers,
  and Rinke]{vansetten2015}
van Setten, M.~J. et~al.
\newblock {GW100}: Benchmarking {G$_0$W$_0$} for molecular systems.
\newblock \emph{J. Chem. Theory Comput.} {\bf 11}, 5665--5687 (2015).

\bibitem[Krause et~al.(2015)Krause, Harding, and Klopper]{krause2015}
Krause, K., Harding, M.~E. \& Klopper, W.
\newblock Coupled-cluster reference values for the {GW27} and {GW100} test sets
  for the assessment of {GW} methods.
\newblock \emph{Mol. Phys.} {\bf 113}, 1952--1960 (2015).

\bibitem[Durante~Rincón et~al.(2001)Durante~Rincón, Hernández, Alonso,
  Garriga, Wasim, Rincón, and León]{rincon2001}
Durante~Rincón, C. et~al.
\newblock Optical transitions near the band edge in bulk
  {CuIn$_x$Ga$_{1-x}$Se$_2$} from ellipsometric measurements.
\newblock \emph{Mater. Chem. Phys.} {\bf 70}, 300--304 (2001).

\bibitem[Zhuo et~al.(2018)Zhuo, Mansouri~Tehrani, and Brgoch]{zhuo2018}
Zhuo, Y., Mansouri~Tehrani, A. \& Brgoch, J.
\newblock Predicting the band gaps of inorganic solids by machine learning.
\newblock \emph{J. Phys. Chem. Lett.} {\bf 9}, 1668--1673 (2018).

\bibitem[Shishkin and Kresse(2007)]{shishkin2007}
Shishkin, M. \& Kresse, G.
\newblock Self-consistent {GW} calculations for semiconductors and insulators.
\newblock \emph{Phys. Rev. B} {\bf 75}, 235102 (2007).

\bibitem[Haastrup et~al.(2018)Haastrup, Strange, Pandey, Deilmann, Schmidt,
  Hinsche, Gjerding, Torelli, Larsen, Riis-Jensen, Gath, Jacobsen,
  Jørgen~Mortensen, Olsen, and Thygesen]{haastrup2018}
Haastrup, S. et~al.
\newblock The {Computational 2D Materials Database}: high-throughput modeling
  and discovery of atomically thin crystals.
\newblock \emph{2D Mater.} {\bf 5}, 042002 (2018).

\bibitem[Hinuma et~al.(2014)Hinuma, Grüneis, Kresse, and Oba]{hinuma2014}
Hinuma, Y., Grüneis, A., Kresse, G. \& Oba, F.
\newblock Band alignment of semiconductors from density-functional theory and
  many-body perturbation theory.
\newblock \emph{Phys. Rev. B} {\bf 90}, 155405 (2014).

\bibitem[Chen and Pasquarello(2014)]{chen2014}
Chen, W. \& Pasquarello, A.
\newblock Band-edge positions in {GW}: Effects of starting point and
  self-consistency.
\newblock \emph{Phys. Rev. B} {\bf 90}, 165133 (2014).

\bibitem[Rangel et~al.(2016)Rangel, Berland, Sharifzadeh, Brown-Altvater, Lee,
  Hyldgaard, Kronik, and Neaton]{rangel2016}
Rangel, T. et~al.
\newblock Structural and excited-state properties of oligoacene crystals from
  first principles.
\newblock \emph{Phys. Rev. B} {\bf 93}, 115206 (2016).

\bibitem[Lof et~al.(1992)Lof, van Veenendaal, Koopmans, Jonkman, and
  Sawatzky]{lof1992}
Lof, R.~W., van Veenendaal, M.~A., Koopmans, B., Jonkman, H.~T. \& Sawatzky,
  G.~A.
\newblock Band gap, excitons, and {Coulomb} interaction in solid {C$_{60}$}.
\newblock \emph{Phys. Rev. Lett.} {\bf 68}, 3924--3927 (1992).

\bibitem[Sutton et~al.(2016)Sutton, Eperon, Miranda, Parrott, Kamino, Patel,
  Hörantner, Johnston, Haghighirad, Moore, and Snaith]{sutton2016}
Sutton, R.~J. et~al.
\newblock Bandgap-tunable cesium lead halide perovskites with high thermal
  stability for efficient solar cells.
\newblock \emph{Adv. Energy Mater.} {\bf 6}, 1502458 (2016).

\bibitem[Lau et~al.(2016)Lau, Deng, Ma, Zheng, Yun, Green, Huang, and
  Ho-Baillie]{lau2016}
Lau, C. F.~J. et~al.
\newblock {CsPbIBr$_2$} perovskite solar cell by spray-assisted deposition.
\newblock \emph{ACS Energy Lett.} {\bf 1}, 573--577 (2016).

\bibitem[Eperon et~al.(2014)Eperon, Stranks, Menelaou, Johnston, Herz, and
  Snaith]{eperon2014}
Eperon, G.~E. et~al.
\newblock Formamidinium lead trihalide: a broadly tunable perovskite for
  efficient planar heterojunction solar cells.
\newblock \emph{Energy Environ. Sci.} {\bf 7}, 982 (2014).

\bibitem[Koh et~al.(2015)Koh, Krishnamoorthy, Yantara, Shi, Leong, Boix,
  Grimsdale, Mhaisalkar, and Mathews]{koh2015}
Koh, T.~M. et~al.
\newblock Formamidinium tin-based perovskite with low {$E_g$} for photovoltaic
  applications.
\newblock \emph{J. Mater. Chem. A} {\bf 3}, 14996--15000 (2015).

\bibitem[Dar et~al.(2016)Dar, Jacopin, Meloni, Mattoni, Arora, Boziki,
  Zakeeruddin, Rothlisberger, and Grätzel]{dar2016}
Dar, M.~I. et~al.
\newblock Origin of unusual bandgap shift and dual emission in
  organic-inorganic lead halide perovskites.
\newblock \emph{Sci. Adv.} {\bf 2}, e1601156 (2016).

\bibitem[Sabino et~al.(2024)Sabino, Zhao, Dalpian, and Zunger]{sabino2024}
Sabino, F.~P., Zhao, X.~G., Dalpian, G.~M. \& Zunger, A.
\newblock Impact of symmetry breaking and spin-orbit coupling on the band gap
  of halide perovskites.
\newblock \emph{Phys. Rev. B} {\bf 110}, 035160 (2024).

\bibitem[Zhao et~al.(2025)Zhao, Chen, Li, Li, Van~de Walle, and
  Zhang]{zhao2025}
Zhao, B.-Q. et~al.
\newblock Distinguishing thermal fluctuations from polaron formation in halide
  perovskites.
\newblock \emph{Phys. Rev. Lett.} {\bf 134}, 226402 (2025).

\bibitem[Cui et~al.(2018)Cui, Wang, Zhang, Xu, and Jiang]{cui2018}
Cui, Z.-H., Wang, Y.-C., Zhang, M.-Y., Xu, X. \& Jiang, H.
\newblock Doubly screened hybrid functional: An accurate first-principles
  approach for both narrow- and wide-gap semiconductors.
\newblock \emph{J. Phys. Chem. Lett.} {\bf 9}, 2338--2345 (2018).

\bibitem[Prasanna et~al.(2017)Prasanna, Gold-Parker, Leijtens, Conings,
  Babayigit, Boyen, Toney, and McGehee]{prasanna2017}
Prasanna, R. et~al.
\newblock Band gap tuning via lattice contraction and octahedral tilting in
  perovskite materials for photovoltaics.
\newblock \emph{J. Am. Chem. Soc.} {\bf 139}, 11117--11124 (2017).

\bibitem[Zunger et~al.(1990)Zunger, Wei, Ferreira, and Bernard]{zunger1990}
Zunger, A., Wei, S.-H., Ferreira, L.~G. \& Bernard, J.~E.
\newblock Special quasirandom structures.
\newblock \emph{Phys. Rev. Lett.} {\bf 65}, 353--356 (1990).

\bibitem[Wu et~al.(2002)Wu, Walukiewicz, Yu, Ager, Haller, Lu, and
  Schaff]{wu2002}
Wu, J. et~al.
\newblock Small band gap bowing in {In$_{1-x}$Ga$_x$N} alloys.
\newblock \emph{Appl. Phys. Lett.} {\bf 80}, 4741--4743 (2002).

\bibitem[Alam et~al.(2020)Alam, Zubialevich, Ghafary, and Parbrook]{alam2020}
Alam, S.~N., Zubialevich, V.~Z., Ghafary, B. \& Parbrook, P.~J.
\newblock Bandgap and refractive index estimates of {InAlN} and related
  nitrides across their full composition ranges.
\newblock \emph{Sci. Rep.} {\bf 10}, 16205 (2020).

\bibitem[DiStefano and Eastman(1971)]{distefano1971}
DiStefano, T. \& Eastman, D.
\newblock The band edge of amorphous {SiO$_2$} by photoinjection and
  photoconductivity measurements.
\newblock \emph{Solid State Commun.} {\bf 9}, 2259--2261 (1971).

\bibitem[Ding et~al.(2014)Ding, Friedrich, Fronk, Gordan, Zahn, Chen,
  Wei~Zhang, Cobet, and Esser]{ding2014}
Ding, L. et~al.
\newblock Correlation of band gap position with composition in high-k films.
\newblock \emph{J. Vac. Sci. Technol. B} {\bf 32}, 03D115 (2014).

\bibitem[Fung et~al.(2008)Fung, Chuang, Nomura, Shieh, Hosono, and
  Kanicki]{fung2008}
Fung, T. et~al.
\newblock Photofield-effect in amorphous {In-Ga-Zn-O} ({a-IGZO}) thin-film
  transistors.
\newblock \emph{J. Inf. Disp.} {\bf 9}, 21--29 (2008).

\bibitem[Perdew et~al.(2017)Perdew, Yang, Burke, Yang, Gross, Scheffler,
  Scuseria, Henderson, Zhang, Ruzsinszky, Peng, Sun, Trushin, and
  Görling]{perdew2017}
Perdew, J.~P. et~al.
\newblock Understanding band gaps of solids in generalized {Kohn}--{Sham}
  theory.
\newblock \emph{Proc. Natl Acad. Sci. USA} {\bf 114}, 2801--2806 (2017).

\bibitem[Dabo et~al.(2010)Dabo, Ferretti, Poilvert, Li, Marzari, and
  Cococcioni]{dabo2010}
Dabo, I. et~al.
\newblock {Koopmans}’ condition for density-functional theory.
\newblock \emph{Phys. Rev. B} {\bf 82}, 115121 (2010).

\bibitem[Sun et~al.(2020)Sun, Zhang, Banerjee, Bao, Barbry, Blunt, Bogdanov,
  Booth, Chen, Cui, Eriksen, Gao, Guo, Hermann, Hermes, Koh, Koval, Lehtola,
  Li, Liu, Mardirossian, McClain, Motta, Mussard, Pham, Pulkin, Purwanto,
  Robinson, Ronca, Sayfutyarova, Scheurer, Schurkus, Smith, Sun, Sun, Upadhyay,
  Wagner, Wang, White, Whitfield, Williamson, Wouters, Yang, Yu, Zhu,
  Berkelbach, Sharma, Sokolov, and Chan]{sun2020}
Sun, Q. et~al.
\newblock Recent developments in the {PySCF} program package.
\newblock \emph{J. Chem. Phys.} {\bf 153}, 024109 (2020).

\bibitem[Nagai et~al.(2020)Nagai, Akashi, and Sugino]{nagai2020}
Nagai, R., Akashi, R. \& Sugino, O.
\newblock Completing density functional theory by machine learning hidden
  messages from molecules.
\newblock \emph{npj Comput. Mater.} {\bf 6}, 43 (2020).

\bibitem[Bystrom and Kozinsky(2022)]{bystrom2022}
Bystrom, K. \& Kozinsky, B.
\newblock {CIDER}: An expressive, nonlocal feature set for machine learning
  density functionals with exact constraints.
\newblock \emph{J. Chem. Theory Comput.} {\bf 18}, 2180--2192 (2022).

\bibitem[Hao et~al.(2012)Hao, Fang, Sun, Csonka, Philipsen, and
  Perdew]{hao2012}
Hao, P. et~al.
\newblock Lattice constants from semilocal density functionals with zero-point
  phonon correction.
\newblock \emph{Phys. Rev. B} {\bf 85}, 014111 (2012).

\bibitem[Wang et~al.(2021)Wang, Kingsbury, McDermott, Horton, Jain, Ong,
  Dwaraknath, and Persson]{wang2021}
Wang, A. et~al.
\newblock A framework for quantifying uncertainty in {DFT} energy corrections.
\newblock \emph{Sci. Rep.} {\bf 11}, 15496 (2021).

\bibitem[Mortensen et~al.(2024)Mortensen, Larsen, Kuisma, Ivanov, Taghizadeh,
  Peterson, Haldar, Dohn, Schäfer, Jónsson, Hermes, Nilsson, Kastlunger,
  Levi, Jónsson, Häkkinen, Fojt, Kangsabanik, Sødequist, Lehtomäki, Heske,
  Enkovaara, Winther, Dulak, Melander, Ovesen, Louhivuori, Walter, Gjerding,
  Lopez-Acevedo, Erhart, Warmbier, Würdemann, Kaappa, Latini, Boland,
  Bligaard, Skovhus, Susi, Maxson, Rossi, Chen, Schmerwitz, Schiøtz, Olsen,
  Jacobsen, and Thygesen]{mortensen2024}
Mortensen, J.~J. et~al.
\newblock {GPAW}: An open {Python} package for electronic structure
  calculations.
\newblock \emph{J. Chem. Phys.} {\bf 160}, 092503 (2024).

\bibitem[Blöchl(1994)]{blochl1994}
Blöchl, P.~E.
\newblock Projector augmented-wave method.
\newblock \emph{Phys. Rev. B} {\bf 50}, 17953--17979 (1994).

\bibitem[Lehtola et~al.(2018)Lehtola, Steigemann, Oliveira, and
  Marques]{lehtola2018}
Lehtola, S., Steigemann, C., Oliveira, M.~J. \& Marques, M.~A.
\newblock Recent developments in libxc --- a comprehensive library of
  functionals for density functional theory.
\newblock \emph{SoftwareX} {\bf 7}, 1--5 (2018).

\bibitem[Takamoto et~al.(2022)Takamoto, Shinagawa, Motoki, Nakago, Li, Kurata,
  Watanabe, Yayama, Iriguchi, Asano, Onodera, Ishii, Kudo, Ono, Sawada,
  Ishitani, Ong, Yamaguchi, Kataoka, Hayashi, Charoenphakdee, and
  Ibuka]{takamoto2022}
Takamoto, S. et~al.
\newblock Towards universal neural network potential for material discovery
  applicable to arbitrary combination of 45 elements.
\newblock \emph{Nat. Commun.} {\bf 13}, 2991 (2022).

\bibitem[mat(2026)]{matlantis}
{Matlantis}, software as a service style material discovery tool.
\newblock \url{https://matlantis.com/} (2026).
\newblock PFP version 9.0.0, released 16 July 2026.

\bibitem[Shinagawa et~al.(2026)Shinagawa, Takamoto, Shintani, Zhuang, Tsuboi,
  Nishimra, Shinohara, Iwase, Tanaka, and Li]{shinagawa2026}
Shinagawa, C. et~al.
\newblock {Matlantis-PFP} v8: universal machine learning interatomic potential
  with better experimental agreements via {r$^2$SCAN} functional.
\newblock Preprint at https://arxiv.org/abs/2603.11063 (2026).

\bibitem[Miglio et~al.(2020)Miglio, Brousseau-Couture, Godbout, Antonius, Chan,
  Louie, Côté, Giantomassi, and Gonze]{miglio2020}
Miglio, A. et~al.
\newblock Predominance of non-adiabatic effects in zero-point renormalization
  of the electronic band gap.
\newblock \emph{npj Comput. Mater.} {\bf 6}, 167 (2020).

\bibitem[Fuchs et~al.(2007)Fuchs, Furthmüller, Bechstedt, Shishkin, and
  Kresse]{fuchs2007}
Fuchs, F., Furthmüller, J., Bechstedt, F., Shishkin, M. \& Kresse, G.
\newblock Quasiparticle band structure based on a generalized {Kohn}-{Sham}
  scheme.
\newblock \emph{Phys. Rev. B} {\bf 76}, 115109 (2007).

\bibitem[Sawatzky and Allen(1984)]{sawatzky1984}
Sawatzky, G.~A. \& Allen, J.~W.
\newblock Magnitude and origin of the band gap in {NiO}.
\newblock \emph{Phys. Rev. Lett.} {\bf 53}, 2339--2342 (1984).

\bibitem[van Elp et~al.(1991{\natexlab{a}})van Elp, Potze, Eskes, Berger, and
  Sawatzky]{vanelp1991mno}
van Elp, J., Potze, R.~H., Eskes, H., Berger, R. \& Sawatzky, G.~A.
\newblock Electronic structure of {MnO}.
\newblock \emph{Phys. Rev. B} {\bf 44}, 1530--1537 (1991).

\bibitem[van Elp et~al.(1991{\natexlab{b}})van Elp, Wieland, Eskes, Kuiper,
  Sawatzky, de~Groot, and Turner]{vanelp1991coo}
van Elp, J. et~al.
\newblock Electronic structure of {CoO}, {Li}-doped {CoO}, and {LiCoO$_2$}.
\newblock \emph{Phys. Rev. B} {\bf 44}, 6090--6103 (1991).

\bibitem[Bowen et~al.(1975)Bowen, Adler, and Auker]{bowen1975}
Bowen, H., Adler, D. \& Auker, B.
\newblock Electrical and optical properties of {FeO}.
\newblock \emph{J. Solid State Chem.} {\bf 12}, 355--359 (1975).

\bibitem[Parkinson(2016)]{parkinson2016}
Parkinson, G.~S.
\newblock Iron oxide surfaces.
\newblock \emph{Surf. Sci. Rep.} {\bf 71}, 272--365 (2016).

\bibitem[Kasikov et~al.(2019)Kasikov, Tarre, and Marandi]{kasikov2019}
Kasikov, A., Tarre, A. \& Marandi, M.
\newblock Dispersion of chromia films (eskolaite) in {UV-VIS}.
\newblock \emph{J. Electr. Eng.} {\bf 70}, 36--43 (2019).

\bibitem[Giannozzi et~al.(2017)]{giannozzi2017}
Giannozzi, P. et~al.
\newblock Advanced capabilities for materials modelling with {Quantum
  ESPRESSO}.
\newblock \emph{J. Phys.: Condens. Matter} {\bf 29}, 465901 (2017).

\end{thebibliography}
\end{document}


\maketitle
\tableofcontents

\paragraph{Abbreviations.} PBE, the generalized gradient approximation of Perdew, Burke and Ernzerhof\cite{perdew1996}; PBE0, its hybrid with 25\% exact exchange; SCAN, the strongly constrained and appropriately normed meta-generalized-gradient approximation\cite{sun2015}, and r$^2$SCAN its regularized form\cite{furness2020}; rVV10, the revised Vydrov--van Voorhis nonlocal dispersion correction\cite{sabatini2013}; LOCUS (LOCalized-shell-Upgraded SCAN), the base functional of the main text, r$^2$SCAN + rVV10 + shell term; LDA, local-density approximation; HSE06, the hybrid functional of Heyd, Scuseria and Ernzerhof\cite{heyd2003,krukau2006}; mBJ, the modified Becke--Johnson potential of Tran and Blaha\cite{tran2009}; DSH, the doubly screened hybrid functional\cite{cui2018}; GW, the approximation that builds the self-energy from the Green's function $G$ and the screened Coulomb interaction $W$; G$_0$W$_0$, GW$_0$ and evGW, one-shot, partially self-consistent and eigenvalue self-consistent GW; G$_0$W$_0$@PBE, G$_0$W$_0$ on PBE in the plasmon-pole approximation; QPG$_0$W$_0$, full-frequency quasiparticle G$_0$W$_0$ with spin--orbit coupling; QSG$\hat{W}$, vertex-corrected quasiparticle self-consistent GW, compared here with spin--orbit coupling; $\alpha$QSGW, quasiparticle self-consistent GW with 80\% of its self-energy mixed with 20\% of the LDA exchange-correlation potential\cite{grossmann2026}; COHSEX, the static Coulomb-hole plus screened-exchange approximation; CCSD(T), coupled cluster with single, double and perturbative triple excitations (UCCSD(T) for open shells); EOM-CCSD, equation-of-motion coupled cluster with single and double excitations; X2C, the exact two-component relativistic Hamiltonian; GW100, the benchmark set of 100 molecules of ref.~\citenum{vansetten2015}; W4-11, the thermochemistry benchmark of ref.~\citenum{karton2011}; VBM and CBM, valence-band maximum and conduction-band minimum; IP and EA, ionization potential and electron affinity; MAE and ME, mean absolute and mean signed error; PAW, projector augmented wave; SCF, self-consistent field calculation; KS and TF (subscripts), Kohn--Sham and Thomas--Fermi; AFM-II, type-II antiferromagnetic order; ICSD, Inorganic Crystal Structure Database (collection codes in Supplementary Table~\ref{tab:gaps}); C2DB, Computational 2D Materials Database\cite{haastrup2018}; CIGS, CuIn$_{1-x}$Ga$_x$Se$_2$; IGZO, InGaZnO$_4$; ALD, atomic layer deposition; PFP, PreFerred Potential, a universal neural-network potential\cite{takamoto2022}; MPI, Message Passing Interface.

\section{Calibration and rule ablations}
\label{sec:calib}

Everything that enters our correction is listed in Table~1 of the main text. This section gives the data the coefficients were fitted to and the effect of each rule that was chosen by inspecting the benchmark.

The three bare-shift coefficients were fitted by least squares through the origin to the exact-exchange corrections $\langle\Sigma_x - v_x^\mathrm{PBE}\rangle$ of the PBE states of ten semiconductors (Supplementary Table~\ref{tab:cal10}), prototypical group-IV, III--V and II--VI semiconductors with experimental gaps from 0.24 to 3.5~eV of which those with semicore d states (Ga, In, Zn, Cd) also fix $k_d$. The ten solids were not selected for performance; Section~\ref{sec:general} tests random calibration sets. The factor $s$ of each base was fitted, with $g = 1$, to the CCSD(T)/def2-TZVPP highest occupied levels of 66 molecules of GW100\cite{vansetten2015,krause2015} (Supplementary Table~\ref{tab:gw100}). Because the molecular levels depend on $k_V$ only through the product $s\,k_V$, only this product is fixed by the molecules; the VBM shift of solids, which is also proportional to $s\,k_V$, therefore does not depend on $k_V$ separately. The only property of the bare-shift coefficients that changes band gaps is the ratio $k_C/k_V$ (Section~\ref{sec:general}).

\begin{table}[H]
\centering
\caption{\textbf{Calibration data of the bare shifts.} Exact-exchange corrections $D_x = \langle\Sigma_x - v_x^\mathrm{PBE}\rangle$ of the VBM, CBM and semicore d states on PBE states, and the bare shifts $\hat{D}$ from the fitted coefficients (eV).}
\label{tab:cal10}
\small
\begin{tabular}{llrrrrr}
\toprule
Solid & Space group & $D_x^\mathrm{VBM}$ & $\hat{D}_\mathrm{VBM}$ & $D_x^\mathrm{CBM}$ & $\hat{D}_\mathrm{CBM}$ & $D_x^\mathrm{d}$ \\
\midrule
Si & $Fd-3m$ & -2.49 & -2.59 & 2.04 & 1.38 & -- \\
Ge & $Fd-3m$ & -2.24 & -2.42 & 2.31 & 2.67 & -- \\
GaAs & $F-43m$ & -2.82 & -2.90 & 2.30 & 2.51 & -15.14 \\
GaP & $F-43m$ & -3.01 & -3.13 & 2.14 & 1.69 & -15.13 \\
InP & $F-43m$ & -2.88 & -3.12 & 2.13 & 2.10 & -10.43 \\
InSb & $F-43m$ & -1.96 & -2.58 & 1.84 & 2.01 & -10.43 \\
GaN & $P6_3mc$ & -4.87 & -4.99 & 2.43 & 2.16 & -15.22 \\
ZnO & $P6_3mc$ & -7.76 & -7.30 & 1.97 & 2.18 & -13.00 \\
CdTe & $F-43m$ & -3.24 & -3.27 & 2.09 & 2.23 & -9.44 \\
ZnSe & $F-43m$ & -4.31 & -4.08 & 2.39 & 2.26 & -13.29 \\
\bottomrule
\end{tabular}
\end{table}

{\small
\begin{longtable}{llrrrr}
\caption{\textbf{Molecules that fix $s$.} Highest occupied levels (eV): CCSD(T)/def2-TZVPP reference\cite{krause2015}, PBE and r$^2$SCAN + rVV10 + shell term (LOCUS) eigenvalues, and the LOCUS level after our correction with $s = 0.658$ and $g = 1$.}\label{tab:gw100}\\
\toprule
Molecule & Formula & CCSD(T) & PBE & LOCUS & LOCUS + $\ES$ \\
\midrule
\endfirsthead
\toprule
Molecule & Formula & CCSD(T) & PBE & LOCUS & LOCUS + $\ES$ \\
\midrule
\endhead
Aluminum trifluoride & AlF$_{3}$ & -15.32 & -9.62 & -10.09 & -15.19 \\
Argon & Ar & -15.54 & -10.18 & -10.63 & -14.59 \\
Diborane & B$_{2}$H$_{6}$ & -12.25 & -7.86 & -8.15 & -12.70 \\
Fluoroborane & BF & -11.09 & -6.75 & -6.98 & -10.90 \\
Borane & BH$_{3}$ & -13.27 & -8.48 & -8.79 & -13.33 \\
Boron nitride & BN & -11.98 & -7.44 & -7.63 & -10.92 \\
Beryllium monoxide & BeO & -9.94 & -6.10 & -6.19 & -9.90 \\
Acetylene & C$_{2}$H$_{2}$ & -11.42 & -7.16 & -7.36 & -10.24 \\
Vinyl chloride & C$_{2}$H$_{3}$Cl & -10.09 & -6.38 & -6.58 & -9.58 \\
Vinyl fluoride & C$_{2}$H$_{3}$F & -10.55 & -6.49 & -6.69 & -9.83 \\
Ethylene & C$_{2}$H$_{4}$ & -10.67 & -6.75 & -6.91 & -9.55 \\
Acetaldehyde & C$_{2}$H$_{4}$O & -10.21 & -5.91 & -6.17 & -10.53 \\
Ethane & C$_{2}$H$_{6}$ & -12.71 & -8.15 & -8.44 & -12.69 \\
Ethanol & C$_{2}$H$_{6}$O & -10.69 & -6.04 & -6.36 & -10.83 \\
Cyclopropane & C$_{3}$H$_{6}$ & -10.87 & -7.03 & -7.23 & -10.04 \\
Propane & C$_{3}$H$_{8}$ & -12.03 & -7.74 & -8.01 & -12.31 \\
Tetracarbon & C$_{4}$ & -11.24 & -7.22 & -7.51 & -11.53 \\
Uracil & C$_{4}$H$_{4}$N$_{2}$O$_{2}$ & -9.48 & -6.20 & -6.51 & -10.67 \\
Pyridine & C$_{5}$H$_{5}$N & -9.73 & -5.89 & -6.19 & -10.15 \\
Cyclopentadiene & C$_{5}$H$_{6}$ & -8.71 & -5.37 & -5.47 & -7.99 \\
Hexafluorobenzene & C$_{6}$F$_{6}$ & -9.93 & -6.57 & -6.81 & -10.15 \\
Benzene & C$_{6}$H$_{6}$ & -9.36 & -6.30 & -6.45 & -8.89 \\
Carbon tetrachloride & CCl$_{4}$ & -11.50 & -7.60 & -7.96 & -11.50 \\
Carbon tetrafluoride & CF$_{4}$ & -16.23 & -10.33 & -10.89 & -16.20 \\
Formaldehyde & CH$_{2}$O & -10.84 & -6.20 & -6.48 & -11.28 \\
Formic acid & CH$_{2}$O$_{2}$ & -11.42 & -6.85 & -7.15 & -11.55 \\
Methane & CH$_{4}$ & -14.37 & -9.45 & -9.78 & -13.82 \\
Urea & CH$_{4}$N$_{2}$O & -10.05 & -5.80 & -6.10 & -10.20 \\
Methanol & CH$_{4}$O & -11.04 & -6.21 & -6.54 & -11.04 \\
Hydrogen cyanide & CHN & -13.72 & -9.00 & -9.25 & -12.48 \\
Carbon monoxide & CO & -14.21 & -9.29 & -9.69 & -14.16 \\
Carbon dioxide & CO$_{2}$ & -13.71 & -9.03 & -9.38 & -13.69 \\
Carbon oxysulfide & COS & -11.17 & -7.44 & -7.72 & -10.99 \\
Carbon disulfide & CS$_{2}$ & -9.98 & -6.77 & -7.00 & -10.02 \\
Chlorine & Cl$_{2}$ & -11.41 & -7.22 & -7.53 & -11.18 \\
Magnesium chloride & Cl$_{2}$Mg & -11.66 & -7.61 & -7.94 & -11.24 \\
Hydrogen chloride & ClH & -12.59 & -7.92 & -8.27 & -11.77 \\
Sodium chloride & ClNa & -9.03 & -5.19 & -5.38 & -8.61 \\
Fluorine & F$_{2}$ & -15.58 & -9.34 & -9.89 & -15.57 \\
Magnesium fluoride & F$_{2}$Mg & -13.71 & -8.29 & -8.64 & -13.50 \\
Sulfur tetrafluoride & F$_{4}$S & -12.59 & -8.08 & -8.42 & -12.89 \\
Hydrogen fluoride & FH & -16.03 & -9.38 & -9.90 & -15.09 \\
Lithium fluoride & FLi & -11.32 & -5.99 & -6.24 & -11.03 \\
Hydrogen & H$_{2}$ & -16.40 & -10.37 & -10.77 & -17.58 \\
Water & H$_{2}$O & -12.56 & -6.99 & -7.39 & -11.80 \\
Hydrogen peroxide & H$_{2}$O$_{2}$ & -11.52 & -6.28 & -6.66 & -11.41 \\
Hydrogen sulfide & H$_{2}$S & -10.31 & -6.20 & -6.46 & -9.59 \\
Ammonia & H$_{3}$N & -10.81 & -5.99 & -6.31 & -10.26 \\
Phosphine & H$_{3}$P & -10.52 & -6.67 & -6.89 & -10.20 \\
Hydrazine & H$_{4}$N$_{2}$ & -9.69 & -5.14 & -5.43 & -9.67 \\
Silane & H$_{4}$Si & -12.80 & -8.52 & -8.84 & -13.07 \\
Disilane & H$_{6}$Si$_{2}$ & -10.64 & -7.29 & -7.52 & -10.23 \\
Lithium hydride & HLi & -7.96 & -4.37 & -4.48 & -9.22 \\
Hydrogen azide & HN$_{3}$ & -10.68 & -6.73 & -6.96 & -10.61 \\
Helium & He & -24.51 & -15.63 & -16.41 & -27.19 \\
Lithium dimer & Li$_{2}$ & -5.27 & -3.24 & -3.21 & -5.43 \\
Magnesium monoxide & MgO & -7.91 & -4.80 & -4.91 & -8.82 \\
Nitrogen & N$_{2}$ & -15.48 & -10.21 & -10.67 & -15.29 \\
Phosphorus mononitride & NP & -11.81 & -7.67 & -8.02 & -11.97 \\
Sodium dimer & Na$_{2}$ & -4.95 & -3.14 & -3.08 & -4.75 \\
Sodium tetramer & Na$_{4}$ & -4.24 & -2.69 & -2.64 & -4.06 \\
Sodium hexamer & Na$_{6}$ & -4.39 & -3.00 & -2.96 & -4.03 \\
Neon & Ne & -21.32 & -13.15 & -13.82 & -19.89 \\
Sulfur dioxide & O$_{2}$S & -12.30 & -7.89 & -8.21 & -12.32 \\
Ozone & O$_{3}$ & -12.71 & -7.83 & -8.22 & -13.01 \\
Phosphorus dimer & P$_{2}$ & -10.53 & -7.07 & -7.33 & -9.72 \\
\bottomrule
\end{longtable}
}

Supplementary Table~\ref{tab:ablation} shows the effect of each rule, re-evaluated from the stored factors of the edge shifts without new calculations. The localized-edge screening hardly changes the MAE of the bulk gaps of the benchmark (0.437~eV, and 0.439~eV without it); it was selected for molecular crystals, whose MAE it lowered from 0.72 to 0.39~eV in the comparison made when the rule was chosen. Spin--orbit coupling is needed for the heavy-element compounds; without it the MAE over all held-out crystals rises to 0.50~eV. A CBM bare shift that includes $e^{-\alpha}$, fitted on the same ten solids, is worse (0.46~eV against 0.44~eV), as expected from its poorer agreement with the exact-exchange corrections. A constant CBM bare shift, set to the mean exact-exchange correction of the CBM of the ten calibration solids on PBE states (2.16~eV), gives the same accuracy on bulk solids (0.43~eV) and on monolayers. In bulk the CBM bare shift therefore carries no more information than a constant, as the exact-exchange corrections show (main text, Methods). The local CBM bare shift is kept because it makes our correction local, which a constant cannot do for states that extend into vacuum or belong to a molecule.

\begin{table}[H]
\centering
\caption{\textbf{Rule ablations.} MAE and ME (eV) of the gaps of the held-out crystals, by range of the experimental gap, and of 20 monolayers against C2DB G$_0$W$_0$.}
\label{tab:ablation}
\small
\begin{tabular}{lrrrrrr}
\toprule
Variant & MAE & ME & 0--3 eV & 3--6 eV & $>$6 eV & monolayers \\
\midrule
adopted & 0.44 & 0.12 & 0.37 & 0.53 & 0.80 & 0.46 \\
localized-edge screening off & 0.44 & 0.09 & 0.37 & 0.52 & 0.90 & 0.43 \\
constant CBM bare shift & 0.43 & 0.12 & 0.37 & 0.50 & 0.77 & 0.46 \\
CBM bare shift with $e^{-\alpha}$ & 0.46 & 0.14 & 0.39 & 0.56 & 0.84 & -- \\
no spin--orbit coupling & 0.50 & 0.26 & 0.46 & 0.55 & 0.73 & -- \\
\bottomrule
\end{tabular}
\end{table}

Supplementary Table~\ref{tab:core} tests the two central factors in the same way (main text, Discussion). Every edge shift of the LOCUS calculations is $s\,g\,\hat{D}$ with the signed bare shift $\hat{D}$ of the edge (main text, Methods); the variants replace $g$, the bare shifts of the VBM and CBM, or both (rigid shift), by their means over the ten calibration solids on LOCUS states ($g = 0.126$, $\hat{D}_\mathrm{VBM} = -3.58$~eV, $\hat{D}_\mathrm{CBM} = 2.13$~eV), so that they use the same calibration information as the adopted form; spin--orbit coupling is unchanged. Below 3~eV, where the screening of most semiconductors is similar, all variants give 0.36--0.37~eV. Above 6~eV, a constant screening raises the MAE from 0.80 to 2.19~eV and constant bare shifts to 1.30~eV, and the surface ionization potentials, which depend on the absolute VBM shift, lose accuracy with every simplification (MAE from 0.24~eV to 0.32--0.43~eV). The material-specific screening and the local bare shift are therefore needed for wide-gap crystals and for absolute levels, not for the gaps of narrow-gap semiconductors.

\begin{table}[H]
\centering
\caption{\textbf{Ablation of the central factors.} MAE and ME (eV) of the gaps of the held-out crystals (all, below 3~eV, above 6~eV) and of the ionization potentials of the ten surfaces with experiment, with the factors of each material replaced by their means over the calibration solids.}
\label{tab:core}
\small
\begin{tabular}{lrrrrrr}
\toprule
Variant & Gaps & $<$3 eV & $>$6 eV & ME & Surface IP & ME \\
\midrule
adopted & 0.44 & 0.37 & 0.80 & +0.12 & 0.24 & $-$0.18 \\
constant g & 0.53 & 0.36 & 2.19 & $-$0.17 & 0.37 & $-$0.37 \\
constant bare shifts & 0.46 & 0.36 & 1.30 & $-$0.05 & 0.32 & $-$0.32 \\
rigid shift & 0.57 & 0.37 & 2.36 & $-$0.25 & 0.43 & $-$0.43 \\
\bottomrule
\end{tabular}
\end{table}

\section{Screening from the Kohn--Sham spectrum}
\label{sec:screen}

The screening of our correction is $g = 1/(\varepsilon_\mathrm{KS} + \varepsilon_\mathrm{TF})$, computed on the states of the base functional with no fitted parameter.

\paragraph{Interband part.} $\varepsilon_\mathrm{KS}$ is the long-wavelength, independent-particle electronic dielectric constant without local fields:
\begin{equation}
  \varepsilon_\mathrm{KS} = 1 + \frac{8\pi}{\Omega} \sum_{\mathbf{k}} w_\mathbf{k} \sum_{n<m} f_{n\mathbf{k}}\,(1 - f_{m\mathbf{k}})\,
  \frac{2\,|\langle\psi_{m\mathbf{k}}|\hat{\mathbf{p}}|\psi_{n\mathbf{k}}\rangle|^2}{3\,\omega_{nm\mathbf{k}}^{3}}\; R_{nm\mathbf{k}},
  \qquad \omega_{nm\mathbf{k}} = \max(\epsilon_{m\mathbf{k}} - \epsilon_{n\mathbf{k}},\ 0.1~\mathrm{eV}),
  \label{eq:epsks}
\end{equation}
in atomic units. $\Omega$ is the cell volume, $w_\mathbf{k}$ are the normalized weights of the irreducible $k$-points, the factor $8\pi$ includes the spin degeneracy, and $|\langle\psi_{m\mathbf{k}}|\hat{\mathbf{p}}|\psi_{n\mathbf{k}}\rangle|^2$ is summed over the three Cartesian components and includes the PAW corrections. For integer occupations the factors $f_{n\mathbf{k}}(1 - f_{m\mathbf{k}})$ equal the $f_{n\mathbf{k}} - f_{m\mathbf{k}}$ of linear response. For partial occupations they differ (two half-filled states give 0.25 instead of 0); the choice is a heuristic that keeps the expression continuous through band crossings, and the intraband screening of metals, which dominates $g$ there, is carried by the Thomas--Fermi term below. The floor on $\omega$ bounds the sum when a gap closes. In terms of the direction-averaged oscillator strengths $f_{nm} = 2|\langle m|\hat{\mathbf{p}}|n\rangle|^2/(3\omega_{nm})$, equation~\eqref{eq:epsks} is the Penn-type sum $\varepsilon - 1 = (8\pi/\Omega)\sum f_{nm}/\omega_{nm}^2$; the sum is truncated at the computed empty bands and extrapolated in their number (see Empty bands, below).

\paragraph{Mesh-cell average at direct-gap minima.} In insulators, at a $k$-point where the direct gap $\omega_0$ is smaller than at all nearest mesh points, the integrand varies on the scale of the mesh cell, and a single $k$-point over-weights the minimum (for InSb on a $7^3$ mesh the $\Gamma$ point alone gave 135 of $\varepsilon = 150$). There each transition $nm$ is averaged over a sphere with the volume of the mesh cell, with a two-band dispersion $\omega_{nm}(k) = (\omega_{nm}^2 + b k^2)^{1/2}$. The curvature $b = (\overline{\omega_0^2} - \omega_0^2)/d^2$ is the same for all transitions at that $k$-point; it comes from the direct gaps at the nearest mesh points, at distance $d$, whose mean is denoted by the bar. With $\langle\cdot\rangle$ the average over the sphere,
\begin{equation}
  R_{nm} = \omega_{nm}^3\langle\omega_{nm}(k)^{-3}\rangle = \frac{3}{X^3}\left[\operatorname{asinh} X - \frac{X}{\sqrt{1 + X^2}}\right], \qquad
  X = \frac{k_c\sqrt{b}}{\omega_{nm}}, \qquad k_c = \left(\frac{3(2\pi)^3}{4\pi\,\Omega\,N_\mathbf{k}}\right)^{1/3},
\end{equation}
with $R_{nm} \to 1$ for $X \to 0$, and $R = 1$ at all other $k$-points; $N_\mathbf{k}$ is the number of points of the full mesh. For the transition at the gap, $\omega_{nm} = \omega_0$.

\paragraph{Empty bands.} The band sum uses empty bands up to twice the number of occupied ones. Its truncation error decreases as a power of the number $N$ of empty bands, with an effective exponent of 1.1--2.6 (median 1.7) in the range used, as found against complete-basis (Sternheimer) calculations for eight solids. The sum is extrapolated with a fixed exponent of 2 from the sums with $N$ and $N/2$ empty bands, $\varepsilon(\infty) = \varepsilon(N) + [\varepsilon(N) - \varepsilon(N/2)]/3$, which agrees with the Sternheimer values within 0.014 in $\varepsilon$, against 0.016--0.038 without extrapolation.

\paragraph{Metals.} The intraband screening of partially filled bands is a Thomas--Fermi term at $q = 1/L$, $\varepsilon_\mathrm{TF} = 4\pi D_F L^2$ with $L = 6$~bohr, where the density of states at the Fermi level $D_F$ is evaluated with a fixed broadening of 0.05~eV, independent of the smearing of the calculation. It vanishes for insulators.

\paragraph{Isolated systems.} A molecule in a box has no $k$ dispersion. With the dispersion-weighted estimate $\varepsilon_\mathrm{disp}$, in which the oscillator strengths are replaced by $f_{nm} = \mathrm{Var}_\mathbf{k}(\omega_{nm})\,\Omega^{2/3}/\overline{\omega}_{nm}$, an isolated molecule has $\varepsilon_\mathrm{disp} = 1$ exactly, while solids, even rare-gas solids, have $\varepsilon_\mathrm{disp} \ge 1.33$. The screening used in our correction is $(1 - h)\,g + h$ with $h = \sigma[(1.15 - \varepsilon_\mathrm{disp})/0.03]$, where $\sigma$ is the logistic function, which gives $h \approx 0.99$ for isolated molecules and $h < 0.003$ for solids; molecules are treated with $g = 1$, as in the calibration of $s$. Because $\varepsilon_\mathrm{disp}$ uses the variation of the transition energies over the $k$-points, a periodic cell computed at the $\Gamma$ point alone would also give $\varepsilon_\mathrm{disp} = 1$. With a single $k$-point the switch is therefore applied only when the cell has a vacuum gap of at least 5~\AA{} between the atomic planes normal to every periodic axis, as a molecule in a box has; otherwise $h = 0$. The only periodic system of this work computed at the $\Gamma$ point alone, a 320-atom cell of CsPbI$_3$ (Section~\ref{sec:pvk}), has no vacuum gap and is therefore treated as a solid ($h = 0$); all others, including the 40- to 112-atom cells of perovskites and amorphous oxides, were computed with several $k$-points, for which this rule has no effect.

\paragraph{Cell representation.} The screening is a property of the crystal and not of the cell used to describe it. For MgO and GaAs computed as primitive cells and as $2\times2\times2$ and $3\times3\times3$ supercells with the equivalent folded $k$-point meshes, $g$ agrees within 0.0005 and the corrected gap within 0.003~eV.

\paragraph{Monolayers.} In a cell with vacuum, the independent-particle dielectric constant of the whole cell, $\varepsilon_\mathrm{cell}$ (its $\varepsilon_\mathrm{KS}$), has $\varepsilon_\mathrm{cell} - 1$ proportional to the fraction of the cell occupied by the layer and depends on the vacuum thickness. For monolayers we use an effective dielectric constant of the layer for the adopted thickness $d$, $\varepsilon_s = 1 + (\varepsilon_\mathrm{cell} - 1)\,L_c/d$, with $L_c$ the cell length normal to the layer and $d$ the atomic thickness plus 3.1~\AA, the interlayer spacing of layered parent compounds, and $g = 1/(\varepsilon_s + \varepsilon_\mathrm{TF})$. Increasing the vacuum between layer images from 15~\AA{} (the production cells) to 20 and 30~\AA{} on the LOCUS base lowers $\varepsilon_\mathrm{cell}$ from 1.55 to 1.27 for BN and from 5.84 to 3.68 for MoTe$_2$, whereas $\varepsilon_s$ varies by less than 3\%; the corrected gap changes by at most 0.05~eV for BN, of which 0.03~eV is the change of the Kohn--Sham gap itself, and by less than 0.001~eV for MoTe$_2$. On the PBE base the changes are at most 0.02 and 0.001~eV. The added 3.1~\AA{} is a model choice common to all monolayers; changing it to 2.6 or 3.6~\AA{} changes the MAE on the 20 monolayers from 0.46~eV to 0.47 and 0.45~eV and the gap of any monolayer by at most 0.13~eV.

\paragraph{Localized band edges.} An s or p band edge localized on one atom is screened less than the long-wavelength $g$ implies ($g_\phi > g$). For each edge the dominant atomic channel (atom and angular momentum, from the PAW projections) is identified, and three quantities are computed: an on-site interaction $U = s\,\langle u\rangle_\mathrm{at}\,g_\phi$, with $\langle u\rangle_\mathrm{at}$ the sum of the occupied and empty bare shifts evaluated on the normalized density of this atomic shell in the free atom (all-electron PBE, scalar-relativistic); the width $W$, from the spread of the levels weighted by their projection on that channel, of the band block at the VBM (all occupied levels down to the first internal gap of more than 1~eV) and of the band complex at each edge (the edge band and the adjacent bands whose energy ranges overlap it within 0.1~eV); and the band-edge effective mass $m^*$, from finite differences of the band energies on the $k$-point grid. The screening seen by this atomic shell is
\begin{equation}
  g_\phi = \frac{\int |F(q)|^2\,\varepsilon^{-1}(q)\,\mathrm{d}q}{\int |F(q)|^2\,\mathrm{d}q},
  \qquad \varepsilon(q) = 1 + \left[\frac{1}{\varepsilon_\infty - 1} + 1.563\,\frac{q^2}{q_\mathrm{TF}^2} + \frac{q^4}{4k_F^2 q_\mathrm{TF}^2}\right]^{-1},
\end{equation}
with $F(q)$ the form factor of the shell density, $\varepsilon_\infty = 1/g$ and the model dielectric function of Cappellini et al.\cite{cappellini1993} ($q_\mathrm{TF}$ and $k_F$ from the density of the outer s and p electrons). The switch $\zeta = \max\{\sigma[(U/W_\mathrm{block} - 2)/0.3],\ \sigma[(U/W_\mathrm{complex} - 2)/0.3]\,\sigma[(m^* - 2.5)/0.25]\}$, the first term at the VBM only, replaces $g$ by $(1 - \zeta)\,g + \zeta\,g_\phi$ for that edge; it is applied to spin-paired systems only, so that the open-shell oxides use the global $g$. The thresholds $U/W = 2$ and $m^* = 2.5$ were selected on the benchmark and on molecular crystals (main text, Table~1; Supplementary Table~\ref{tab:ablation}).

\paragraph{Comparison with experiment.} On LOCUS states, $1/g$ agrees with the experimental $\varepsilon_\infty$ of 13 solids with a root-mean-square deviation of 6\% in $\ln\varepsilon$ (main text, Fig.~1c).

\section{The shell term}
\label{sec:shell}

The base functional LOCUS is r$^2$SCAN\cite{furness2020} + rVV10\cite{sabatini2013} ($b = 12$) + shell term. Our correction was evaluated for the solids of this work, except the 320-atom CsPbI$_3$ cells (Section~\ref{sec:pvk}), on an earlier variant of LOCUS, whose shell gate has a sigmoid instead of the $C^2$ upper cut and which contains additional terms that act only in metals; on 40 semiconductors and insulators computed with both, the corrected gaps differ by 0.001~eV on average and by at most 0.02~eV. The formation enthalpies of Supplementary Table~\ref{tab:energetics} use the final form, and the molecular values of that table use the gate before the $C^2$ upper cut; adding the cut changes them by less than 0.01~kcal~mol$^{-1}$. In the plane-wave calculations the functional is blended into PBE at low density, $e = w\,e_\mathrm{LOCUS} + (1 - w)\,e_\mathrm{PBE}$ with $w$ a smoothstep in $\log_{10} n$ from 0 at $n = 10^{-4}$ to 1 at $n = 10^{-3}$~bohr$^{-3}$ of the total density; this keeps the meta-GGA kinetic-energy term stable in vacuum, acts only in vacuum regions and the voids of molecular crystals, and leaves dense solids unchanged. In the molecular calculations (PySCF) the dispersion term is PySCF's VV10 nonlocal correlation with the rVV10 parameters ($b = 12$, $C = 0.0093$), and the shell descriptor is evaluated from the density fitted in the overlap metric to an even-tempered auxiliary basis. The shell term adds to the exchange--correlation energy
\begin{equation}
  E_\mathrm{sh} = c \int \Theta(\mathbf{r})\, e_x^\mathrm{LDA}\big(n(\mathbf{r})\big)\, \mathrm{d}\mathbf{r}, \qquad
  e_x^\mathrm{LDA}(n) = -\tfrac{3}{4}\left(\tfrac{3}{\pi}\right)^{1/3} n^{4/3}, \qquad c = 0.2,
\end{equation}
evaluated on the total density; both spin channels therefore experience the same potential, and the term introduces no explicit spin dependence. Exchange splittings can still change indirectly, through the different spatial distributions of the two spin densities and the self-consistent response. The gate $\Theta$ is built from the density $n$ (bohr$^{-3}$), the reduced gradient $p = |\nabla n|^2/[4(3\pi^2)^{2/3} n^{8/3}]$ and the shell descriptor $x = x_c(4) = (G_\ell * n)/n$, the density averaged by a Gaussian $G_\ell$ of width $\ell = \min(\max(4/k_F, 0.3), 6)$~bohr, divided by the local density ($x = 1$ in the uniform gas; the argument 4 is the width in units of $1/k_F$). This is a nonlocal density descriptor of the kind introduced for machine-learned functionals\cite{nagai2020}, in the form used for the CIDER features\cite{bystrom2022}. With $\sigma$ the logistic function,
\begin{align}
  \Theta = {}& \sigma\!\left(\frac{\ln n}{0.3}\right) \sigma\!\left(\frac{0.15 - p}{0.08}\right) \sigma\!\left(\frac{x - 0.33}{0.02}\right) S(x) \nonumber \\
        & \times \left[1 - W_\mathrm{N}(x)\left(1 - \sigma\!\left(\frac{\ln 6 - \ln n}{0.3}\right)\right)\right]
        \left[1 - W_\mathrm{M}(x)\left(1 - \sigma\!\left(\frac{\ln n - \ln 2.5}{0.3}\right)\right)\right], \\
  S(x) = {}& t^3 (10 - 15t + 6t^2), \quad t = \min\!\left(\max\!\left(\frac{1 - x}{1 - 0.92}, 0\right), 1\right), \\
  W_\mathrm{N}(x) = {}& \sigma\!\left(\frac{x - 0.62}{0.03}\right) \sigma\!\left(\frac{0.92 - x}{0.03}\right), \qquad
  W_\mathrm{M}(x) = \sigma\!\left(\frac{0.60 - x}{0.03}\right).
\end{align}
The first three factors select dense regions of small gradient inside a shell maximum, which are the closed d shells; the second-row L shells, with $p \approx 1$, remain outside. $S(x)$ is a $C^2$ upper cut that makes the term and its first two derivatives vanish in the uniform electron gas ($x = 1$). $W_\mathrm{N}$ caps the density inside the range of the N shells (principal quantum number 4), keeping the outer 4d shell and dropping the complete shells beneath it. $W_\mathrm{M}$ sets a density floor inside the range of the M shells (principal quantum number 3), which retains the 3d shells (densities 4--10~bohr$^{-3}$) and excludes the molecular L shells.

The strength $c$ was set from the Koopmans condition\cite{dabo2010} on six d$^{10}$ atoms and ions (Supplementary Table~\ref{tab:koop}). For each we computed the error of the energy along a fractional d hole, $\mathrm{err}(\delta) = E(N-\delta) - E(N) - \delta\,\mathrm{IP}_d$, on fractional-hole PBE0 densities (all-electron x2c-TZVPall basis, scalar-relativistic X2C), with $\mathrm{IP}_d$ from $\Delta$-UCCSD(T)/x2c-QZVPPall. The slope of $\mathrm{err}$ at $\delta \to 0$ is of similar magnitude for r$^2$SCAN in all six ($-5.7$ to $-8.8$~eV per electron). The shell term changes it by an amount per unit $c$ that depends on how much of the shell the gate covers. $c$ was chosen so that, in the four systems whose shell the gate covers (Zn, Ga$^+$, Cd, In$^+$), the term removes a fraction of the error comparable to the screening $1/\varepsilon_\infty$ of typical d compounds ($\varepsilon_\infty \approx 5$--6): $c = 0.2$ removes 15--20\%, whereas removing a quarter would need $c = 0.25$--0.33. The diffuse shells of Cu$^+$ and Ag$^+$ are only partly covered. The gate was checked on 27 solids with d bands, which include the 13 solids of the d-band comparison of the main text; that comparison is therefore not an independent test of the d-band positions.

\begin{table}[H]
\centering
\caption{\textbf{Koopmans analysis of the shell term.} Slope of the fractional-hole error of r$^2$SCAN at $\delta \to 0$ (eV per electron), slope of the shell term per unit $c$, the $c$ that would remove the whole error and a quarter of it, and the fraction of the error removed by $c = 0.2$. Computed with the gate before the $C^2$ upper cut; adding the cut changes the shell energy of these atoms and ions by less than 0.02\%.}
\label{tab:koop}
\small
\begin{tabular}{lrrrrr}
\toprule
System & r$^2$SCAN slope & shell slope / $c$ & $c$ (full) & $c$ (1/4) & removed by $c = 0.2$ \\
\midrule
Cu$^+$ & $-6.77$ & 2.25 & 3.01 & 0.75 & 7\% \\
Zn & $-6.89$ & 5.18 & 1.33 & 0.33 & 15\% \\
Ga$^+$ & $-8.84$ & 8.48 & 1.04 & 0.26 & 19\% \\
Ag$^+$ & $-5.92$ & 2.87 & 2.07 & 0.52 & 10\% \\
Cd & $-5.68$ & 4.98 & 1.14 & 0.29 & 18\% \\
In$^+$ & $-6.81$ & 6.72 & 1.01 & 0.25 & 20\% \\
\bottomrule
\end{tabular}
\end{table}

The range boundaries ($x = 0.60$, 0.62 and 0.92), the upper cut ($x = 0.92$--1) and the lower cut $x_t = 0.33$ lie in empty intervals of the distribution of shell maxima of all-electron atoms from Li to Bi, whose $x$ values cluster by principal quantum number (M shells $\le 0.46$, N shells 0.76--0.87, O shells $\ge 1.04$). A closed shell that lies entirely inside or outside the gate contributes only an element-specific constant, whereas a shell that straddles a boundary contributes an amount that varies with its environment; the boundaries are therefore placed where no shell maximum lies. Only the gradient threshold $p_t$ and the density cap of the N-shell range cross continuous distributions and are choices. Supplementary Table~\ref{tab:shellpar} lists every parameter of the term and how it was set, and Supplementary Table~\ref{tab:gatesens} shows the effect of these two choices and of other details of the gate.

\begin{table}[H]
\centering
\caption{\textbf{Parameters of the shell term.} Widths of the logistic factors in parentheses.}
\label{tab:shellpar}
\small
\begin{tabular}{>{\raggedright\arraybackslash}p{0.27\textwidth}>{\raggedright\arraybackslash}p{0.22\textwidth}>{\raggedright\arraybackslash}p{0.43\textwidth}}
\toprule
Quantity & Value & How it was set \\
\midrule
\multicolumn{3}{l}{\emph{Fitted to theory}}\\
strength $c$ & 0.2 & screened Koopmans condition on six d$^{10}$ atoms and ions (Supplementary Table~\ref{tab:koop}) \\
\multicolumn{3}{l}{\emph{Fixed by construction}}\\
lower cut in $x$ & 0.33 (0.02) & below the 3d shells, above the K and L shells of the 3d series \\
range boundaries in $x$ & M shells: $< 0.60$; N shells: 0.62--0.92 (0.03) & intervals between the shell maxima of free atoms from Li to Bi (Supplementary Fig.~\ref{fig:gate}) \\
density floors & 1 bohr$^{-3}$ (0.3 in $\ln n$); 2.5 bohr$^{-3}$ in the M-shell range & below the closed d shells, above the bonding regions of molecules \\
upper cut & $C^2$, $x = 0.92$--1 & term and its first two derivatives vanish in the uniform gas \\
\multicolumn{3}{l}{\emph{Chosen (Supplementary Table~\ref{tab:gatesens})}}\\
gradient threshold & $p_t = 0.15$ (0.08) & flat tops of the shell maxima; checked on 27 solids with d bands \\
density cap of the N-shell range & 6 bohr$^{-3}$ (0.3 in $\ln n$) & keeps the outer 4d shell, drops the complete shells beneath it \\
\bottomrule
\end{tabular}
\end{table}

\begin{figure}[h]
\centering
\includegraphics[width=\textwidth]{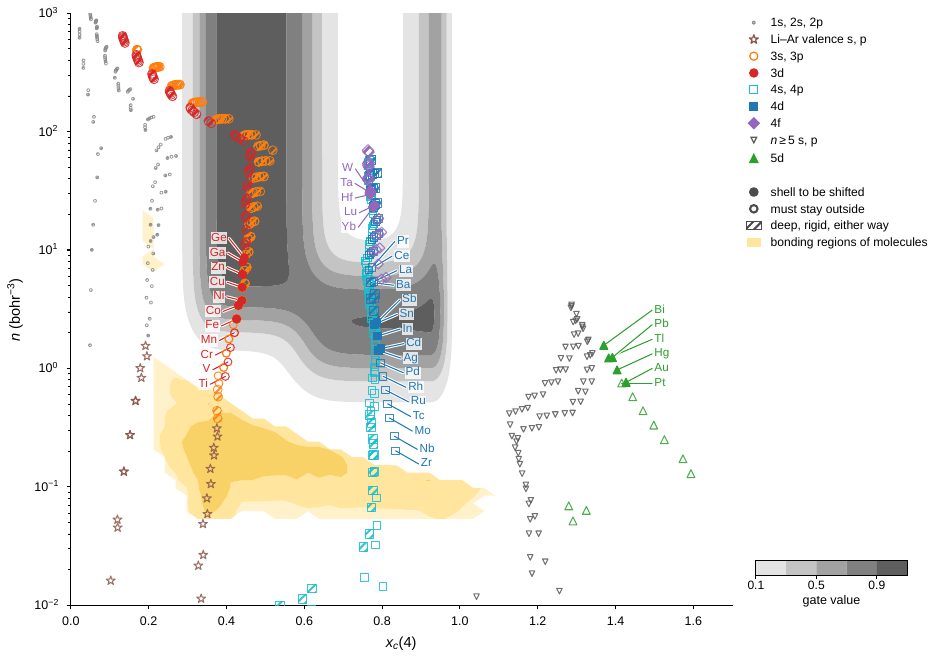}
\caption{\textbf{The shell-term gate and the shells of free atoms.} Grey, the gate $\Theta$ of LOCUS in the plane of the shell descriptor $x = x_c(4)$ and the density $n$, with the gradient factor set to one. Markers, radial-density maxima of the occupied shells of spherically averaged all-electron atoms from Li to Bi (PBE, scalar-relativistic X2C, x2c-TZVPall). Solid markers, the outermost d or f shells whose levels the term is meant to shift (the filled 3d shells of Cu to Ge and 4d shells of Ag to Sb, and the 3d shells of Fe to Ni; level within 35~eV of the vacuum); hollow markers, shells that must remain outside (valence shells, the d shells of Sc to Mn and Y to Pd, and the shells at the radius of open 4f shells); hatched markers, deep filled shells for which the gate only adds an element constant. Yellow, electron-weighted distribution of the bonding regions of 88 molecules of GMTKN55\cite{goerigk2017} (PBE0 densities) that pass the $x$ and gradient factors.}
\label{fig:gate}
\end{figure}

Supplementary Fig.~\ref{fig:gate} shows these maxima in the plane of $x$ and $n$ together with the gate. The shells the term is meant to shift, the 3d shells of Fe to Ge and the filled 4d shells of Ag to Sb, lie inside the gate, those of Fe to Ni and Ag to Cd where the gate is below its full value. The partially filled 3d shells of Ti to Mn and the 4d shells of Zr to Ru lie below the density floors, as do the bonding regions of molecules. Two groups of filled shells remain outside: the 5d shells of Au to Bi, at $x > 1.2$, whose levels are left to our correction, and the shallow 4f$^{14}$ shells of Yb to W, which lie above the density cap of the N-shell range; the same cap excludes the open 4f shells of La to Tm. The 4d shells of Rh and Pd, which are meant to stay outside, lie on the lower edge of the gate, with 12 and 15\% of the electrons of a free-atom 4d orbital inside it (gate-weighted, gradient factor included), the only place where the descriptors do not follow this classification. The 5d and 4f shells are left out for the following reasons. The 5d shells have no flat top at their maxima ($p \ge 0.24$ in the free atoms, above the threshold 0.15), so a gate could enclose only their flanks, and a shell on the edge of the gate responds to its environment through the edge, as a pressure and a level shift that depend on the bonding, instead of receiving a constant correction. The descriptors also do not separate the 5d shells of Pt and Au, whereas r$^2$SCAN places the d bands of Pt metal too deep and those of Au metal too shallow, so a term that only lowers levels would worsen Pt. The density cap of the N-shell range is set to keep out the N shells of La to Tm, which share their radius with the empty or partly filled 4f orbitals: because the term shifts all states in the regions it covers, including them would lower the empty 4f conduction bands and narrow the gaps of oxides such as CeO$_2$ (Supplementary Table~\ref{tab:gatesens}). At the adopted cap the 4d shell of La still lies inside the gate and those of Ce to Nd on its edge, so some of this narrowing remains, a limitation of a gate on the density alone. The filled 4f$^{14}$ shells of Yb to W lie in the same density range and are excluded with them. Extending the term to the 5d and 4f shells, for example with a 5d window in $x$ and $n$ without the gradient factor, or with descriptors that separate filled from empty 4f shells, is left to future work.

\begin{table}[H]
\centering
\caption{\textbf{Sensitivity of the shell gate} (base functional alone, gate before the $C^2$ upper cut). Depth of the semicore d band (eV; experiment CdTe 10.5, GaAs 18.9), Kohn--Sham gaps (eV), lattice-constant errors (\%), first-order changes of the formation energies of 11 oxides (eV per atom; the largest change or a bound) and the W4-11 atomization energies (MAE, kcal mol$^{-1}$). First-order changes are differences of the shell-term energy from that of the adopted gate, evaluated on fixed PBE densities. Formation energies are per atom of the oxide minus its elemental solids (Ag$_2$O, BaO, Cu$_2$O, Fe$_2$O$_3$, HfO$_2$, MoO$_3$, NiO, SnO$_2$, WO$_3$, ZnO and ZrO$_2$; the O$_2$ reference is omitted). First-order $dE/d\ln a$ is the change of the derivative of the energy with respect to the logarithm of the lattice constant, from lattices scaled by 0.99 and 1.01; positive values contract the lattice. The variants of the $x$-cut width, $p_t$ and the density cap also include the merged range boundaries.}
\label{tab:gatesens}
\footnotesize
\begin{tabularx}{\textwidth}{@{}>{\raggedright\arraybackslash}X cc>{\raggedright\arraybackslash}p{3.3cm}>{\raggedright\arraybackslash}p{2.0cm}c@{}}
\toprule
Change & d band & Gaps & Lattice error (\%) & Formation & W4-11 \\
 & CdTe / GaAs & CdTe / InSb / CeO$_2$ & Ag / Pd / Mo / Fe & (eV per atom) & \\
\midrule
adopted & 9.16 / 17.09 & 1.32 / 0.74 / 1.89 & +1.47 / +1.01 / $-0.49$ / +0.49 & -- & 4.10 \\
range boundaries merged (0.61 / 1.0) & 9.16 / 17.09 & 1.32 / 0.74 / 1.90 & +1.47 / -- / -- / +0.54 & $\le 0.004$ & 4.11 \\
width of the $x$ cut 0.02 $\to$ 0.03 & unchanged & unchanged & Fe: first-order $dE/d\ln a$ +0.25 eV per atom & $\le 0.012$ (NiO) & 4.11 \\
$p_t$ 0.15 $\to$ 0.12 / 0.20 & -- & -- & Nb--Rh: first-order $dE/d\ln a$ $\mp$0.15--0.38 eV per atom & NiO +0.032 / $-0.027$, ZnO +0.021 / $-0.040$ & 4.11 / 4.12 \\
$p$ cap 0.25 in the N-shell range & 9.40 / 17.10 & 1.27 / 0.69 / 1.83 & +1.45 / +0.84 / $-0.75$ / -- & $\le 0.009$ & 4.12 \\
N-shell range: density cap 6 $\to$ 10 / 4.5 & -- & CeO$_2$ 1.74 / -- & Sn $-0.12$, Pd $-0.04$ (cap 10) & SnO$_2$ $-0.025$ / +0.022 & 4.11 \\
density floor 1 $\to$ 0.6 & 9.27 / 17.13 & 1.31 / 0.73 / 1.91 & +1.66 / +1.15 / $-0.89$ / -- & -- & 4.18 \\
\bottomrule
\end{tabularx}
\end{table}

The shell term changes the structures little. On 13 solids (Al, Cu, Mo, Pd, Si, MgO, Ir, Pt, Ag, Au, Ca, Na, W), self-consistent equations of state (Birch--Murnaghan fits to five volumes between 0.94 and 1.06 of the experimental one) give mean absolute errors of the lattice constants of 0.66\% for LOCUS and 0.62\% for r$^2$SCAN + rVV10 against experiment corrected for zero-point anharmonic expansion\cite{hao2012} (Supplementary Table~\ref{tab:latt}). The shell term expands the lattices of Ag and Cu and contracts that of Mo, and changes the others by less than 0.1\%.

\begin{table}[H]
\centering
\caption{\textbf{Lattice constants of 13 solids.} Error (\%) of the self-consistent lattice constant against experiment corrected for zero-point anharmonic expansion\cite{hao2012}.}
\label{tab:latt}
\small
\begin{tabular}{lrr}
\toprule
Solid & r$^2$SCAN + rVV10 & LOCUS \\
\midrule
Ag & $+0.88$ & $+1.50$ \\
Al & $-0.66$ & $-0.66$ \\
Au & $+1.93$ & $+1.87$ \\
Ca & $-0.07$ & $-0.11$ \\
Cu & $-0.21$ & $+0.26$ \\
Ir & $+0.36$ & $+0.29$ \\
MgO & $+0.86$ & $+0.86$ \\
Mo & $+0.37$ & $-0.51$ \\
Na & $+0.15$ & $+0.15$ \\
Pd & $+1.00$ & $+1.00$ \\
Pt & $+0.87$ & $+0.80$ \\
Si & $+0.38$ & $+0.37$ \\
W & $+0.31$ & $+0.24$ \\
\midrule
MAE & 0.62 & 0.66 \\
\bottomrule
\end{tabular}
\end{table} 

The shell term also leaves the thermochemistry of its parent functional nearly unchanged (Supplementary Table~\ref{tab:energetics}). For 542 compounds of a curated set of experimental formation enthalpies\cite{wang2021}, restricted to values with an uncertainty of at most 0.05~eV per atom or a confirming NIST-JANAF table, self-consistent GPAW calculations (PAW, 600~eV, Materials Project\cite{jain2013} structures of the measured polymorphs, several magnetic orders tested for open-shell 3d compounds) give an MAE of 0.081~eV per atom for LOCUS, against 0.077~eV for r$^2$SCAN + rVV10, 0.076~eV for r$^2$SCAN and 0.133~eV for PBE. The shell term lowers the errors of the 3d and ternary oxides (0.066 and 0.063~eV per atom, against 0.089 and 0.077~eV) and raises those of the transition-metal intermetallics (0.097 against 0.085~eV). For the W4-11, BH76, G21IP and G21EA subsets of GMTKN55\cite{goerigk2017}, evaluated non-self-consistently on PBE0 densities, LOCUS and r$^2$SCAN + rVV10 differ by at most 0.02~kcal~mol$^{-1}$ in the MAE.

\begin{table}[H]
\centering
\caption{\textbf{Thermochemistry of the base functionals.} MAE of formation enthalpies (eV per atom) against the curated experimental set of ref.~\citenum{wang2021}, by class of compound (TM, transition metal), and MAE of four GMTKN55 subsets\cite{goerigk2017}. The molecular values of LOCUS use the gate before its $C^2$ upper cut, as in Supplementary Table~\ref{tab:gatesens}; the final upper cut changes any of these reaction energies by less than 0.01~kcal~mol$^{-1}$.}
\label{tab:energetics}
\small
\begin{tabular}{lrrrrr}
\toprule
Set & $n$ & PBE & r$^2$SCAN & r$^2$SCAN + rVV10 & LOCUS \\
\midrule
3d oxides & 16 & 0.330 & 0.087 & 0.089 & 0.066 \\
4d/5d oxides & 15 & 0.157 & 0.096 & 0.112 & 0.106 \\
ternary oxides & 59 & 0.299 & 0.083 & 0.077 & 0.063 \\
TM fluorides and chlorides & 44 & 0.262 & 0.092 & 0.116 & 0.122 \\
TM chalcogenides & 28 & 0.090 & 0.064 & 0.068 & 0.073 \\
TM nitrides, carbides, borides & 38 & 0.096 & 0.087 & 0.093 & 0.095 \\
TM silicides and germanides & 55 & 0.049 & 0.057 & 0.059 & 0.072 \\
TM--TM intermetallics & 110 & 0.075 & 0.086 & 0.085 & 0.097 \\
TM--sp intermetallics & 76 & 0.057 & 0.062 & 0.060 & 0.065 \\
other ternaries (Heusler etc.) & 39 & 0.087 & 0.076 & 0.076 & 0.089 \\
TM hydrides & 2 & 0.026 & 0.041 & 0.064 & 0.068 \\
compounds without TM & 60 & 0.179 & 0.066 & 0.053 & 0.053 \\
\midrule
all & 542 & 0.133 & 0.076 & 0.077 & 0.081 \\
\midrule
\multicolumn{6}{l}{\emph{Molecules, MAE (kcal mol$^{-1}$), non-self-consistent on PBE0 densities}} \\
W4-11 & 140 & -- & 3.78 & 4.08 & 4.10 \\
BH76 & 76 & -- & 7.10 & 7.37 & 7.37 \\
G21IP & 36 & -- & 4.76 & 4.82 & 4.82 \\
G21EA & 25 & -- & 3.86 & 4.02 & 4.01 \\
\bottomrule
\end{tabular}
\end{table}

\paragraph{Open shells and oxidation states.} The gate depends on the density, the reduced gradient and the shell descriptor, not on orbital occupations or oxidation states. It was designed for the compact d shells whose energies semilocal functionals overestimate: it covers fully the filled 3d shells of Cu to Ge and 4d shells of In to Sb, and partly the shells that lie on its inner slope, the 3d shells of Fe to Ni, closed or open, and the 4d shells of Ag and Cd (Supplementary Fig.~\ref{fig:gate}); the open 3d shells of Ti to Mn lie below its density floor in the free atoms, but in oxides, where the cation shells contract, they reach into the gate. The strength $c$, fixed on d$^{10}$ atoms and ions, is used unchanged for every shell the gate covers. Supplementary Table~\ref{tab:oxshell} compares, for oxides of several oxidation states and d counts, the base functional with and without the shell term (structures and magnetic orders of Supplementary Table~\ref{tab:os} for NiO, MnO, Fe$_2$O$_3$ and Cr$_2$O$_3$, of the benchmark for ZnO and Cu$_2$O, and of the formation-enthalpy set for MnO$_2$). For the filled 3d shell of ZnO the term lowers the Zn 3d band by 0.84~eV relative to the VBM and opens the Kohn--Sham gap by 0.50~eV. In Cu$_2$O, whose VBM is formed by the Cu 3d band, the gap and the d centre relative to the VBM change by less than 0.06~eV. In the open-shell oxides it moves the occupied metal-d centres by at most 0.15~eV relative to the VBM and the empty ones by at most 0.37~eV (NiO); it changes their Kohn--Sham gaps by $-0.27$~eV (NiO) to $+0.39$~eV (MnO) and their site moments by less than 0.08~$\mu_\mathrm{B}$. On the oxides of the formation-enthalpy set the shell term changes the oxidation energies between oxides of the same metal by at most 0.27~eV per metal atom (Supplementary Table~\ref{tab:oxsteps}); the MAE of the 14 oxidation steps with grade A or B references is 0.23~eV with the term and 0.24~eV without it (PBE 0.45~eV). The largest changes are those of Cr$_2$O$_3$ $\to$ CrO$_3$ and Cu$_2$O $\to$ CuO, which the term moves away from experiment, and those of SnO $\to$ SnO$_2$ and FeO $\to$ Fe$_3$O$_4$, which it moves towards it. On the 11 magnetic oxides of the same set the site moments change by at most 0.04~$\mu_\mathrm{B}$.

\begin{table}[H]
\centering
\caption{\textbf{Shell term on oxides of several oxidation states.} Base functional without (r$^2$SCAN + rVV10) and with the shell term (LOCUS): Kohn--Sham gap (eV), site moment of the metal (PAW sphere, $\mu_\mathrm{B}$), and the centres of the occupied (VBM $-$ 25~eV to VBM) and empty (CBM to CBM + 6~eV) metal-d states relative to the VBM (eV, d projections). Plane waves of 600~eV, the $k$-meshes and setups of the benchmark.}
\label{tab:oxshell}
\small
\begin{tabular}{llrrrrrrrr}
\toprule
 & & \multicolumn{2}{c}{Gap} & \multicolumn{2}{c}{Moment} & \multicolumn{2}{c}{Occupied d} & \multicolumn{2}{c}{Empty d} \\
Oxide & Metal & w/o & with & w/o & with & w/o & with & w/o & with \\
\midrule
ZnO & Zn$^{2+}$ d$^{10}$ & 1.13 & 1.63 & -- & -- & $-$4.45 & $-$5.29 & 5.88 & 6.49 \\
Cu$_{2}$O & Cu$^{1+}$ d$^{10}$ & 0.74 & 0.73 & -- & -- & $-$2.37 & $-$2.42 & 3.17 & 3.32 \\
NiO & Ni$^{2+}$ d$^{8}$ & 2.32 & 2.05 & 1.60 & 1.54 & $-$1.85 & $-$1.97 & 3.08 & 2.71 \\
Fe$_{2}$O$_{3}$ & Fe$^{3+}$ d$^{5}$ & 1.62 & 1.46 & 4.04 & 3.96 & $-$4.04 & $-$4.19 & 2.50 & 2.28 \\
MnO & Mn$^{2+}$ d$^{5}$ & 1.49 & 1.89 & 4.48 & 4.50 & $-$2.38 & $-$2.45 & 4.13 & 4.16 \\
MnO$_{2}$ & Mn$^{4+}$ d$^{3}$ & 0.52 & 0.39 & 2.72 & 2.76 & $-$4.81 & $-$4.77 & 2.65 & 2.57 \\
Cr$_{2}$O$_{3}$ & Cr$^{3+}$ d$^{3}$ & 2.55 & 2.51 & 2.76 & 2.80 & $-$2.71 & $-$2.63 & 3.99 & 4.02 \\
\bottomrule
\end{tabular}
\end{table}

\begin{table}[H]
\centering
\caption{\textbf{Oxidation steps.} Lower oxide + O$_2$ $\to$ higher oxide (eV per metal atom): experiment\cite{wang2021} and the error of each functional (calculation $-$ experiment), from the self-consistent calculations of Supplementary Table~\ref{tab:energetics}; the grade of a step is the lower grade of its two compounds.}
\label{tab:oxsteps}
\small
\begin{tabular}{llrrrr}
\toprule
Step & Grade & Exp. & PBE & r$^2$SCAN + rVV10 & LOCUS \\
\midrule
Cr$_{2}$O$_{3}$ $\to$ CrO$_{3}$ & B & $-$0.20 & $-$0.94 & +0.08 & +0.35 \\
Cu$_{2}$O $\to$ CuO & B & $-$0.73 & +0.18 & +0.04 & +0.17 \\
FeO $\to$ Fe$_{3}$O$_{4}$ & C & $-$1.05 & $-$0.21 & $-$0.30 & $-$0.14 \\
Fe$_{3}$O$_{4}$ $\to$ Fe$_{2}$O$_{3}$ & B & $-$0.41 & +0.16 & $-$0.02 & +0.07 \\
MoO$_{2}$ $\to$ MoO$_{3}$ & A & $-$1.63 & $-$0.17 & $-$0.24 & $-$0.09 \\
Na$_{2}$O $\to$ Na$_{2}$O$_{2}$ & A & $-$0.51 & +0.20 & +0.12 & +0.12 \\
OsO$_{2}$ $\to$ OsO$_{4}$ & B & $-$1.03 & $-$1.51 & $-$1.10 & $-$1.10 \\
ReO$_{2}$ $\to$ ReO$_{3}$ & B & $-$1.85 & $-$0.41 & $-$0.21 & $-$0.20 \\
SnO $\to$ SnO$_{2}$ & A & $-$3.08 & +1.08 & +0.44 & +0.27 \\
Ti$_{2}$O$_{3}$ $\to$ Ti$_{3}$O$_{5}$ & B & $-$0.57 & $-$0.09 & $-$0.11 & $-$0.10 \\
Ti$_{3}$O$_{5}$ $\to$ TiO$_{2}$ & B & $-$1.33 & +0.17 & $-$0.02 & +0.01 \\
VO $\to$ V$_{2}$O$_{3}$ & A & $-$1.84 & $-$0.56 & $-$0.36 & $-$0.26 \\
V$_{2}$O$_{3}$ $\to$ VO$_{2}$ & A & $-$1.08 & $-$0.18 & $-$0.07 & +0.05 \\
VO$_{2}$ $\to$ V$_{2}$O$_{5}$ & A & $-$0.64 & $-$0.34 & $-$0.19 & $-$0.14 \\
WO$_{2}$ $\to$ WO$_{3}$ & B & $-$2.62 & $-$0.25 & $-$0.33 & $-$0.33 \\
MAE (grades A and B, 14) & & & 0.45 & 0.24 & 0.23 \\
\bottomrule
\end{tabular}
\end{table}

\paragraph{Forces and molecular dynamics.} The forces of LOCUS contain, besides the terms of GPAW, the explicit dependence of the shell term's one-centre evaluation on the atomic positions: at its one-centre points it takes the Gaussian-convolved fields from the grid at $\mathbf{R}_a + \mathbf{r}$, and the pseudized one-centre tails contribute to those fields at $\mathbf{R}_a$. With these terms the analytic force equals the central finite difference of the energy ($h = 0.01$~\AA) to $1.6 \times 10^{-4}$~eV~\AA$^{-1}$ for GaAs with a displaced Ga atom (500~eV, $4 \times 4 \times 4$ $k$-points; $5 \times 10^{-5}$~eV~\AA$^{-1}$ for r$^2$SCAN + rVV10) and to $4 \times 10^{-5}$~eV~\AA$^{-1}$ for antiferromagnetic NiO with a displaced Ni atom (600~eV, $5 \times 5 \times 3$ $k$-points; $9 \times 10^{-5}$~eV~\AA$^{-1}$ for r$^2$SCAN + rVV10). Because the term takes grid fields at the one-centre points, the energy depends slightly on the position of the atoms relative to the grid: under a rigid translation of GaAs it changes with a slope of 0.015~eV~\AA$^{-1}$ at 500~eV and 0.0015~eV~\AA$^{-1}$ at 800~eV, and the analytic forces reproduce this slope. The factors of the gate are smooth (logistic functions and a $C^2$ upper cut), so that the energy and the forces vary smoothly with the positions. In NVE molecular dynamics of antiferromagnetic NiO (4 atoms, $4 \times 4 \times 3$ $k$-points, velocity Verlet with a 1~fs step for 0.2~ps, velocities drawn at 1200~K, mean temperature 420~K), the total energy varies by 0.05~meV per atom while the potential energy varies by 103~meV per atom, and the total energy drifts by less than 0.01~meV per atom and ps, as without the shell term (variation 0.06~meV per atom).

\section{The factor \texorpdfstring{$s$}{s} and the quasiparticle renormalization}
\label{sec:sz}

In the linearized quasiparticle equation, the G$_0$W$_0$ correction of a band edge is $\delta^\mathrm{GW} = Z(A + B)$ with $A = \langle\Sigma_x - v_x\rangle$ and $B = \langle\Sigma_c - v_c\rangle$. Supplementary Table~\ref{tab:gw} lists these quantities for the band edges of 34 solids (G$_0$W$_0$@PBE, plasmon-pole model, 100~eV response cutoff). The fraction of exchange that survives correlation, $1 + B/A$, is on average proportional to the screening $g$ (slope 0.99); this proportionality corresponds to the static screening assumed in dielectric-dependent hybrids and COHSEX. A fit of $\delta^\mathrm{GW}$ to $\lambda\,g\,A$ gives $\lambda = 0.676$, against $s = 0.708$ fixed on molecules for the same PBE base. The edges have $Z = 0.82$ on average (VBM 0.80, CBM 0.84), and $\delta^\mathrm{GW} = \lambda' Z g A$ gives $\lambda' = 0.79$. Hence $s \approx 0.8\,Z$: the factor represents mainly the dynamical renormalization $Z$, together with the stronger cancellation, on average, in localized states with large exchange, such as 2p-type VBMs. A likely reason why a factor fitted to molecules ($g = 1$) transfers to solids is that the product $\lambda' Z$ ($\approx s$) is similar for the frontier orbitals of molecules and for the band edges of solids. Fitted separately, the slopes of $\delta^\mathrm{GW}$ against $g\,A$ are 0.62 for the VBM and 1.07 for the CBM (root-mean-square residuals 0.42 and 0.27~eV); a single factor can therefore only be a compromise; $s$ is closer to the VBM slope, and our correction underestimates the CBM shift, which on LOCUS is compensated by the opening of the generalized Kohn--Sham gap (main text, Discussion). Full-frequency G$_0$W$_0$ for ten of these solids agrees with the plasmon-pole values within 0.06~eV per edge and 0.04~eV in the gap. For the d$^0$ oxides TiO$_2$, SrTiO$_3$ and KTaO$_3$, full-frequency G$_0$W$_0$ makes the GW correction of the VBM 0.05--0.06~eV more negative, giving $-0.10$ to $-0.28$~eV, against $-0.73$ to $-0.90$~eV from $\ES$ (Supplementary Table~\ref{tab:gw}); in full-frequency G$_0$W$_0$ the fraction of exchange that survives correlation at their VBM is 0.02--0.06, against 0.15--0.33 for CaO, MgO, LiF, ZnO and GaAs. The overestimated gaps of the d$^0$ oxides on LOCUS (Section~\ref{sec:limits}) therefore originate at the VBM.

{\small
\begin{longtable}{llrrrrrrr}
\caption{\textbf{Band-edge decomposition of G$_0$W$_0$@PBE.} $g$, screening of our correction; $A$, $B$, exchange and correlation parts (eV); $Z$, renormalization factor; $\delta^\mathrm{GW}$, quasiparticle correction; $\delta^{\ES}$, shift of our correction on the PBE base (eV).}\label{tab:gw}\\
\toprule
Solid & Edge & $g$ & $A$ & $B$ & $1 + B/A$ & $Z$ & $\delta^\mathrm{GW}$ & $\delta^{\ES}$ \\
\midrule
\endfirsthead
\toprule
Solid & Edge & $g$ & $A$ & $B$ & $1 + B/A$ & $Z$ & $\delta^\mathrm{GW}$ & $\delta^{\ES}$ \\
\midrule
\endhead
AgCl & CBM & 0.194 & 2.52 & -1.88 & 0.25 & 0.83 & 0.53 & 0.28 \\
AgCl & VBM & 0.194 & -6.35 & 5.57 & 0.12 & 0.79 & -0.62 & -0.83 \\
AlAs & CBM & 0.095 & 1.97 & -1.86 & 0.05 & 0.80 & 0.09 & 0.09 \\
AlAs & VBM & 0.095 & -3.09 & 2.37 & 0.23 & 0.79 & -0.56 & -0.19 \\
AlN & CBM & 0.219 & 2.82 & -1.71 & 0.39 & 0.83 & 0.92 & 0.28 \\
AlN & VBM & 0.219 & -5.18 & 4.41 & 0.15 & 0.80 & -0.61 & -0.72 \\
AlP & CBM & 0.114 & 2.03 & -1.84 & 0.09 & 0.80 & 0.15 & 0.10 \\
AlP & VBM & 0.114 & -3.23 & 2.43 & 0.25 & 0.78 & -0.62 & -0.24 \\
Ar & CBM & 0.512 & 2.75 & -1.44 & 0.48 & 0.91 & 1.20 & 0.71 \\
Ar & VBM & 0.512 & -6.93 & 3.27 & 0.53 & 0.86 & -3.15 & -2.58 \\
BN & CBM & 0.213 & 3.29 & -2.11 & 0.36 & 0.86 & 1.01 & 0.27 \\
BN & VBM & 0.213 & -4.81 & 4.37 & 0.09 & 0.83 & -0.37 & -0.70 \\
BP & CBM & 0.105 & 2.57 & -2.13 & 0.17 & 0.81 & 0.36 & 0.11 \\
BP & VBM & 0.105 & -2.81 & 2.40 & 0.15 & 0.80 & -0.33 & -0.22 \\
C & CBM & 0.168 & 3.74 & -2.39 & 0.36 & 0.85 & 1.14 & 0.23 \\
C & VBM & 0.168 & -3.60 & 3.44 & 0.04 & 0.84 & -0.14 & -0.45 \\
CaF$_{2}$ & CBM & 0.432 & 2.87 & -1.70 & 0.41 & 0.89 & 1.04 & 0.64 \\
CaF$_{2}$ & VBM & 0.432 & -8.87 & 6.47 & 0.27 & 0.81 & -1.95 & -2.98 \\
CaO & CBM & 0.261 & 3.98 & -2.28 & 0.43 & 0.85 & 1.44 & 0.39 \\
CaO & VBM & 0.261 & -6.51 & 5.58 & 0.14 & 0.79 & -0.73 & -1.62 \\
CdTe & CBM & 0.099 & 2.09 & -1.57 & 0.25 & 0.81 & 0.42 & 0.16 \\
CdTe & VBM & 0.099 & -3.24 & 2.53 & 0.22 & 0.79 & -0.56 & -0.23 \\
CuBr & CBM & 0.141 & 2.51 & -2.06 & 0.18 & 0.82 & 0.37 & 0.20 \\
CuBr & VBM & 0.141 & -8.86 & 7.79 & 0.12 & 0.74 & -0.79 & -0.75 \\
CuCl & CBM & 0.176 & 2.66 & -2.13 & 0.20 & 0.82 & 0.43 & 0.25 \\
CuCl & VBM & 0.176 & -9.42 & 8.20 & 0.13 & 0.74 & -0.91 & -0.98 \\
GaAs & CBM & 0.055 & 2.30 & -1.53 & 0.33 & 0.80 & 0.61 & 0.10 \\
GaAs & VBM & 0.055 & -2.82 & 2.30 & 0.18 & 0.79 & -0.41 & -0.11 \\
GaN & CBM & 0.166 & 2.43 & -1.41 & 0.42 & 0.82 & 0.83 & 0.25 \\
GaN & VBM & 0.166 & -4.87 & 4.42 & 0.09 & 0.79 & -0.35 & -0.59 \\
GaP & CBM & 0.087 & 2.14 & -1.93 & 0.10 & 0.80 & 0.18 & 0.10 \\
GaP & VBM & 0.087 & -3.01 & 2.37 & 0.21 & 0.78 & -0.50 & -0.19 \\
InP & CBM & 0.071 & 2.13 & -1.72 & 0.19 & 0.79 & 0.33 & 0.11 \\
InP & VBM & 0.071 & -2.88 & 2.34 & 0.19 & 0.77 & -0.42 & -0.16 \\
KCl & CBM & 0.382 & 2.47 & -1.61 & 0.35 & 0.88 & 0.76 & 0.48 \\
KCl & VBM & 0.382 & -5.73 & 3.29 & 0.43 & 0.82 & -2.01 & -1.78 \\
KTaO$_{3}$ & CBM & 0.200 & 3.29 & -2.06 & 0.37 & 0.83 & 1.02 & 0.39 \\
KTaO$_{3}$ & VBM & 0.200 & -6.40 & 6.10 & 0.05 & 0.77 & -0.23 & -0.90 \\
Kr & CBM & 0.456 & 2.54 & -1.46 & 0.42 & 0.90 & 0.97 & 0.60 \\
Kr & VBM & 0.456 & -6.10 & 3.01 & 0.51 & 0.86 & -2.66 & -1.93 \\
LiCl & CBM & 0.309 & 2.87 & -1.78 & 0.38 & 0.85 & 0.93 & 0.38 \\
LiCl & VBM & 0.309 & -5.48 & 3.49 & 0.36 & 0.81 & -1.62 & -1.08 \\
LiF & CBM & 0.493 & 3.39 & -1.42 & 0.58 & 0.89 & 1.76 & 0.77 \\
LiF & VBM & 0.493 & -8.94 & 6.01 & 0.33 & 0.83 & -2.42 & -3.04 \\
MgO & CBM & 0.313 & 2.87 & -1.42 & 0.51 & 0.86 & 1.25 & 0.46 \\
MgO & VBM & 0.313 & -6.72 & 5.41 & 0.19 & 0.81 & -1.06 & -1.37 \\
NaCl & CBM & 0.367 & 2.59 & -1.59 & 0.39 & 0.87 & 0.87 & 0.47 \\
NaCl & VBM & 0.367 & -5.64 & 3.35 & 0.41 & 0.82 & -1.88 & -1.37 \\
NaF & CBM & 0.540 & 2.91 & -1.29 & 0.55 & 0.91 & 1.46 & 0.84 \\
NaF & VBM & 0.540 & -8.99 & 5.92 & 0.34 & 0.83 & -2.53 & -3.64 \\
Ne & CBM & 0.787 & 3.26 & -0.86 & 0.74 & 0.96 & 2.30 & 1.30 \\
Ne & VBM & 0.787 & -11.80 & 5.33 & 0.55 & 0.88 & -5.72 & -5.56 \\
SiC & CBM & 0.140 & 2.29 & -1.79 & 0.22 & 0.83 & 0.41 & 0.14 \\
SiC & VBM & 0.140 & -3.86 & 3.42 & 0.12 & 0.80 & -0.36 & -0.35 \\
Si & CBM & 0.070 & 2.04 & -1.94 & 0.05 & 0.79 & 0.08 & 0.07 \\
Si & VBM & 0.070 & -2.49 & 1.91 & 0.23 & 0.77 & -0.45 & -0.13 \\
SrO & CBM & 0.258 & 2.93 & -1.94 & 0.34 & 0.86 & 0.85 & 0.34 \\
SrO & VBM & 0.258 & -6.51 & 5.55 & 0.15 & 0.78 & -0.75 & -1.90 \\
SrTiO$_{3}$ & CBM & 0.169 & 5.93 & -3.79 & 0.36 & 0.78 & 1.66 & 0.37 \\
SrTiO$_{3}$ & VBM & 0.169 & -6.30 & 6.16 & 0.02 & 0.77 & -0.10 & -0.76 \\
TiO$_{2}$ & CBM & 0.162 & 6.07 & -3.92 & 0.35 & 0.78 & 1.68 & 0.36 \\
TiO$_{2}$ & VBM & 0.162 & -6.16 & 6.12 & 0.01 & 0.76 & -0.04 & -0.73 \\
ZnO & CBM & 0.188 & 1.97 & -1.28 & 0.35 & 0.83 & 0.58 & 0.29 \\
ZnO & VBM & 0.188 & -7.76 & 6.66 & 0.14 & 0.78 & -0.86 & -0.98 \\
ZnS & CBM & 0.158 & 2.69 & -2.06 & 0.23 & 0.82 & 0.51 & 0.23 \\
ZnS & VBM & 0.158 & -4.81 & 3.76 & 0.22 & 0.79 & -0.82 & -0.52 \\
ZnSe & CBM & 0.123 & 2.39 & -1.76 & 0.26 & 0.82 & 0.51 & 0.20 \\
ZnSe & VBM & 0.123 & -4.31 & 3.41 & 0.21 & 0.79 & -0.72 & -0.36 \\
\bottomrule
\end{longtable}
}

\section{Molecules}
\label{sec:mol}

\begin{figure}[h]
\centering
\includegraphics[width=\textwidth]{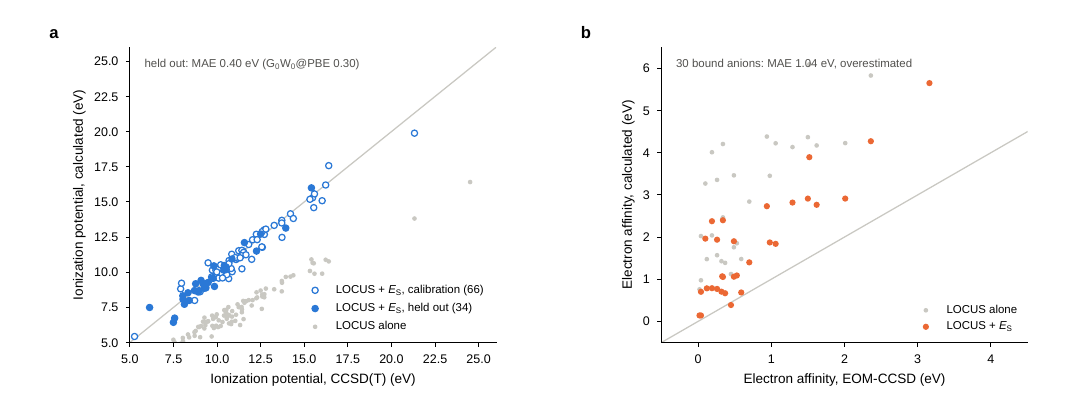}
\caption{\textbf{Molecular ionization potentials and electron affinities.} \textbf{a}, Ionization potentials of GW100 molecules against CCSD(T)\cite{krause2015}: LOCUS with our correction for the 66 molecules that fix $s$ (open) and the 34 held out (filled), and LOCUS alone (grey). \textbf{b}, Electron affinities of the bound anions against EOM-CCSD\cite{lange2018}, with our correction (orange) and LOCUS alone (grey). Grey diagonals, $y = x$.}
\label{fig:mol}
\end{figure}

Isolated molecules are the limit $g = 1$ of our correction (Section~\ref{sec:screen}). Supplementary Fig.~\ref{fig:mol} compares the corrected levels of the GW100 molecules computed with PySCF (def2-TZVPP) with coupled-cluster references: CCSD(T) ionization potentials\cite{krause2015} and EOM-CCSD electron affinities\cite{lange2018}, the latter only for bound anions (reference EA $> 0$). The 66 molecules that fix $s$ are shown separately from the 34 held out. On the held-out molecules the corrected ionization potentials have an MAE of 0.40~eV with a mean error of $-0.12$~eV, compared with 0.30~eV for G$_0$W$_0$@PBE\cite{vansetten2015} and 3.07~eV for LOCUS alone (Supplementary Table~\ref{tab:molsum}). Over the 30 bound anions, the MAE of the electron affinities falls from 2.31~eV for LOCUS alone to 1.04~eV, but the affinities remain overestimated in 29 of the 30 cases (KBr is the exception): the CBM bare shift is insufficient for the diffuse lowest unoccupied orbitals, the same deficit as in wide-gap insulators (Section~\ref{sec:wide}). Refitting $s$ on the unoccupied levels would double our correction of the CBM in solids and degrade the accuracy on the benchmark of solids; the electron affinities of molecules are therefore a limitation of the present form. Supplementary Table~\ref{tab:molall} lists every molecule.

\begin{table}[H]
\centering
\caption{\textbf{Molecules: summary.} MAE and mean error (ME, eV) of ionization potentials against CCSD(T) and of electron affinities of bound anions against EOM-CCSD.}
\label{tab:molsum}
\small
\begin{tabular}{llrrr}
\toprule
Set & Method & $n$ & MAE & ME \\
\midrule
IP, calibration (66) & LOCUS alone & 66 & 4.09 & -4.09 \\
IP, calibration (66) & LOCUS + $E_\mathrm{S}$ & 66 & 0.47 & -0.12 \\
IP, calibration (66) & PBE + $E_\mathrm{S}$ & 66 & 0.49 & -0.11 \\
IP, calibration (66) & G$_0$W$_0$@PBE & 66 & 0.44 & -0.44 \\
IP, held out (34) & LOCUS alone & 34 & 3.07 & -3.07 \\
IP, held out (34) & LOCUS + $E_\mathrm{S}$ & 34 & 0.40 & -0.12 \\
IP, held out (34) & PBE + $E_\mathrm{S}$ & 34 & 0.50 & 0.01 \\
IP, held out (34) & G$_0$W$_0$@PBE & 34 & 0.30 & -0.26 \\
EA, bound anions & LOCUS alone & 30 & 2.31 & 2.31 \\
EA, bound anions & LOCUS + $E_\mathrm{S}$ & 30 & 1.04 & 1.04 \\
EA, bound anions & PBE + $E_\mathrm{S}$ & 0 & -- & -- \\
\bottomrule
\end{tabular}
\end{table}

{\footnotesize
\begin{longtable}{llcrrrrr}
\caption{\textbf{Molecules: every system} (eV). IP: CCSD(T), LOCUS with our correction and G$_0$W$_0$@PBE; EA: EOM-CCSD and LOCUS with our correction, for bound anions. ``cal.'': used to fix $s$.}\label{tab:molall}\\
\toprule
Molecule & Formula & & IP ref. & IP $+\ES$ & IP G$_0$W$_0$ & EA ref. & EA $+\ES$ \\
\midrule
\endfirsthead
\toprule
Molecule & Formula & & IP ref. & IP $+\ES$ & IP G$_0$W$_0$ & EA ref. & EA $+\ES$ \\
\midrule
\endhead
Adenine & C$_{5}$H$_{5}$N$_{5}$ &  & 8.33 & 8.54 & 8.15 & -- & -- \\
Aluminum triiodide & AlI$_{3}$ &  & 9.72 & 9.57 & 9.58 & 0.33 & 1.07 \\
Aniline & C$_{6}$H$_{7}$N &  & 8.04 & 8.08 & 7.79 & -- & -- \\
Arsenic dimer & As$_{2}$ &  & 9.85 & 8.99 & 9.66 & 0.26 & 1.94 \\
Arsine & H$_{3}$As &  & 10.40 & 10.14 & 10.20 & -- & -- \\
Bromine & Br$_{2}$ &  & 10.52 & 10.36 & 10.56 & 0.94 & 2.73 \\
Butane & C$_{4}$H$_{10}$ &  & 11.57 & 12.11 & 11.59 & -- & -- \\
Carbon oxyselenide & COSe &  & 10.47 & 10.22 & 10.43 & -- & -- \\
Carbon tetrabromide & CBr$_{4}$ &  & 10.41 & 10.47 & 10.22 & 0.49 & 1.90 \\
Carbon tetraiodide & CI$_{4}$ &  & 9.19 & 9.19 & 9.05 & 1.62 & 2.76 \\
Copper cyanide & CCuN &  & 10.85 & 10.94 & 9.52 & 0.98 & 1.87 \\
Copper dimer & Cu$_{2}$ &  & 7.57 & 6.74 & 6.63 & 0.34 & 1.06 \\
Cyclooctatetraene & C$_{8}$H$_{8}$ &  & 8.40 & 7.99 & 8.18 & -- & -- \\
Cytosine & C$_{4}$H$_{5}$N$_{3}$O &  & 8.77 & 9.18 & 8.45 & -- & -- \\
Dipotassium & K$_{2}$ &  & 4.07 & 3.77 & 4.08 & 0.32 & 0.71 \\
Dirubidium & Rb$_{2}$ &  & 3.94 & 3.56 & 3.88 & 0.37 & 0.67 \\
Ethoxyethane & C$_{4}$H$_{10}$O &  & 9.82 & 10.45 & 9.43 & -- & -- \\
Ethylbenzene & C$_{8}$H$_{10}$ &  & 8.92 & 8.59 & 8.66 & -- & -- \\
Gallium monochloride & ClGa &  & 9.77 & 9.56 & 9.74 & -- & -- \\
Germane & H$_{4}$Ge &  & 12.50 & 12.71 & 12.12 & -- & -- \\
Guanine & C$_{5}$H$_{5}$N$_{5}$O &  & 8.03 & 8.32 & 7.87 & -- & -- \\
Iodine & I$_{2}$ &  & 9.48 & 9.27 & 9.58 & 1.50 & 2.91 \\
Krypton & Kr &  & 13.94 & 13.14 & 13.89 & -- & -- \\
Pentasilane & H$_{12}$Si$_{5}$ &  & 9.27 & 8.91 & 9.05 & -- & -- \\
Phenol & C$_{6}$H$_{6}$O &  & 8.70 & 8.70 & 8.51 & -- & -- \\
Potassium bromide & BrK &  & 8.13 & 7.72 & 7.61 & 0.45 & 0.39 \\
Potassium hydride & HK &  & 6.13 & 7.49 & 4.97 & 0.04 & 0.14 \\
Silver dimer & Ag$_{2}$ &  & 7.49 & 6.44 & 7.97 & 0.70 & 1.40 \\
Thymine & C$_{5}$H$_{6}$N$_{2}$O$_{2}$ &  & 9.08 & 9.42 & 8.87 & -- & -- \\
Titanium tetrafluoride & F$_{4}$Ti &  & 15.41 & 16.00 & 14.08 & 1.06 & 1.84 \\
Toluene & C$_{7}$H$_{8}$ &  & 8.97 & 8.69 & 8.72 & -- & -- \\
Vinyl bromide & C$_{2}$H$_{3}$Br &  & 9.27 & 8.86 & 9.14 & -- & -- \\
Vinyl iodide & C$_{2}$H$_{3}$I &  & 9.33 & 8.93 & 9.19 & -- & -- \\
Xenon & Xe &  & 12.26 & 11.50 & 12.33 & -- & -- \\
Acetaldehyde & C$_{2}$H$_{4}$O & cal. & 10.21 & 10.53 & 9.66 & -- & -- \\
Acetylene & C$_{2}$H$_{2}$ & cal. & 11.42 & 10.24 & 11.09 & -- & -- \\
Aluminum trifluoride & AlF$_{3}$ & cal. & 15.32 & 15.19 & 14.48 & -- & -- \\
Ammonia & H$_{3}$N & cal. & 10.81 & 10.26 & 10.39 & -- & -- \\
Argon & Ar & cal. & 15.54 & 14.59 & 15.28 & -- & -- \\
Benzene & C$_{6}$H$_{6}$ & cal. & 9.36 & 8.89 & 9.10 & -- & -- \\
Beryllium monoxide & BeO & cal. & 9.94 & 9.90 & 8.51 & 2.01 & 2.91 \\
Borane & H$_{3}$B & cal. & 13.27 & 13.33 & 12.96 & -- & -- \\
Boron nitride & BN & cal. & 11.98 & 10.92 & 11.09 & 3.16 & 5.65 \\
Carbon dioxide & CO$_{2}$ & cal. & 13.71 & 13.69 & 13.46 & -- & -- \\
Carbon disulfide & CS$_{2}$ & cal. & 9.98 & 10.02 & 9.95 & -- & -- \\
Carbon monoxide & CO & cal. & 14.21 & 14.16 & 13.71 & -- & -- \\
Carbon oxysulfide & COS & cal. & 11.17 & 10.99 & 11.11 & -- & -- \\
Carbon tetrachloride & CCl$_{4}$ & cal. & 11.50 & 11.50 & 11.21 & -- & -- \\
Carbon tetrafluoride & CF$_{4}$ & cal. & 16.23 & 16.20 & 15.60 & -- & -- \\
Chlorine & Cl$_{2}$ & cal. & 11.41 & 11.18 & 11.31 & 0.19 & 2.38 \\
Cyclopentadiene & C$_{5}$H$_{6}$ & cal. & 8.71 & 7.99 & 8.45 & -- & -- \\
Cyclopropane & C$_{3}$H$_{6}$ & cal. & 10.87 & 10.04 & 10.65 & -- & -- \\
Diborane & H$_{6}$B$_{2}$ & cal. & 12.25 & 12.70 & 11.93 & -- & -- \\
Disilane & H$_{6}$Si$_{2}$ & cal. & 10.64 & 10.23 & 10.41 & -- & -- \\
Ethane & C$_{2}$H$_{6}$ & cal. & 12.71 & 12.69 & 12.46 & -- & -- \\
Ethanol & C$_{2}$H$_{6}$O & cal. & 10.69 & 10.83 & 10.27 & -- & -- \\
Ethylene & C$_{2}$H$_{4}$ & cal. & 10.67 & 9.55 & 10.40 & -- & -- \\
Fluorine & F$_{2}$ & cal. & 15.58 & 15.57 & 15.11 & -- & -- \\
Fluoroborane & BF & cal. & 11.09 & 10.90 & 10.73 & -- & -- \\
Formaldehyde & CH$_{2}$O & cal. & 10.84 & 11.28 & 10.46 & -- & -- \\
Formic acid & CH$_{2}$O$_{2}$ & cal. & 11.42 & 11.55 & 10.87 & -- & -- \\
Helium & He & cal. & 24.51 & 27.19 & 23.49 & -- & -- \\
Hexafluorobenzene & C$_{6}$F$_{6}$ & cal. & 9.93 & 10.15 & 9.74 & -- & -- \\
Hydrazine & H$_{4}$N$_{2}$ & cal. & 9.69 & 9.67 & 9.37 & -- & -- \\
Hydrogen & H$_{2}$ & cal. & 16.40 & 17.58 & 15.85 & -- & -- \\
Hydrogen azide & HN$_{3}$ & cal. & 10.68 & 10.61 & 10.55 & -- & -- \\
Hydrogen chloride & HCl & cal. & 12.59 & 11.77 & 12.36 & -- & -- \\
Hydrogen cyanide & CHN & cal. & 13.72 & 12.48 & 13.32 & -- & -- \\
Hydrogen fluoride & HF & cal. & 16.03 & 15.09 & 15.37 & -- & -- \\
Hydrogen peroxide & H$_{2}$O$_{2}$ & cal. & 11.52 & 11.41 & 11.10 & -- & -- \\
Hydrogen sulfide & H$_{2}$S & cal. & 10.31 & 9.59 & 10.13 & -- & -- \\
Lithium dimer & Li$_{2}$ & cal. & 5.27 & 5.43 & 5.06 & 0.12 & 0.78 \\
Lithium fluoride & FLi & cal. & 11.32 & 11.03 & 10.27 & 0.02 & 0.14 \\
Lithium hydride & HLi & cal. & 7.96 & 9.22 & 6.59 & -- & -- \\
Magnesium chloride & Cl$_{2}$Mg & cal. & 11.66 & 11.24 & 11.19 & 0.19 & 0.79 \\
Magnesium fluoride & F$_{2}$Mg & cal. & 13.71 & 13.50 & 12.51 & 0.04 & 0.70 \\
Magnesium monoxide & MgO & cal. & 7.91 & 8.82 & 6.75 & 1.29 & 2.82 \\
Methane & CH$_{4}$ & cal. & 14.37 & 13.82 & 14.00 & -- & -- \\
Methanol & CH$_{4}$O & cal. & 11.04 & 11.04 & 10.67 & -- & -- \\
Neon & Ne & cal. & 21.32 & 19.89 & 20.33 & -- & -- \\
Nitrogen & N$_{2}$ & cal. & 15.48 & 15.29 & 15.05 & -- & -- \\
Ozone & O$_{3}$ & cal. & 12.71 & 13.01 & 11.51 & 1.52 & 3.89 \\
Phosphine & H$_{3}$P & cal. & 10.52 & 10.20 & 10.35 & -- & -- \\
Phosphorus dimer & P$_{2}$ & cal. & 10.53 & 9.72 & 10.38 & 0.10 & 1.96 \\
Phosphorus mononitride & NP & cal. & 11.81 & 11.97 & 11.30 & -- & -- \\
Propane & C$_{3}$H$_{8}$ & cal. & 12.03 & 12.31 & 11.89 & -- & -- \\
Pyridine & C$_{5}$H$_{5}$N & cal. & 9.73 & 10.15 & 9.18 & -- & -- \\
Silane & H$_{4}$Si & cal. & 12.80 & 13.07 & 12.39 & -- & -- \\
Sodium chloride & ClNa & cal. & 9.03 & 8.61 & 8.35 & 0.59 & 0.69 \\
Sodium dimer & Na$_{2}$ & cal. & 4.95 & 4.75 & 4.91 & 0.26 & 0.77 \\
Sodium hexamer & Na$_{6}$ & cal. & 4.39 & 4.03 & 4.34 & 0.49 & 1.06 \\
Sodium tetramer & Na$_{4}$ & cal. & 4.24 & 4.06 & 4.21 & 0.53 & 1.09 \\
Sulfur dioxide & O$_{2}$S & cal. & 12.30 & 12.32 & 12.06 & 0.34 & 2.40 \\
Sulfur tetrafluoride & F$_{4}$S & cal. & 12.59 & 12.89 & 12.39 & -- & -- \\
Tetracarbon & C$_{4}$ & cal. & 11.24 & 11.53 & 10.91 & 2.36 & 4.27 \\
Uracil & C$_{4}$H$_{4}$N$_{2}$O$_{2}$ & cal. & 9.48 & 10.67 & 9.38 & -- & -- \\
Urea & CH$_{4}$N$_{2}$O & cal. & 10.05 & 10.20 & 9.46 & -- & -- \\
Vinyl chloride & C$_{2}$H$_{3}$Cl & cal. & 10.09 & 9.58 & 9.90 & -- & -- \\
Vinyl fluoride & C$_{2}$H$_{3}$F & cal. & 10.55 & 9.83 & 10.32 & -- & -- \\
Water & H$_{2}$O & cal. & 12.56 & 11.80 & 12.05 & -- & -- \\
\bottomrule
\end{longtable}
}

\section{Results for every system}
\label{sec:all}

Supplementary Table~\ref{tab:flags} gives the statistics for all systems, without the systems with each flag and without those with either flag (flags on the experimental references; main text, Methods). The GW value used for the low-reference flag is, in order of availability, $\alpha$QSGW, QPG$_0$W$_0$, or G$_0$W$_0$@PBE scaled by 1.069, the median ratio of $\alpha$QSGW to G$_0$W$_0$@PBE on the 128 solids with both\cite{grossmann2026}. The literature HSE06, mBJ and G$_0$W$_0$@PBE gaps do not include spin--orbit coupling, whereas the corrected gaps do. Shifted by the spin--orbit change of the PBE gap from ref.~\citenum{grossmann2026} (PBE + SOC $-$ PBE; median magnitude 0.04~eV), the literature HSE06, mBJ and G$_0$W$_0$@PBE gaps have MAEs of 0.51, 0.49 and 0.57~eV on the 438, 437 and 281 solids with these values, against 0.43, 0.43 and 0.42~eV for our correction on the same solids. Below 3~eV the shifted HSE06 gaps have an MAE of 0.34~eV, against 0.36~eV for our correction. Supplementary Table~\ref{tab:gaps} lists every held-out crystal, and Supplementary Tables~\ref{tab:surf}--\ref{tab:molcr} the other sets.

In each gap range (main text, Fig.~2b), on the systems with values of each GW method, the MAE of our correction is below that of G$_0$W$_0$@PBE (0.34 against 0.51~eV below 3~eV, 0.51 against 0.70~eV between 3 and 6~eV, 0.75 against 0.93~eV above 6~eV) and above that of QSG$\hat{W}$ (0.26 against 0.18~eV, 0.39 against 0.28~eV, 0.72 against 0.33~eV).

Supplementary Table~\ref{tab:gwallsum} compares all methods on the same 81 solids, those for which ref.~\citenum{grossmann2026} gives values of HSE06, mBJ, G$_0$W$_0$@PBE, QPG$_0$W$_0$ and QSG$\hat{W}$ (Supplementary Table~\ref{tab:gwall} lists them; QPG$_0$W$_0$ and QSG$\hat{W}$ include spin--orbit coupling, the other three do not). On these solids the MAE of our correction is 0.43~eV, against 0.88~eV for HSE06, 0.60~eV for mBJ and 0.55~eV for G$_0$W$_0$@PBE, and 0.28 and 0.24~eV for QPG$_0$W$_0$ and QSG$\hat{W}$. The advantage over HSE06 comes from the gaps above 3~eV (0.41 against 0.64~eV between 3 and 6~eV, 0.72 against 2.16~eV above 6~eV), which HSE06 underestimates; below 3~eV the two are close (0.26 and 0.23~eV).

\begin{table}[H]
\centering
\caption{\textbf{All methods on the same solids.} MAE and, in parentheses, mean error (eV) against experiment on the 81 held-out solids with values of every method, by range of the experimental gap. LOCUS alone without spin--orbit coupling; our correction with it; literature values from ref.~\citenum{grossmann2026} (QPG$_0$W$_0$ and QSG$\hat{W}$ with spin--orbit coupling, HSE06, mBJ and G$_0$W$_0$@PBE without it).}
\label{tab:gwallsum}
\small
\begin{tabular}{lrrrr}
\toprule
 & Below 3~eV & 3--6~eV & Above 6~eV & All \\
\midrule
Solids & 39 & 19 & 23 & 81 \\ \midrule
LOCUS alone & 0.44 (-0.19) & 1.13 (-1.12) & 3.03 (-3.03) & 1.34 (-1.21) \\
LOCUS $+\ES$ & 0.26 (0.09) & 0.41 (-0.17) & 0.72 (-0.71) & 0.43 (-0.20) \\
HSE06 & 0.23 (0.01) & 0.64 (-0.53) & 2.16 (-2.16) & 0.88 (-0.73) \\
mBJ & 0.24 (0.03) & 0.66 (-0.62) & 1.15 (-1.15) & 0.60 (-0.46) \\
G$_0$W$_0$@PBE & 0.37 (0.05) & 0.66 (-0.35) & 0.76 (-0.57) & 0.55 (-0.22) \\
QPG$_0$W$_0$ & 0.17 (-0.04) & 0.35 (-0.05) & 0.39 (-0.06) & 0.28 (-0.05) \\
QSG$\hat{W}$ & 0.18 (0.02) & 0.25 (0.03) & 0.33 (-0.01) & 0.24 (0.01) \\
\bottomrule
\end{tabular}
\end{table}

{\small
\begin{longtable}{lrrrrrrrr}
\caption{\textbf{Gaps of the 81 solids of Supplementary Table~\ref{tab:gwallsum}} (eV). Base, LOCUS alone (without spin--orbit coupling); $+\ES$, our correction on LOCUS (with spin--orbit coupling); the other columns from ref.~\citenum{grossmann2026}.}\label{tab:gwall}\\
\toprule
Solid & Exp. & Base & $+\ES$ & HSE06 & mBJ & G$_0$W$_0$@PBE & QPG$_0$W$_0$ & QSG$\hat{W}$ \\
\midrule
\endfirsthead
\toprule
Solid & Exp. & Base & $+\ES$ & HSE06 & mBJ & G$_0$W$_0$@PBE & QPG$_0$W$_0$ & QSG$\hat{W}$ \\
\midrule
\endhead
AgBr & 2.71 & 1.89 & 2.90 & 2.05 & 2.42 & 1.84 & 2.18 & 2.40 \\
AgCl & 3.25 & 2.04 & 3.38 & 2.43 & 2.83 & 2.35 & 2.73 & 2.98 \\
AgI & 2.91 & 2.16 & 2.89 & 2.50 & 2.72 & 2.92 & 2.30 & 2.49 \\
AgInO$_{2}$ & 1.90 & 0.99 & 2.30 & 1.69 & 1.23 & 1.09 & 2.07 & 2.28 \\
AlAs & 2.23 & 1.90 & 2.14 & 2.07 & 2.15 & 2.12 & 2.16 & 2.09 \\
AlN & 6.11 & 4.86 & 5.96 & 5.50 & 5.30 & 5.53 & 6.29 & 6.25 \\
AlP & 2.45 & 1.87 & 2.26 & 2.28 & 2.32 & 2.50 & 2.51 & 2.42 \\
AlSb & 1.69 & 1.39 & 1.40 & 1.70 & 1.80 & 2.21 & 1.67 & 1.60 \\
Ar & 14.15 & 9.50 & 12.66 & 10.34 & 12.55 & 13.41 & 14.01 & 13.64 \\
BN & 5.96 & 4.76 & 6.06 & 5.62 & 5.96 & 5.77 & 6.52 & 6.54 \\
BN & 6.36 & 4.93 & 5.95 & 5.79 & 5.77 & 6.71 & 6.65 & 6.50 \\
BP & 2.10 & 1.40 & 1.76 & 2.02 & 1.91 & 2.18 & 2.11 & 2.05 \\
BaS & 3.81 & 2.59 & 3.35 & 3.03 & 3.12 & 3.54 & 3.69 & 3.78 \\
BeSe & 3.80 & 3.11 & 3.57 & 3.47 & 3.53 & 4.05 & 3.92 & 3.83 \\
BeTe & 2.80 & 2.24 & 2.34 & 2.61 & 2.64 & 3.87 & 2.76 & 2.67 \\
C & 5.50 & 4.35 & 5.06 & 5.34 & 4.96 & 5.96 & 5.83 & 5.66 \\
CaF$_{2}$ & 12.10 & 7.78 & 11.29 & 9.34 & 10.45 & 10.61 & 11.39 & 12.13 \\
CaO & 6.88 & 4.23 & 6.22 & 5.31 & 5.38 & 5.26 & 6.51 & 6.95 \\
CaS & 4.43 & 2.85 & 3.66 & 3.42 & 3.34 & 3.83 & 4.33 & 4.46 \\
CaSe & 3.85 & 2.59 & 3.17 & 2.99 & 2.86 & 3.31 & 3.73 & 3.84 \\
CdS & 2.48 & 1.87 & 2.68 & 2.29 & 2.63 & 2.04 & 2.39 & 2.56 \\
CdSe & 1.74 & 1.41 & 1.93 & 1.72 & 2.00 & 1.33 & 1.67 & 1.88 \\
CdSe & 1.73 & 1.43 & 1.99 & 1.74 & 2.02 & 1.30 & 1.70 & 1.96 \\
CuBr & 3.07 & 1.34 & 2.77 & 2.18 & 1.47 & 1.39 & 2.20 & 2.56 \\
CuCl & 3.40 & 1.40 & 3.04 & 2.35 & 1.54 & 1.61 & 2.43 & 2.77 \\
CuI & 3.12 & 2.10 & 3.05 & 2.64 & 2.12 & 2.27 & 2.63 & 2.77 \\
CuScO$_{2}$ & 3.30 & 2.96 & 4.38 & 3.94 & 2.43 & 3.72 & 3.96 & 3.61 \\
CuTaN$_{2}$ & 1.50 & 1.06 & 1.46 & 2.05 & 1.56 & 1.26 & 1.68 & 1.74 \\
HfS$_{2}$ & 1.96 & 1.48 & 1.99 & 1.81 & 1.64 & 1.44 & 1.67 & 1.74 \\
HgS & 0.54 & 1.19 & 1.31 & 0.50 & 0.68 & 0.00 & 0.07 & 0.09 \\
InAs & 0.42 & 0.32 & 0.41 & 0.55 & 0.64 & 0.00 & 0.32 & 0.54 \\
InPS$_{4}$ & 3.12 & 3.22 & 3.97 & 3.59 & 3.31 & 3.55 & 4.03 & 4.09 \\
KBr & 7.80 & 4.91 & 6.66 & 5.51 & 6.46 & 6.93 & 7.18 & 7.41 \\
KCl & 8.69 & 5.74 & 7.92 & 6.52 & 7.78 & 8.16 & 8.40 & 8.77 \\
Kr & 11.59 & 8.05 & 10.37 & 8.67 & 10.64 & 11.03 & 10.98 & 10.92 \\
LiBr & 7.60 & 5.78 & 6.87 & 6.17 & 6.49 & 7.14 & 7.56 & 7.44 \\
LiCl & 9.40 & 7.16 & 8.67 & 7.74 & 8.27 & 9.01 & 9.51 & 9.33 \\
LiF & 13.60 & 10.00 & 13.76 & 11.39 & 12.08 & 13.63 & 14.61 & 14.52 \\
LiZnAs & 1.61 & 1.30 & 1.54 & 1.52 & 1.63 & 1.07 & 1.54 & 1.66 \\
LiZnN & 1.91 & 1.30 & 2.13 & 1.89 & 1.91 & 1.32 & 2.06 & 2.11 \\
LiZnP & 2.04 & 1.77 & 2.13 & 1.95 & 2.06 & 2.07 & 2.03 & 2.02 \\
Mg$_{2}$Ge & 0.69 & 0.55 & 0.63 & 0.53 & 0.61 & 0.83 & 0.56 & 0.54 \\
Mg$_{2}$Si & 0.78 & 0.56 & 0.70 & 0.59 & 0.61 & 0.64 & 0.63 & 0.59 \\
Mg$_{2}$Sn & 0.30 & 0.14 & 0.09 & 0.25 & 0.34 & 0.46 & 0.16 & 0.25 \\
MgO & 7.67 & 5.55 & 7.49 & 6.44 & 6.76 & 7.10 & 8.02 & 7.91 \\
MoSe$_{2}$ & 1.09 & 1.21 & 1.56 & 1.31 & 0.98 & 1.17 & 1.22 & 1.17 \\
Na$_{2}$KSb & 1.00 & 1.36 & 1.54 & 1.36 & 1.77 & 1.83 & 1.54 & 1.66 \\
NaBr & 7.10 & 4.85 & 6.22 & 5.32 & 6.21 & 6.59 & 6.92 & 6.97 \\
NaCl & 8.75 & 5.81 & 7.66 & 6.46 & 7.71 & 8.07 & 8.47 & 8.44 \\
NaF & 11.70 & 7.03 & 11.35 & 8.41 & 10.25 & 10.77 & 11.67 & 11.92 \\
NaI & 5.80 & 4.33 & 5.09 & 4.62 & 5.18 & 6.51 & 5.70 & 5.65 \\
Ne & 21.48 & 12.92 & 19.45 & 14.30 & 18.85 & 19.59 & 22.97 & 22.70 \\
P & 0.31 & 0.50 & 0.79 & 0.25 & 0.07 & 0.60 & 0.06 & 0.22 \\
PbS & 0.29 & 0.91 & 0.85 & 0.72 & 1.00 & 0.75 & 0.29 & 0.26 \\
PbSe & 0.15 & 1.02 & 0.82 & 0.63 & 0.83 & 0.53 & 0.19 & 0.17 \\
PbTe & 0.19 & 1.19 & 0.40 & 1.04 & 1.10 & 1.24 & 0.36 & 0.34 \\
RbBr & 7.20 & 4.86 & 6.61 & 5.51 & 6.44 & 6.84 & 7.14 & 7.15 \\
RbCl & 8.30 & 5.53 & 7.71 & 6.36 & 7.58 & 7.88 & 8.15 & 8.37 \\
RbF & 10.40 & 6.26 & 10.10 & 7.64 & 8.78 & 9.00 & 9.86 & 10.01 \\
RbI & 6.37 & 4.41 & 5.50 & 4.89 & 5.51 & 6.91 & 6.03 & 5.98 \\
ScN & 1.10 & 0.39 & 0.92 & 0.91 & 0.92 & 1.13 & 0.90 & 1.28 \\
ScNiSb & 0.11 & 0.25 & 0.56 & 0.44 & 0.33 & 0.13 & 0.43 & 0.40 \\
ScPdSb & 0.23 & 0.19 & 0.40 & 0.52 & 0.27 & 0.93 & 0.29 & 0.34 \\
SiC & 2.42 & 1.72 & 2.25 & 2.25 & 2.29 & 2.32 & 2.47 & 2.39 \\
SnO$_{2}$ & 3.60 & 1.98 & 3.63 & 3.13 & 3.23 & 2.35 & 3.81 & 4.12 \\
SnTe & 0.36 & 0.64 & 0.43 & 0.10 & 0.19 & 0.46 & 0.19 & 0.27 \\
Sn & 0.09 & 0.19 & 0.15 & 0.00 & 0.00 & 0.00 & 0.00 & 0.00 \\
SrCl$_{2}$ & 7.50 & 5.82 & 7.36 & 6.66 & 7.24 & 7.68 & 7.92 & 8.07 \\
SrO & 6.10 & 3.79 & 5.99 & 4.75 & 4.93 & 4.82 & 5.64 & 5.81 \\
SrS & 4.32 & 2.95 & 3.78 & 3.43 & 3.51 & 4.04 & 4.20 & 4.30 \\
SrSe & 3.81 & 2.72 & 3.32 & 3.04 & 3.08 & 3.54 & 3.66 & 3.72 \\
TlBr & 2.66 & 2.26 & 2.25 & 2.54 & 2.83 & 3.39 & 2.49 & 2.41 \\
TlCl & 3.23 & 2.54 & 2.91 & 2.97 & 3.36 & 3.91 & 3.32 & 3.27 \\
TlI & 2.92 & 2.21 & 2.44 & 2.52 & 2.76 & 3.87 & 2.65 & 2.59 \\
Xe & 9.29 & 6.77 & 8.01 & 7.38 & 8.15 & 10.30 & 8.95 & 8.75 \\
YNiSb & 0.18 & 0.27 & 0.54 & 0.46 & 0.36 & 0.42 & 0.51 & 0.50 \\
YPtSb & 0.16 & 0.66 & 0.53 & 0.92 & 0.61 & 0.34 & 0.49 & 0.58 \\
ZnS & 3.72 & 2.92 & 3.74 & 3.38 & 3.65 & 3.29 & 3.67 & 3.74 \\
ZnS & 3.91 & 2.96 & 3.79 & 3.41 & 3.67 & 3.35 & 3.72 & 3.82 \\
ZrS$_{2}$ & 1.68 & 1.27 & 1.77 & 1.58 & 1.29 & 1.95 & 1.79 & 1.87 \\
ZrSe$_{2}$ & 1.18 & 0.70 & 0.88 & 1.02 & 0.82 & 1.18 & 0.99 & 1.08 \\
\bottomrule
\end{longtable}
}

\begin{table}[H]
\centering
\caption{\textbf{Reference-quality flags.} MAE and mean error (eV) of our correction, and MAE of our correction and of HSE06\cite{grossmann2026} on the systems with HSE06 data.}
\label{tab:flags}
\small
\begin{tabular}{lrrrrr}
\toprule
Set & $n$ & MAE & ME & MAE (HSE06 systems) & HSE06 \\
\midrule
all & 438 & 0.43 & 0.13 & 0.43 & 0.52 \\
without low-reference flag & 402 & 0.36 & 0.05 & 0.36 & 0.49 \\
without exciton flag & 380 & 0.35 & 0.10 & 0.35 & 0.39 \\
without either flag & 357 & 0.31 & 0.05 & 0.31 & 0.37 \\
\bottomrule
\end{tabular}
\end{table}

{\footnotesize
\begin{longtable}{llrrrrrrrr}
\caption{\textbf{Band gaps of the held-out crystals} (eV). ICSD, collection code (for the systems added for this work, the structure source: exp, experimental structure; cub, orth, gamma, cubic, orthorhombic and $\gamma$ phase; mp, Materials Project identifier). Base, LOCUS alone (without spin--orbit coupling); $+\ES$ on that base, on PBE and on r$^2$SCAN; literature HSE06 and mBJ (without spin--orbit coupling) and QSG$\hat{W}$ (with spin--orbit coupling)\cite{grossmann2026}. All $+\ES$ columns include spin--orbit coupling. $^a$ low-reference flag, $^b$ exciton flag, $^c$ added for this work (not in the Borlido benchmark\cite{borlido2019}; the perovskites among them are in their high-symmetry crystal structures and compared with room-temperature gaps).}\label{tab:gaps}\\
\toprule
Solid & ICSD & Exp. & Base & $+\ES$ & PBE$+\ES$ & r$^2$SCAN$+\ES$ & HSE06 & mBJ & QSG$\hat{W}$ \\
\midrule
\endfirsthead
\toprule
Solid & ICSD & Exp. & Base & $+\ES$ & PBE$+\ES$ & r$^2$SCAN$+\ES$ & HSE06 & mBJ & QSG$\hat{W}$ \\
\midrule
\endhead
Ag$_{2}$GeS$_{3}$ & 41711 & 1.98 & 1.23 & 1.91 & 1.16 & -- & 1.08 & 1.07 & -- \\
Ag$_{2}$MoO$_{4}$$^{b}$ & 238013 & 3.40 & 2.89 & 4.14 & 3.00 & 3.60 & 3.93 & 3.91 & -- \\
Ag$_{2}$O & 247821 & 1.30 & 0.28 & 1.24 & 0.40 & 1.01 & 1.20 & 0.47 & -- \\
Ag$_{2}$PdO$_{2}$ & 51498 & 0.18 & 0.57 & 1.34 & 0.72 & 1.11 & 1.22 & 0.96 & -- \\
Ag$_{2}$S & 44507 & 0.85 & 1.37 & 2.00 & 1.51 & 1.93 & 1.34 & 1.04 & -- \\
Ag$_{2}$Se & 260148 & 0.15 & 0.03 & 0.15 & 0.35 & 0.39 & 0.01 & 0.05 & -- \\
Ag$_{3}$AsS$_{3}$ & 61806 & 2.13 & 1.38 & 2.00 & 1.56 & 1.91 & 1.84 & 1.57 & -- \\
Ag$_{3}$SbS$_{3}$ & 64986 & 1.93 & 1.20 & 1.77 & 1.28 & 1.63 & 1.68 & 1.49 & -- \\
Ag$_{5}$IO$_{6}$ & 415893 & 1.40 & 0.34 & 1.48 & 0.17 & -- & 1.03 & 0.72 & -- \\
AgAlO$_{2}$ & 99688 & 2.95 & 1.64 & 3.32 & 2.20 & 3.43 & 2.42 & 2.27 & -- \\
AgAlS$_{2}$ & 604692 & 3.13 & 2.85 & 3.73 & 2.78 & 3.47 & 3.33 & 3.19 & -- \\
AgAlSe$_{2}$ & 604704 & 2.55 & 2.12 & 2.81 & 1.90 & 2.61 & 2.07 & 2.07 & -- \\
AgAlTe$_{2}$ & 28746 & 2.35 & 2.08 & 2.39 & 1.60 & 2.26 & 2.04 & 1.96 & -- \\
AgBiP$_{2}$S$_{6}$ & 170639 & 1.70 & 1.55 & 1.69 & 1.31 & -- & 1.95 & 1.59 & -- \\
AgBiP$_{2}$Se$_{6}$ & 195334 & 1.40 & 1.77 & 1.94 & 1.47 & -- & 2.08 & 1.97 & -- \\
AgBr & 65061 & 2.71 & 1.89 & 2.90 & 1.49 & 2.33 & 2.05 & 2.42 & 2.40 \\
AgCl & 64734 & 3.25 & 2.04 & 3.38 & 1.95 & 2.74 & 2.43 & 2.83 & 2.98 \\
AgF & 18008 & 2.80 & 0.70 & 2.64 & 0.13 & 1.27 & 1.34 & 2.15 & -- \\
AgGaS$_{2}$ & 605187 & 2.71 & 1.92 & 2.71 & 1.73 & 2.37 & 2.42 & 2.50 & -- \\
AgGaSe$_{2}$ & 52570 & 1.81 & 1.19 & 1.78 & 0.83 & 1.46 & 1.61 & 1.75 & -- \\
AgGaTe$_{2}$ & 71007 & 1.32 & 1.13 & 1.35 & 0.53 & 1.10 & 1.16 & 1.15 & -- \\
AgI & 56552 & 2.91 & 2.16 & 2.89 & 1.93 & 2.60 & 2.50 & 2.72 & 2.49 \\
AgI & 56553 & 3.02 & 2.21 & 2.98 & 2.01 & 2.68 & 2.60 & 2.60 & -- \\
AgInO$_{2}$ & 202429 & 1.90 & 0.99 & 2.30 & 1.44 & 2.05 & 1.69 & 1.23 & 2.28 \\
AgInS$_{2}$ & 51617 & 2.04 & 1.16 & 1.92 & 0.98 & 1.58 & 1.59 & 1.68 & -- \\
AgInSe$_{2}$ & 52583 & 1.24 & 0.72 & 1.26 & 0.04 & 0.94 & 1.04 & 1.16 & -- \\
AgInTe$_{2}$ & 236191 & 0.96 & 0.91 & 1.12 & 0.31 & -- & 1.12 & 1.11 & -- \\
AgScP$_{2}$Se$_{6}$$^{b}$ & 195336 & 1.55 & 2.04 & 3.04 & 2.62 & -- & 2.50 & 1.95 & -- \\
Al$_{2}$O$_{3}$$^{b}$ & 31545 & 8.80 & 7.05 & 8.97 & 7.98 & 9.13 & 8.05 & 7.86 & -- \\
AlAs & 656315 & 2.23 & 1.90 & 2.14 & 1.72 & 2.18 & 2.07 & 2.15 & 2.09 \\
AlN & 602460 & 6.11 & 4.86 & 5.96 & 5.19 & 6.07 & 5.50 & 5.30 & 6.25 \\
AlPO$_{4}$$^{b}$ & 50100 & 7.89 & 6.48 & 9.61 & 9.11 & -- & 7.80 & 8.26 & -- \\
AlP & 52649 & 2.45 & 1.87 & 2.26 & 1.91 & 2.33 & 2.28 & 2.32 & 2.42 \\
AlSb & 609288 & 1.69 & 1.39 & 1.40 & 1.17 & 1.52 & 1.70 & 1.80 & 1.60 \\
Ar$^{b}$ & 24788 & 14.15 & 9.50 & 12.66 & 11.93 & 12.94 & 10.34 & 12.55 & 13.64 \\
As$_{2}$S$_{3}$ & 15239 & 2.60 & 2.27 & 2.88 & 2.59 & -- & 2.79 & 2.43 & -- \\
As$_{2}$Se$_{3}$ & 44058 & 2.15 & 1.67 & 2.12 & 1.80 & -- & 1.95 & 1.67 & -- \\
As$_{2}$Te$_{3}$ & 196146 & 0.90 & 0.62 & 0.56 & 0.44 & 0.61 & 0.75 & 0.57 & -- \\
AsI$_{3}$ & 56571 & 2.47 & 2.49 & 2.83 & 2.59 & 2.94 & 2.80 & 2.41 & -- \\
B$_{12}$P$_{2}$ & 62748 & 3.35 & 2.65 & 3.04 & 2.90 & 3.09 & 3.15 & 2.64 & -- \\
BN & 241875 & 5.96 & 4.76 & 6.06 & 5.89 & 6.31 & 5.62 & 5.96 & 6.54 \\
BN & 77271 & 6.36 & 4.93 & 5.95 & 5.41 & 6.05 & 5.79 & 5.77 & 6.50 \\
BP & 602964 & 2.10 & 1.40 & 1.76 & 1.58 & 1.81 & 2.02 & 1.91 & 2.05 \\
B & 431643 & 2.00 & 1.48 & 1.83 & 1.75 & -- & 1.50 & 1.23 & -- \\
Ba$_{2}$BiInS$_{5}$$^{a}$ & 261678 & 1.55 & 1.88 & 2.35 & 1.92 & -- & 2.12 & 2.05 & -- \\
Ba$_{2}$InTaO$_{6}$$^{a}$$^{b}$ & 261480 & 4.17 & 4.85 & 6.24 & 5.66 & 6.35 & 5.64 & 5.29 & -- \\
Ba$_{3}$MgTa$_{2}$O$_{9}$ & 240279 & 4.28 & 3.78 & 5.12 & 4.56 & 5.21 & 4.79 & 4.58 & -- \\
BaAg$_{2}$GeS$_{4}$ & 10040 & 2.02 & 1.51 & 2.27 & 1.41 & 1.99 & 1.70 & 1.71 & -- \\
BaAg$_{2}$SnS$_{4}$ & 41898 & 1.77 & 1.24 & 2.02 & 1.31 & 1.81 & 1.39 & 1.31 & -- \\
BaBiBS$_{4}$ & 248222 & 2.34 & 2.22 & 2.40 & 2.05 & 2.44 & 2.76 & 2.51 & -- \\
BaBiO$_{2}$Cl$^{a}$$^{b}$ & 79532 & 3.71 & 3.77 & 4.74 & 4.25 & 4.81 & 4.41 & 4.17 & -- \\
BaCeO$_{3}$$^{c}$ & exp & 4.10 & 2.08 & 3.49 & -- & 4.19 & -- & -- & -- \\
BaCl$_{2}$$^{b}$ & 262674 & 7.00 & 5.50 & 6.84 & 6.33 & 7.04 & 6.33 & 6.33 & -- \\
BaCu$_{2}$GeS$_{4}$ & 10006 & 2.47 & 1.59 & 2.40 & 1.62 & -- & 2.62 & 1.92 & -- \\
BaCu$_{2}$S$_{2}$ & 89573 & 2.10 & 1.51 & 2.28 & 1.58 & -- & 2.11 & 1.50 & -- \\
BaCu$_{2}$SnSe$_{4}$ & 170857 & 1.72 & 0.91 & 1.55 & 0.90 & -- & 1.71 & 0.99 & -- \\
BaCuTeF & 245624 & 2.30 & 1.47 & 2.03 & 1.40 & -- & 1.98 & 1.70 & -- \\
BaF$_{2}$$^{b}$ & 64717 & 9.10 & 7.29 & 10.87 & 10.58 & -- & 8.73 & 9.48 & -- \\
BaGe$_{2}$ & 409260 & 0.97 & 0.92 & 1.11 & 0.88 & -- & 0.93 & 0.71 & -- \\
BaHgS$_{2}$ & 251739 & 1.93 & 1.54 & 2.18 & 1.87 & 2.25 & 2.07 & 1.85 & -- \\
BaO & 616005 & 4.10 & 2.31 & 3.60 & 3.31 & 3.84 & 3.24 & 3.36 & -- \\
BaS & 52690 & 3.81 & 2.59 & 3.35 & 3.01 & 3.49 & 3.03 & 3.12 & 3.78 \\
BaSbBS$_{4}$$^{a}$ & 248221 & 2.70 & 2.79 & 3.46 & 3.13 & -- & 3.25 & 2.91 & -- \\
BaSe & 52696 & 3.42 & 2.36 & 2.90 & 2.49 & 2.99 & 2.71 & 2.74 & -- \\
BaSi$_{2}$ & 185375 & 1.15 & 1.14 & 1.35 & 1.14 & -- & 1.19 & 0.93 & -- \\
BaSnO$_{3}$ & 239582 & 3.70 & 1.82 & 3.25 & 2.25 & 3.24 & 2.70 & 2.77 & -- \\
BaZn$_{2}$As$_{2}$ & 417000 & 0.23 & 0.18 & 0.31 & 0.12 & 0.05 & 0.21 & 0.28 & -- \\
BaZnOS & 171239 & 3.90 & 2.71 & 3.63 & 3.16 & 3.68 & 3.37 & 3.30 & -- \\
BeO$^{b}$ & 391224 & 10.59 & 8.41 & 10.35 & 9.45 & 10.55 & 9.23 & 9.17 & -- \\
BeSe & 616419 & 3.80 & 3.11 & 3.57 & 3.01 & 3.59 & 3.47 & 3.53 & 3.83 \\
BeTe & 616439 & 2.80 & 2.24 & 2.34 & 2.02 & 2.46 & 2.61 & 2.64 & 2.67 \\
Bi$_{2}$O$_{2}$CO$_{3}$ & 252588 & 3.42 & 1.78 & 2.55 & 2.01 & 2.60 & 0.39 & 0.59 & -- \\
Bi$_{2}$O$_{3}$ & 189995 & 2.28 & 1.88 & 2.67 & 2.22 & -- & 2.56 & 2.34 & -- \\
Bi$_{2}$O$_{3}$$^{a}$ & 94229 & 2.50 & 2.58 & 3.43 & 2.99 & -- & 3.39 & 3.23 & -- \\
Bi$_{2}$S$_{3}$ & 153946 & 1.45 & 1.68 & 1.44 & 1.19 & -- & 2.05 & 1.96 & -- \\
Bi$_{2}$Se$_{3}$ & 165226 & 0.22 & 0.68 & 0.33 & 0.45 & 0.45 & 0.65 & 0.68 & -- \\
Bi$_{2}$Te$_{3}$ & 193330 & 0.15 & 0.84 & 0.24 & 0.29 & 0.30 & 0.62 & 0.54 & -- \\
BiCuOSe & 189174 & 0.82 & 0.80 & 0.93 & 0.42 & 0.67 & 1.23 & 0.79 & -- \\
BiF$_{3}$$^{a}$ & 25567 & 3.94 & 4.62 & 5.25 & 4.57 & -- & 5.39 & 5.19 & -- \\
BiI$_{3}$ & 53634 & 1.67 & 2.92 & 2.10 & 1.80 & 2.14 & 3.11 & 2.88 & -- \\
BiNSr$_{3}$ & 152053 & 0.89 & 0.57 & 0.67 & 0.39 & 0.76 & 0.87 & 0.86 & -- \\
BiOBr & 61225 & 2.92 & 2.59 & 2.98 & 2.48 & 3.00 & 3.14 & 3.11 & -- \\
BiOCl & 195115 & 3.46 & 2.96 & 3.46 & 2.96 & 3.51 & 3.64 & 3.60 & -- \\
BiOI & 391354 & 1.89 & 1.98 & 2.03 & 1.63 & 2.09 & 2.40 & 2.43 & -- \\
BiSBr & 31389 & 1.95 & 2.36 & 2.07 & 1.76 & 2.08 & 2.71 & 2.58 & -- \\
BiSCl & 100173 & 1.89 & 2.29 & 2.07 & 1.74 & -- & 2.60 & 2.51 & -- \\
BiSI & 23631 & 1.56 & 2.27 & 1.65 & 1.34 & 1.69 & 2.50 & 2.30 & -- \\
BiSeBr & 76649 & 1.50 & 2.04 & 1.60 & 1.26 & 1.56 & 1.69 & 1.51 & -- \\
BiSeI & 280311 & 1.30 & 2.05 & 1.40 & 1.11 & 1.40 & 2.21 & 2.03 & -- \\
C & 190650 & 5.50 & 4.35 & 5.06 & 4.83 & 5.11 & 5.34 & 4.96 & 5.66 \\
Ca$_{2}$Pb & 58920 & 0.46 & 0.27 & 0.21 & 0.12 & -- & 0.44 & 0.43 & -- \\
Ca$_{2}$Si & 158275 & 1.90 & 0.49 & 0.70 & 0.44 & 0.74 & 0.77 & 0.65 & -- \\
Ca$_{2}$SnS$_{4}$$^{a}$ & 429695 & 2.32 & 2.85 & 3.72 & 3.20 & -- & 3.56 & 3.36 & -- \\
Ca$_{2}$Sn & 659611 & 0.90 & 0.28 & 0.44 & 0.20 & 0.51 & 0.48 & 0.43 & -- \\
CaB$_{6}$ & 196516 & 1.15 & 0.02 & 0.08 & 0.11 & -- & 0.54 & 0.26 & -- \\
CaCO$_{3}$$^{a}$$^{b}$ & 52151 & 6.00 & 5.77 & 8.78 & 8.22 & -- & 7.14 & 7.14 & -- \\
CaCl$_{2}$$^{b}$ & 246416 & 6.90 & 5.95 & 7.75 & 7.13 & 7.95 & 6.97 & 6.72 & -- \\
CaF$_{2}$$^{b}$ & 82707 & 12.10 & 7.78 & 11.29 & 10.87 & 11.70 & 9.34 & 10.45 & 12.13 \\
CaH$_{2}$ & 260873 & 4.40 & 3.45 & 4.69 & 4.12 & 4.82 & 4.04 & 4.25 & -- \\
CaMg$_{2}$N$_{2}$ & 79123 & 3.25 & 2.67 & 3.53 & 2.86 & 3.62 & 3.14 & 3.37 & -- \\
CaO & 261847 & 6.88 & 4.23 & 6.22 & 5.72 & 6.47 & 5.31 & 5.38 & 6.95 \\
CaS & 603165 & 4.43 & 2.85 & 3.66 & 3.16 & 3.76 & 3.42 & 3.34 & 4.46 \\
CaSe & 52788 & 3.85 & 2.59 & 3.17 & 2.59 & 3.21 & 2.99 & 2.86 & 3.84 \\
CaSnO$_{3}$ & 193658 & 4.40 & 3.24 & 4.84 & 4.02 & -- & 4.29 & 4.17 & -- \\
CaZnOS$^{a}$ & 245309 & 3.70 & 3.48 & 4.46 & 3.65 & 4.38 & 4.00 & 4.35 & -- \\
CdAl$_{2}$S$_{4}$$^{a}$ & 83526 & 3.58 & 3.56 & 4.42 & 3.70 & 4.39 & 4.09 & 4.00 & -- \\
CdAs$_{2}$ & 609931 & 1.00 & 0.98 & 1.24 & 0.68 & 1.09 & 0.99 & 0.89 & -- \\
CdGa$_{2}$S$_{4}$ & 31354 & 3.31 & 2.97 & 3.72 & 2.98 & 3.63 & 3.15 & 3.15 & -- \\
CdGa$_{2}$Se$_{4}$ & 30908 & 2.16 & 2.16 & 2.64 & 1.92 & 2.55 & 2.47 & 2.54 & -- \\
CdGa$_{2}$Te$_{4}$$^{a}$ & 25646 & 1.45 & 1.73 & 1.83 & 1.33 & 1.82 & 2.17 & 2.10 & -- \\
CdGeAs$_{2}$ & 42098 & 0.67 & 0.30 & 0.49 & 0.35 & 0.39 & 0.61 & 0.67 & -- \\
CdGeP$_{2}$ & 100467 & 1.73 & 1.44 & 1.79 & 1.23 & 1.71 & 1.84 & 1.80 & -- \\
CdI$_{2}$ & 43852 & 3.47 & 3.11 & 3.60 & 2.91 & 3.50 & 3.33 & 3.28 & -- \\
CdIn$_{2}$S$_{4}$ & 252372 & 2.44 & 1.92 & 2.60 & 1.74 & 2.45 & 2.25 & 2.44 & -- \\
CdIn$_{2}$Se$_{4}$ & 151953 & 1.55 & 1.55 & 2.04 & 1.41 & 1.99 & 1.87 & 1.88 & -- \\
CdIn$_{2}$Se$_{4}$ & 151954 & 1.46 & 1.64 & 2.11 & 1.46 & 2.05 & 1.83 & 1.88 & -- \\
CdO$^{c}$ & exp & 0.84 & 0.45 & 1.59 & -- & -- & -- & -- & -- \\
CdP$_{4}$ & 25605 & 1.15 & 0.68 & 0.98 & 0.75 & 0.99 & 0.92 & 0.69 & -- \\
CdS & 154186 & 2.48 & 1.87 & 2.68 & 1.91 & 2.50 & 2.29 & 2.63 & 2.56 \\
CdSb$_{2}$Se$_{3}$Br$_{2}$ & 159464 & 1.40 & 1.43 & 1.76 & 1.43 & -- & 1.69 & 1.52 & -- \\
CdSb & 52830 & 0.46 & 0.47 & 0.60 & 0.11 & -- & 0.71 & 0.59 & -- \\
CdSe & 187310 & 1.74 & 1.41 & 1.93 & 1.07 & 1.72 & 1.72 & 2.00 & 1.88 \\
CdSe & 415785 & 1.73 & 1.43 & 1.99 & 1.14 & 1.79 & 1.74 & 2.02 & 1.96 \\
CdSiAs$_{2}$ & 22187 & 1.55 & 1.31 & 1.58 & 0.95 & 1.51 & 1.69 & 1.67 & -- \\
CdSiP$_{2}$ & 23696 & 2.09 & 1.60 & 1.93 & 1.66 & 1.97 & 1.98 & 1.96 & -- \\
CdSnAs$_{2}$ & 44258 & 0.26 & 0.17 & 0.34 & 0.30 & 0.04 & 0.43 & 0.51 & -- \\
CdSnO$_{3}$ & 181930 & 3.10 & 2.04 & 3.54 & 2.56 & 3.37 & 2.71 & 2.93 & -- \\
CdSnP$_{2}$ & 44257 & 1.23 & 1.05 & 1.40 & 0.78 & 1.27 & 1.39 & 1.40 & -- \\
CeO$_{2}$$^{c}$ & exp & 3.00 & 1.88 & 2.89 & -- & 3.48 & -- & -- & -- \\
CrSi$_{2}$ & 161434 & 0.35 & 0.70 & 0.92 & 0.69 & 0.94 & 1.50 & 0.63 & -- \\
Cs$_{2}$AgBiBr$_{6}$ & 291597 & 2.19 & 1.98 & 2.64 & 1.83 & 2.37 & 2.32 & 2.30 & -- \\
Cs$_{2}$AgBiCl$_{6}$ & 291598 & 2.77 & 2.49 & 3.25 & 2.48 & 2.87 & 3.01 & 2.91 & -- \\
Cs$_{2}$AgVS$_{4}$$^{b}$ & 50460 & 1.77 & 1.59 & 2.71 & 2.42 & -- & 2.41 & 1.72 & -- \\
Cs$_{2}$Au$_{2}$Br$_{6}$ & 170696 & 1.60 & 0.73 & 1.25 & 1.14 & 1.32 & 1.39 & 0.81 & -- \\
Cs$_{2}$Au$_{2}$Cl$_{6}$ & 6061 & 2.04 & 0.83 & 1.45 & 1.34 & 1.54 & 1.71 & 0.90 & -- \\
Cs$_{2}$Au$_{2}$I$_{6}$ & 59269 & 1.31 & 0.91 & 1.31 & 1.14 & 1.36 & 1.21 & 0.77 & -- \\
Cs$_{2}$Cd$_{3}$Te$_{4}$ & 90369 & 2.48 & 2.18 & 2.53 & 1.89 & -- & 2.55 & 2.65 & -- \\
Cs$_{2}$Hg$_{3}$I$_{8}$ & 4074 & 2.56 & 2.13 & 2.82 & 2.42 & 2.94 & 2.81 & 2.63 & -- \\
Cs$_{2}$SnBr$_{6}$ & 158957 & 2.70 & 2.07 & 3.29 & 2.73 & 3.32 & 2.27 & 2.19 & -- \\
Cs$_{2}$SnCl$_{6}$$^{a}$$^{b}$ & 9023 & 3.90 & 3.43 & 5.12 & 4.55 & 5.50 & 4.33 & 4.24 & -- \\
Cs$_{2}$TeBr$_{6}$ & 65058 & 2.20 & 2.28 & 3.11 & 2.81 & 3.16 & 2.59 & 2.44 & -- \\
Cs$_{2}$TeI$_{6}$ & 38105 & 1.52 & 1.56 & 1.90 & 1.72 & 1.98 & 1.85 & 1.69 & -- \\
Cs$_{2}$TiAg$_{2}$S$_{4}$ & 280645 & 2.44 & 2.07 & 3.12 & 2.68 & -- & 2.39 & 1.93 & -- \\
Cs$_{3}$Bi$_{2}$Br$_{9}$ & 1142 & 2.50 & 2.89 & 2.88 & 2.47 & 2.93 & 3.27 & 3.17 & -- \\
Cs$_{3}$Bi$_{2}$I$_{9}$$^{a}$ & 23124 & 1.80 & 2.51 & 2.58 & 2.26 & -- & 2.71 & 2.58 & -- \\
Cs$_{3}$Sb$_{2}$Br$_{9}$ & 39824 & 2.30 & 2.16 & 2.95 & 2.62 & 2.97 & 2.18 & 2.06 & -- \\
Cs$_{3}$Sb$_{2}$I$_{9}$ & 84989 & 1.89 & 1.97 & 2.63 & 2.36 & -- & 2.15 & 2.03 & -- \\
Cs$_{3}$Sb & 53243 & 1.60 & 1.33 & 1.60 & 1.04 & 1.70 & 1.41 & 1.63 & -- \\
CsAu & 58427 & 2.50 & 1.96 & 2.46 & 2.14 & 2.50 & 1.04 & 1.14 & -- \\
CsLaHgSe$_{3}$ & 281441 & 2.46 & 1.89 & 2.59 & 2.14 & -- & 2.41 & 2.39 & -- \\
CsPbBr$_{3}$ & 97851 & 2.32 & 2.40 & 2.25 & 1.71 & -- & 2.42 & 2.58 & -- \\
CsPbBr$_{3}$$^{c}$ & cub & 2.31 & 1.91 & 1.48 & -- & -- & -- & -- & -- \\
CsPbBr$_{3}$$^{c}$ & orth & 2.31 & 2.34 & 2.17 & -- & -- & -- & -- & -- \\
CsPbCl$_{3}$$^{c}$ & cub & 2.99 & 2.24 & 2.04 & -- & -- & -- & -- & -- \\
CsPbI$_{3}$$^{c}$ & cub & 1.72 & 1.57 & 0.78 & -- & -- & -- & -- & -- \\
CsPbI$_{3}$$^{c}$ & gamma & 1.72 & 1.77 & 1.23 & -- & -- & -- & -- & -- \\
CsSnBr$_{3}$ & 4071 & 1.75 & 2.00 & 2.56 & 2.34 & 2.60 & 0.75 & 0.86 & -- \\
CsSnBr$_{3}$$^{c}$ & cub & 1.81 & 0.57 & 0.69 & -- & -- & -- & -- & -- \\
CsSnCl$_{3}$$^{a}$$^{b}$ & 14199 & 2.80 & 3.44 & 4.69 & 4.19 & -- & 3.62 & 3.39 & -- \\
CsSnI$_{3}$ & 69996 & 1.31 & 0.58 & 0.75 & 0.60 & -- & 0.84 & 0.86 & -- \\
CsSnI$_{3}$$^{c}$ & cub & 1.25 & 0.39 & 0.32 & -- & -- & -- & -- & -- \\
CsSnI$_{3}$$^{c}$ & gamma & 1.25 & 0.45 & 0.62 & -- & -- & -- & -- & -- \\
CsYCdSe$_{3}$$^{a}$ & 281433 & 2.48 & 2.72 & 3.31 & 2.73 & 3.35 & 3.16 & 3.02 & -- \\
CsYHgSe$_{3}$ & 281440 & 2.54 & 1.96 & 2.52 & 2.04 & 2.57 & 2.43 & 2.33 & -- \\
CsYZnSe$_{3}$$^{a}$ & 280847 & 2.29 & 2.63 & 3.24 & 2.69 & 3.28 & 3.08 & 2.92 & -- \\
Cu$_{2}$CdGeS$_{4}$ & 26150 & 2.05 & 1.09 & 1.89 & 1.10 & 1.58 & 1.91 & 1.34 & -- \\
Cu$_{2}$CdGeSe$_{4}$ & 95235 & 1.20 & 0.37 & 0.79 & 0.39 & 0.41 & 1.00 & 0.60 & -- \\
Cu$_{2}$GeS$_{3}$ & 85138 & 1.53 & 0.57 & 1.12 & 0.49 & 0.84 & 1.41 & 0.78 & -- \\
Cu$_{2}$GeSe$_{3}$ & 160386 & 0.82 & 0.03 & 0.04 & 0.08 & -- & 0.00 & 0.00 & -- \\
Cu$_{2}$GeSe$_{3}$ & 192171 & 0.78 & 0.18 & 0.47 & 0.40 & 0.43 & 0.45 & 0.15 & -- \\
Cu$_{2}$O & 172174 & 2.17 & 0.73 & 1.84 & 1.34 & 1.75 & 2.02 & 0.77 & -- \\
Cu$_{2}$S & 16550 & 1.04 & 1.00 & 1.71 & 1.04 & 1.44 & 1.25 & 0.75 & -- \\
Cu$_{2}$SnS$_{3}$ & 91762 & 0.86 & 0.14 & 0.48 & 0.32 & 0.21 & 0.89 & 0.35 & -- \\
Cu$_{2}$SnSe$_{3}$ & 97966 & 0.84 & 0.10 & 0.45 & 0.36 & -- & 0.34 & 0.08 & -- \\
Cu$_{2}$Te & 77055 & 0.90 & 0.10 & 0.03 & -0.01 & -- & 0.00 & 0.00 & -- \\
Cu$_{2}$ZnGeSe$_{4}$ & 93409 & 1.29 & 0.24 & 0.59 & 0.25 & 0.04 & 0.97 & 0.55 & -- \\
Cu$_{2}$ZnSiS$_{4}$ & 261367 & 2.97 & 2.12 & 3.03 & 2.22 & 2.76 & 2.58 & 1.93 & -- \\
Cu$_{2}$ZnSnS$_{4}$ & 184478 & 1.51 & 0.63 & 1.33 & 0.09 & 0.96 & 1.75 & 1.16 & -- \\
Cu$_{3}$AsS$_{4}$ & 413350 & 1.24 & 0.38 & 0.82 & 0.05 & 0.47 & 1.36 & 0.65 & -- \\
Cu$_{3}$AsSe$_{4}$ & 610361 & 0.88 & 0.15 & 0.21 & 0.07 & -- & 0.35 & 0.00 & -- \\
Cu$_{3}$PS$_{4}$ & 412240 & 2.35 & 1.61 & 2.35 & 1.80 & 2.19 & 2.52 & 1.63 & -- \\
Cu$_{3}$PSe$_{4}$ & 95412 & 1.40 & 0.56 & 1.01 & 0.26 & 0.73 & 1.35 & 0.82 & -- \\
Cu$_{3}$SbS$_{4}$ & 412239 & 0.88 & 0.20 & 0.26 & 0.03 & -- & 0.88 & 0.23 & -- \\
Cu$_{3}$SbSe$_{4}$ & 400652 & 0.29 & 0.08 & 0.17 & 0.08 & -- & 0.28 & 0.00 & -- \\
CuAlO$_{2}$ & 60844 & 2.99 & 2.38 & 4.14 & 3.42 & 3.94 & 3.49 & 1.98 & -- \\
CuAlS$_{2}$ & 42124 & 3.45 & 2.50 & 3.50 & 2.58 & 3.25 & 3.28 & 2.47 & -- \\
CuAlSe$_{2}$ & 28734 & 2.68 & 1.79 & 2.62 & 1.71 & 2.37 & 2.43 & 1.88 & -- \\
CuBS$_{2}$ & 156413 & 3.61 & 2.11 & 2.87 & 2.39 & 2.76 & 2.73 & 2.06 & -- \\
CuBiP$_{2}$Se$_{6}$ & 195341 & 1.20 & 1.10 & 1.86 & 1.36 & -- & 1.87 & 1.12 & -- \\
CuBr & 78274 & 3.07 & 1.34 & 2.77 & 1.36 & 2.12 & 2.18 & 1.47 & 2.56 \\
CuCl & 78270 & 3.40 & 1.40 & 3.04 & 1.69 & 2.35 & 2.35 & 1.54 & 2.77 \\
CuGaO$_{2}$ & 60846 & 2.55 & 1.49 & 3.12 & 2.32 & 2.86 & 2.41 & 1.07 & -- \\
CuGaS$_{2}$ & 66864 & 2.51 & 1.58 & 2.45 & 1.50 & 2.05 & 2.21 & 1.63 & -- \\
CuGaSe$_{2}$ & 42097 & 1.70 & 0.95 & 1.57 & 0.63 & 1.22 & 1.35 & 0.96 & -- \\
CuGaTe$_{2}$ & 74456 & 1.23 & 1.05 & 1.26 & 0.45 & 1.00 & 1.64 & 1.28 & -- \\
CuI & 33724 & 3.12 & 2.10 & 3.05 & 1.95 & 2.60 & 2.64 & 2.12 & 2.77 \\
CuInS$_{2}$ & 186714 & 1.53 & 0.68 & 1.40 & 0.02 & 0.92 & 1.61 & 0.98 & -- \\
CuInSe$_{2}$ & 252780 & 1.04 & 0.35 & 0.82 & 0.05 & 0.02 & 0.95 & 0.51 & -- \\
CuInTe$_{2}$ & 74460 & 1.06 & 0.76 & 0.96 & 0.25 & 0.70 & 1.01 & 0.71 & -- \\
CuSbSe$_{2}$ & 418754 & 1.04 & 0.89 & 1.15 & 0.75 & -- & 1.30 & 0.81 & -- \\
CuScO$_{2}$ & 55689 & 3.30 & 2.96 & 4.38 & 3.61 & 4.02 & 3.94 & 2.43 & 3.61 \\
CuTaN$_{2}$ & 71136 & 1.50 & 1.06 & 1.46 & 0.90 & 1.29 & 2.05 & 1.56 & 1.74 \\
FAPbI$_{3}$$^{c}$ & cub & 1.48 & 1.58 & 0.94 & -- & -- & -- & -- & -- \\
FeP$_{2}$ & 15027 & 0.37 & 0.54 & 0.92 & 0.73 & 1.09 & 1.82 & 0.90 & -- \\
FeS$_{2}$ & 109377 & 0.95 & 1.03 & 1.44 & 0.98 & 1.45 & 2.22 & 1.30 & -- \\
Ga$_{2}$O$_{3}$$^{c}$ & mp886 & 4.85 & 3.19 & 4.87 & -- & -- & -- & -- & -- \\
Ga$_{2}$S$_{3}$ & 409550 & 3.44 & 2.59 & 3.34 & 2.65 & 3.30 & 3.06 & 2.97 & -- \\
GaS & 173940 & 2.54 & 1.98 & 2.59 & 2.31 & 2.71 & 2.33 & 2.24 & -- \\
GaSb & 635318 & 0.82 & 0.73 & 0.67 & 0.07 & 0.50 & 0.96 & 0.88 & -- \\
GaSe & 63122 & 2.10 & 1.69 & 2.18 & 1.68 & 2.25 & 1.85 & 1.89 & -- \\
GaTe & 8249 & 1.80 & 1.78 & 2.06 & 1.64 & 2.06 & 1.62 & 1.46 & -- \\
GeAs$_{2}$ & 23872 & 1.06 & 1.11 & 1.39 & 1.10 & -- & 1.14 & 1.00 & -- \\
GeAs & 86361 & 0.65 & 1.06 & 1.33 & 1.07 & 1.36 & 0.96 & 0.84 & -- \\
GeO$_{2}$ & 92551 & 5.35 & 2.31 & 3.97 & 3.36 & 4.22 & 3.86 & 3.99 & -- \\
GeSe & 41738 & 1.08 & 0.89 & 1.16 & 0.87 & 1.15 & 1.13 & 1.01 & -- \\
HfNBr & 95720 & 3.10 & 2.53 & 3.57 & 3.03 & 3.60 & 3.21 & 3.15 & -- \\
HfO$_{2}$$^{b}$ & 173158 & 5.50 & 4.47 & 5.94 & 5.35 & -- & 5.62 & 5.69 & -- \\
HfS$_{2}$ & 601164 & 1.96 & 1.48 & 1.99 & 1.57 & 2.03 & 1.81 & 1.64 & 1.74 \\
HfSe$_{2}$ & 195308 & 1.13 & 0.86 & 1.06 & 0.57 & 1.02 & 1.16 & 1.06 & -- \\
Hg$_{2}$I$_{2}$ & 262368 & 3.00 & 2.32 & 2.78 & 2.38 & 2.86 & 2.48 & 2.09 & -- \\
Hg$_{4}$As$_{2}$CdI$_{4}$ & 416973 & 1.45 & 0.81 & 1.26 & 0.86 & -- & 1.77 & 1.75 & -- \\
HgGa$_{2}$S$_{4}$ & 67220 & 2.79 & 2.26 & 2.94 & 2.41 & 2.94 & 2.81 & 2.76 & -- \\
HgGa$_{2}$Se$_{4}$ & 83712 & 1.99 & 1.60 & 2.02 & 1.47 & 2.00 & 2.02 & 2.01 & -- \\
HgI$_{2}$ & 181575 & 2.40 & 1.64 & 2.09 & 1.69 & 2.20 & 2.11 & 1.99 & -- \\
HgIn$_{2}$Se$_{4}$ & 25649 & 1.16 & 1.20 & 1.62 & 1.07 & 1.60 & 1.50 & 1.52 & -- \\
HgIn$_{2}$Te$_{4}$ & 25652 & 0.90 & 1.19 & 1.28 & 0.77 & 1.28 & 1.49 & 1.39 & -- \\
HgO & 14124 & 1.90 & 1.63 & 2.81 & 2.44 & 2.93 & 2.34 & 1.80 & -- \\
HgS & 81917 & 0.54 & 1.19 & 1.31 & 1.21 & 1.34 & 0.50 & 0.68 & 0.09 \\
HgS$^{a}$ & 81923 & 2.03 & 0.01 & -0.00 & -0.00 & -- & 2.55 & 2.29 & -- \\
HgSnO$_{3}$ & 260029 & 1.60 & 0.55 & 1.63 & 1.49 & 1.63 & 1.55 & 1.55 & -- \\
In$_{2}$O$_{3}$$^{c}$ & exp & 2.93 & 1.86 & 3.39 & -- & -- & -- & -- & -- \\
In$_{2}$Se$_{3}$ & 5419 & 1.36 & 0.02 & -0.16 & -0.17 & -- & 0.00 & 0.00 & -- \\
In$_{2}$Se$_{3}$ & 640498 & 1.31 & 0.11 & 0.30 & 0.26 & -- & 0.00 & 0.00 & -- \\
In$_{6}$S$_{7}$ & 15317 & 0.93 & 0.52 & 0.95 & 0.72 & -- & 0.86 & 0.63 & -- \\
In$_{6}$Se$_{7}$ & 9135 & 2.20 & 0.30 & 0.62 & 0.33 & -- & 0.48 & 0.32 & -- \\
InAs & 41444 & 0.42 & 0.32 & 0.41 & 1.00 & 0.28 & 0.55 & 0.64 & 0.54 \\
InBi$_{2}$S$_{4}$Cl$^{a}$ & 484 & 1.50 & 2.10 & 2.19 & 1.79 & 2.18 & 2.36 & 2.26 & -- \\
InBi$_{2}$Se$_{4}$Br & 159465 & 1.21 & 1.66 & 1.36 & 1.05 & 1.36 & 1.92 & 1.78 & -- \\
InGaZnO$_{4}$$^{c}$ & mp1215688 & 3.50 & 2.07 & 3.57 & -- & -- & -- & -- & -- \\
InN & 109463 & 0.70 & 0.22 & 0.63 & 1.07 & 0.36 & 0.74 & 0.89 & -- \\
InPS$_{4}$$^{a}$ & 1699 & 3.12 & 3.22 & 3.97 & 3.42 & 4.04 & 3.59 & 3.31 & 4.09 \\
InS & 81338 & 2.09 & 1.99 & 2.50 & 1.98 & 2.49 & 1.96 & 1.82 & -- \\
InSb$_{2}$S$_{4}$Br & 159467 & 1.50 & 1.84 & 2.22 & 1.88 & 2.23 & 2.17 & 1.98 & -- \\
InSb$_{2}$S$_{4}$Cl & 159468 & 1.80 & 1.84 & 2.23 & 1.90 & 2.23 & 2.21 & 2.02 & -- \\
InSb$_{2}$Se$_{4}$Br & 159466 & 1.21 & 1.40 & 1.65 & 1.31 & 1.62 & 1.61 & 1.48 & -- \\
InTe & 169418 & 1.16 & 0.08 & 0.30 & 0.28 & 0.37 & 0.16 & 0.06 & -- \\
K$_{2}$AgVS$_{4}$$^{b}$ & 66840 & 1.85 & 1.59 & 2.58 & 2.24 & 2.63 & 2.32 & 1.70 & -- \\
K$_{2}$Au$_{2}$SnS$_{4}$ & 74022 & 2.75 & 1.99 & 2.88 & 2.43 & -- & 2.84 & 2.75 & -- \\
K$_{2}$AuPS$_{4}$ & 85679 & 2.63 & 1.88 & 2.85 & 2.49 & 2.89 & 2.35 & 2.33 & -- \\
K$_{2}$CsSb$^{a}$ & 53237 & 1.00 & 1.59 & 1.85 & 1.22 & 1.94 & 1.66 & 2.07 & -- \\
K$_{2}$Hg$_{3}$Ge$_{2}$S$_{8}$ & 281506 & 2.70 & 1.98 & 2.78 & 2.41 & 2.88 & 2.50 & 2.32 & -- \\
K$_{2}$PbGe$_{2}$S$_{6}$$^{b}$ & 170601 & 2.85 & 2.46 & 3.27 & 2.96 & -- & 3.21 & 2.94 & -- \\
K$_{2}$TeBr$_{6}$$^{a}$ & 22078 & 2.17 & 2.47 & 3.26 & 2.96 & -- & 2.83 & 2.63 & -- \\
K$_{2}$TeI$_{6}$ & 23649 & 1.38 & 1.85 & 2.20 & 1.99 & -- & 2.23 & 1.99 & -- \\
K$_{2}$ZnTe$_{2}$ & 420088 & 3.00 & 2.63 & 3.08 & 2.53 & 3.06 & 2.92 & 3.00 & -- \\
K$_{3}$Sb & 44677 & 1.40 & 1.24 & 1.51 & 0.91 & 1.58 & 1.30 & 1.74 & -- \\
K$_{3}$Sb & 656327 & 1.10 & 0.78 & 1.52 & 1.19 & -- & 0.98 & 1.47 & -- \\
K$_{6}$CdTe$_{4}$ & 420087 & 2.51 & 2.04 & 2.60 & 2.17 & -- & 2.38 & 2.94 & -- \\
KAg$_{2}$PS$_{4}$ & 420033 & 3.02 & 2.27 & 3.12 & 2.23 & 2.83 & 2.30 & 2.29 & -- \\
KAg$_{2}$SbS$_{4}$ & 82143 & 1.71 & 1.45 & 2.22 & 1.42 & 1.97 & 1.37 & 1.17 & -- \\
KBiSiS$_{4}$ & 421485 & 2.25 & 2.41 & 2.43 & 2.11 & -- & 2.96 & 2.76 & -- \\
KBr$^{b}$ & 187220 & 7.80 & 4.91 & 6.66 & 6.07 & 6.86 & 5.51 & 6.46 & 7.41 \\
KCl$^{b}$ & 165593 & 8.69 & 5.74 & 7.92 & 7.42 & 8.19 & 6.52 & 7.78 & 8.77 \\
KCu$_{3}$S$_{2}$ & 100001 & 1.64 & 0.95 & 1.76 & 1.02 & 1.50 & 1.56 & 1.21 & -- \\
KCu$_{4}$AsS$_{4}$ & 420017 & 1.90 & 1.63 & 2.47 & 2.08 & -- & 2.45 & 1.65 & -- \\
KTaO$_{3}$ & 280424 & 3.64 & 2.61 & 3.85 & 3.29 & 3.94 & 3.53 & 3.59 & -- \\
Kr$^{b}$ & 43726 & 11.59 & 8.05 & 10.37 & 9.55 & 10.52 & 8.67 & 10.64 & 10.92 \\
La$_{2}$O$_{3}$$^{b}$ & 56771 & 5.34 & 4.03 & 5.35 & 5.11 & -- & 5.34 & 4.67 & -- \\
LaCuOSe & 96758 & 2.80 & 1.85 & 2.72 & 2.19 & -- & 2.75 & 1.86 & -- \\
LaCuOTe & 416522 & 2.31 & 1.42 & 2.09 & 1.75 & -- & 2.21 & 1.54 & -- \\
LaF$_{3}$$^{b}$ & 252346 & 10.10 & 5.91 & 9.01 & 9.40 & -- & 9.68 & 6.25 & -- \\
LaZnAsO & 420204 & 1.50 & 1.06 & 1.47 & 0.88 & -- & 1.32 & 1.29 & -- \\
LaZnPO & 85777 & 1.70 & 1.08 & 1.53 & 1.02 & -- & 1.48 & 1.41 & -- \\
LiAsS$_{2}$ & 419061 & 1.60 & 1.58 & 2.06 & 1.83 & 2.12 & 1.73 & 1.42 & -- \\
LiAsSe$_{2}$ & 248116 & 1.11 & 0.91 & 1.16 & 0.97 & 1.18 & 1.23 & 0.98 & -- \\
LiBr & 52236 & 7.60 & 5.78 & 6.87 & 5.89 & 6.98 & 6.17 & 6.49 & 7.44 \\
LiCl$^{b}$ & 26909 & 9.40 & 7.16 & 8.67 & 7.73 & 8.84 & 7.74 & 8.27 & 9.33 \\
LiF$^{b}$ & 62361 & 13.60 & 10.00 & 13.76 & 12.87 & 14.08 & 11.39 & 12.08 & 14.52 \\
LiGaTe$_{2}$ & 162555 & 2.65 & 2.11 & 2.37 & 1.93 & 2.38 & 2.51 & 2.36 & -- \\
LiH & 173177 & 4.94 & 3.65 & 5.03 & 4.24 & 5.14 & 3.99 & 4.65 & 5.65 \\
LiIO$_{3}$$^{b}$ & 35473 & 6.10 & 4.08 & 5.50 & 5.00 & -- & 1.31 & 1.05 & -- \\
LiInI$_{4}$O$_{12}$$^{b}$ & 422056 & 4.27 & 3.54 & 4.88 & 4.28 & -- & 4.83 & 4.49 & -- \\
LiInSe$_{2}$ & 60838 & 3.01 & 2.21 & 2.82 & 2.22 & 2.83 & 2.77 & 2.80 & -- \\
LiNbO$_{3}$$^{a}$$^{b}$ & 28294 & 3.78 & 3.69 & 4.89 & 4.59 & -- & 4.82 & 4.22 & -- \\
LiTaO$_{3}$$^{b}$ & 239372 & 4.70 & 4.30 & 5.61 & 5.03 & 5.69 & 4.67 & 4.67 & -- \\
LiZnAs & 74504 & 1.61 & 1.30 & 1.54 & 0.75 & 1.37 & 1.52 & 1.63 & 1.66 \\
LiZnN & 16790 & 1.91 & 1.30 & 2.13 & 1.30 & 1.85 & 1.89 & 1.91 & 2.11 \\
LiZnP & 642242 & 2.04 & 1.77 & 2.13 & 1.75 & 2.16 & 1.95 & 2.06 & 2.02 \\
MAPbI$_{3}$$^{c}$ & orth & 1.65 & 1.96 & 1.63 & -- & -- & -- & -- & -- \\
Mg$_{2}$Ge & 81735 & 0.69 & 0.55 & 0.63 & 0.30 & 0.66 & 0.53 & 0.61 & 0.54 \\
Mg$_{2}$Si & 180944 & 0.78 & 0.56 & 0.70 & 0.45 & 0.75 & 0.59 & 0.61 & 0.59 \\
Mg$_{2}$Sn & 104870 & 0.30 & 0.14 & 0.09 & 0.02 & -- & 0.25 & 0.34 & 0.25 \\
Mg$_{3}$Sb$_{2}$ & 245692 & 0.80 & 0.64 & 0.74 & 0.30 & 0.82 & 0.62 & 0.83 & -- \\
MgF$_{2}$$^{b}$ & 8120 & 12.40 & 8.01 & 11.73 & 10.92 & 12.07 & 9.47 & 10.53 & -- \\
MgO & 9863 & 7.67 & 5.55 & 7.49 & 6.52 & 7.65 & 6.44 & 6.76 & 7.91 \\
MgSiP$_{2}$ & 22189 & 2.41 & 1.57 & 1.95 & 1.66 & 2.02 & 1.94 & 1.97 & -- \\
MgTe & 642882 & 3.49 & 3.22 & 3.58 & 2.87 & 3.71 & 3.39 & 3.54 & -- \\
MoS$_{2}$ & 95569 & 1.23 & 1.39 & 1.83 & 1.72 & 1.91 & 1.47 & 1.08 & -- \\
MoSe$_{2}$ & 601045 & 1.09 & 1.21 & 1.56 & 1.45 & 1.63 & 1.31 & 0.98 & 1.17 \\
MoTe$_{2}$ & 644476 & 1.03 & 1.05 & 1.30 & 1.20 & 1.36 & 1.17 & 0.82 & -- \\
Na$_{2}$CdSnS$_{4}$$^{a}$ & 281233 & 1.52 & 2.73 & 3.68 & 3.06 & 3.68 & 3.06 & 3.32 & -- \\
Na$_{2}$KSb$^{a}$ & 44332 & 1.00 & 1.36 & 1.54 & 0.82 & 1.61 & 1.36 & 1.77 & 1.66 \\
Na$_{2}$TeSe$_{3}$ & 430312 & 1.95 & 1.68 & 2.21 & 1.85 & -- & 1.96 & 1.70 & -- \\
Na$_{2}$Ti$_{3}$O$_{7}$$^{a}$$^{b}$ & 250000 & 3.51 & 3.48 & 5.02 & 4.56 & -- & 4.77 & 3.98 & -- \\
Na$_{3}$Sb & 26882 & 1.10 & 0.88 & 1.04 & 0.44 & 1.11 & 0.93 & 1.32 & -- \\
NaAsS$_{2}$ & 854 & 2.23 & 1.49 & 2.15 & 1.93 & 2.22 & 2.19 & 1.94 & -- \\
NaBr & 41440 & 7.10 & 4.85 & 6.22 & 5.38 & 6.36 & 5.32 & 6.21 & 6.97 \\
NaCl$^{b}$ & 181148 & 8.75 & 5.81 & 7.66 & 6.90 & 7.88 & 6.46 & 7.71 & 8.44 \\
NaF$^{b}$ & 262837 & 11.70 & 7.03 & 11.35 & 10.74 & 11.66 & 8.41 & 10.25 & 11.92 \\
NaI & 61502 & 5.80 & 4.33 & 5.09 & 4.31 & 5.27 & 4.62 & 5.18 & 5.65 \\
NaLa$_{2}$TaO$_{6}$$^{a}$$^{b}$ & 159206 & 4.50 & 4.07 & 5.68 & 5.45 & -- & 5.79 & 4.72 & -- \\
NaSbP$_{2}$S$_{6}$$^{a}$$^{b}$ & 155270 & 2.25 & 2.43 & 2.94 & 2.58 & -- & 2.91 & 2.65 & -- \\
NaTaO$_{3}$ & 239690 & 4.00 & 3.22 & 4.58 & 4.00 & -- & 4.10 & 4.18 & -- \\
NaYbP$_{2}$S$_{6}$$^{b}$ & 93758 & 1.85 & 0.11 & 1.45 & 0.15 & -- & 0.00 & -- & -- \\
Ne$^{b}$ & 43427 & 21.48 & 12.92 & 19.45 & 18.40 & 19.79 & 14.30 & 18.85 & 22.70 \\
OsAs$_{2}$ & 238253 & 0.92 & 0.82 & 1.06 & 0.88 & 1.07 & 1.34 & 0.92 & -- \\
OsP$_{2}$ & 238252 & 1.13 & 0.96 & 1.19 & 1.04 & 1.21 & 1.48 & 1.02 & -- \\
OsSb$_{2}$ & 238254 & 0.21 & 0.67 & 0.81 & 0.61 & 0.87 & 0.98 & 0.62 & -- \\
P$_{3}$N$_{5}$ & 56876 & 5.87 & 4.17 & 5.51 & 4.95 & -- & 5.10 & 5.14 & -- \\
P & 417180 & 0.31 & 0.50 & 0.79 & 0.64 & 0.83 & 0.25 & 0.07 & 0.22 \\
PbBr$_{2}$ & 36170 & 3.98 & 3.58 & 3.58 & 3.08 & 3.48 & 4.07 & 4.13 & -- \\
PbCl$_{2}$ & 27736 & 4.86 & 4.28 & 4.40 & 3.92 & 4.35 & 4.85 & 4.90 & -- \\
PbCrO$_{4}$ & 40920 & 2.25 & 1.73 & 2.72 & -- & -- & 2.71 & 1.98 & -- \\
PbCuSbS$_{3}$ & 427413 & 1.30 & 1.17 & 1.48 & 1.04 & -- & 1.55 & 0.98 & -- \\
PbI$_{2}$ & 68819 & 2.28 & 2.63 & 2.29 & 1.89 & 2.25 & 2.87 & 2.96 & -- \\
PbS & 62190 & 0.29 & 0.91 & 0.85 & 0.69 & -- & 0.72 & 1.00 & 0.26 \\
PbSeO$_{4}$$^{a}$ & 40921 & 4.30 & 3.90 & 5.34 & 4.77 & -- & 5.11 & 5.07 & -- \\
PbSe & 62195 & 0.15 & 1.02 & 0.82 & 0.76 & 0.82 & 0.63 & 0.83 & 0.17 \\
PbTe & 63098 & 0.19 & 1.19 & 0.40 & 0.41 & 0.40 & 1.04 & 1.10 & 0.34 \\
PdS$_{2}$ & 648747 & 0.75 & 0.37 & 0.83 & 0.79 & -- & 0.90 & 0.25 & -- \\
PdS & 61063 & 0.50 & 0.28 & 0.45 & 0.17 & -- & 0.83 & 0.13 & -- \\
PdSe$_{2}$ & 170327 & 0.40 & 0.33 & 0.72 & 0.64 & -- & 0.63 & 0.12 & -- \\
PtSb$_{2}$ & 43105 & 0.11 & 0.06 & 0.04 & 0.08 & -- & 0.12 & 0.00 & -- \\
Rb$_{2}$AgVS$_{4}$$^{b}$ & 66841 & 1.83 & 1.60 & 2.68 & 2.35 & -- & 2.37 & 1.74 & -- \\
Rb$_{2}$Cu$_{2}$SnS$_{4}$ & 74020 & 2.08 & 1.38 & 2.47 & 1.71 & -- & 2.42 & 1.67 & -- \\
Rb$_{2}$CuNbS$_{4}$$^{b}$ & 84304 & 2.64 & 2.22 & 3.39 & 3.11 & -- & 3.05 & 2.21 & -- \\
Rb$_{2}$CuNbSe$_{4}$$^{b}$ & 84305 & 2.19 & 1.89 & 2.82 & 2.50 & 2.91 & 2.58 & 1.89 & -- \\
Rb$_{2}$CuVS$_{4}$$^{b}$ & 84302 & 1.85 & 1.48 & 2.61 & 2.31 & -- & 2.31 & 1.53 & -- \\
Rb$_{2}$TeBr$_{6}$ & 49521 & 2.19 & 2.16 & 2.97 & 2.66 & 3.01 & 2.52 & 2.37 & -- \\
Rb$_{2}$TeI$_{6}$ & 36009 & 1.43 & 1.78 & 2.16 & 1.92 & -- & 2.00 & 1.81 & -- \\
Rb$_{2}$TiCu$_{2}$S$_{4}$$^{b}$ & 280644 & 2.19 & 2.04 & 3.11 & 2.66 & -- & 2.57 & 1.81 & -- \\
Rb$_{3}$Sb$_{2}$Br$_{9}$ & 39823 & 2.48 & 2.10 & 2.89 & 2.55 & 2.90 & 2.35 & 2.28 & -- \\
RbAg$_{2}$SbS$_{4}$ & 82145 & 2.30 & 1.31 & 2.10 & 1.28 & -- & 1.91 & 1.60 & -- \\
RbBr$^{b}$ & 22167 & 7.20 & 4.86 & 6.61 & 6.07 & 6.81 & 5.51 & 6.44 & 7.15 \\
RbCl$^{b}$ & 18016 & 8.30 & 5.53 & 7.71 & 7.25 & 7.97 & 6.36 & 7.58 & 8.37 \\
RbCu$_{2}$VS$_{4}$$^{b}$ & 280516 & 1.45 & 1.34 & 2.36 & 2.11 & 2.44 & 2.12 & 1.30 & -- \\
RbF$^{b}$ & 53828 & 10.40 & 6.26 & 10.10 & 9.67 & 10.39 & 7.64 & 8.78 & 10.01 \\
RbI & 22168 & 6.37 & 4.41 & 5.50 & 5.00 & 5.77 & 4.89 & 5.51 & 5.98 \\
ReS$_{2}$ & 81814 & 1.37 & 1.38 & 1.66 & 1.52 & -- & 1.84 & 1.38 & -- \\
ReSe$_{2}$ & 81813 & 1.19 & 1.34 & 1.61 & 1.46 & 1.64 & 1.64 & 1.25 & -- \\
RuS$_{2}$ & 68472 & 1.45 & 1.07 & 2.08 & 1.03 & 1.28 & 1.87 & 1.25 & -- \\
RuSe$_{2}$ & 68473 & 0.76 & 0.62 & 0.91 & 0.62 & 0.82 & 1.35 & 0.81 & -- \\
RuTe$_{2}$ & 65169 & 0.49 & 0.25 & 0.44 & 0.27 & -- & 0.69 & 0.33 & -- \\
Sb$_{2}$O$_{3}$ & 240206 & 4.00 & 3.49 & 4.44 & 4.12 & -- & 4.17 & 4.11 & -- \\
Sb$_{2}$O$_{3}$ & 240207 & 3.25 & 2.36 & 3.31 & 3.01 & -- & 2.98 & 2.80 & -- \\
Sb$_{2}$S$_{3}$ & 171850 & 1.74 & 1.54 & 1.82 & 1.59 & -- & 1.80 & 1.64 & -- \\
Sb$_{2}$Se$_{3}$ & 194836 & 1.11 & 1.17 & 1.35 & 1.07 & -- & 1.47 & 1.31 & -- \\
Sb$_{2}$Te$_{3}$ & 193341 & 0.28 & 0.43 & 0.15 & 0.19 & 0.18 & 0.23 & 0.21 & -- \\
SbF$_{3}$$^{a}$$^{b}$ & 16142 & 4.30 & 4.92 & 6.68 & 6.22 & -- & 5.72 & 5.61 & -- \\
SbI$_{3}$$^{b}$ & 30906 & 2.41 & 2.60 & 2.92 & 2.55 & 3.02 & 3.11 & 2.75 & -- \\
SbNSr$_{3}$ & 152052 & 1.15 & 0.53 & 0.95 & 0.53 & 1.01 & 0.90 & 0.84 & -- \\
SbSBr & 88584 & 2.27 & 2.03 & 2.42 & 2.08 & 2.45 & 2.35 & 2.16 & -- \\
SbSI & 85301 & 1.88 & 2.04 & 2.20 & 1.87 & 2.22 & 2.28 & 2.06 & -- \\
SbSeI & 35470 & 1.70 & 1.78 & 1.97 & 1.62 & 1.97 & 2.05 & 1.82 & -- \\
ScN & 644666 & 1.10 & 0.39 & 0.92 & 0.20 & 0.94 & 0.91 & 0.92 & 1.28 \\
ScNiSb & 40296 & 0.11 & 0.25 & 0.56 & 0.63 & 0.73 & 0.44 & 0.33 & 0.40 \\
ScPdSb & 415944 & 0.23 & 0.19 & 0.40 & 0.50 & -- & 0.52 & 0.27 & 0.34 \\
Si$_{3}$N$_{4}$$^{a}$ & 170004 & 5.10 & 5.03 & 6.23 & 5.46 & -- & 5.78 & 5.73 & -- \\
SiC & 156190 & 2.86 & 2.39 & 2.93 & 2.50 & 2.99 & 2.92 & 2.89 & -- \\
SiC & 24170 & 3.20 & 2.60 & 3.14 & 2.70 & 3.21 & 3.17 & 3.11 & -- \\
SiC & 603798 & 2.42 & 1.72 & 2.25 & 1.83 & 2.32 & 2.25 & 2.29 & 2.39 \\
SiC & 86253 & 3.33 & 2.80 & 3.34 & 2.91 & 3.41 & 3.26 & 3.37 & -- \\
SiO$_{2}$$^{b}$ & 154289 & 8.90 & 6.59 & 9.09 & 8.45 & -- & 7.87 & 8.23 & -- \\
SiP$_{2}$ & 43098 & 1.89 & 1.51 & 1.91 & 1.61 & -- & 1.39 & 1.09 & -- \\
Sn$_{2}$SbS$_{2}$I$_{3}$ & 23344 & 1.50 & 0.81 & 1.02 & 0.80 & 1.03 & 0.86 & 0.68 & -- \\
SnBr$_{2}$ & 411177 & 3.45 & 2.73 & 3.38 & 2.96 & 3.41 & 3.04 & 2.92 & -- \\
SnCl$_{2}$ & 15452 & 3.80 & 3.31 & 4.16 & 3.80 & 4.24 & 3.91 & 3.81 & -- \\
SnO$_{2}$ & 154960 & 3.60 & 1.98 & 3.63 & 2.69 & 3.65 & 3.13 & 3.23 & 4.12 \\
SnS$_{2}$ & 100610 & 2.48 & 2.13 & 2.76 & 2.29 & 2.75 & 2.31 & 2.05 & -- \\
SnS & 52108 & 1.23 & 1.05 & 1.34 & 1.18 & -- & 1.07 & 1.01 & -- \\
SnSe$_{2}$ & 43594 & 0.97 & 1.18 & 1.60 & 1.14 & 1.56 & 1.40 & 1.25 & -- \\
SnSe & 186650 & 0.90 & 0.97 & 1.19 & 0.94 & -- & 1.03 & 1.01 & -- \\
SnTe & 52489 & 0.36 & 0.64 & 0.43 & 0.51 & 0.42 & 0.10 & 0.19 & 0.27 \\
Sn & 40039 & 0.09 & 0.19 & 0.15 & 0.09 & -- & 0.00 & 0.00 & 0.00 \\
Sr$_{2}$PbO$_{4}$$^{a}$ & 4418 & 1.75 & 1.79 & 3.15 & 2.82 & -- & 2.71 & 2.65 & -- \\
Sr$_{2}$TiO$_{4}$ & 194713 & 3.41 & 2.38 & 3.79 & 3.33 & 3.95 & 3.50 & 2.82 & -- \\
Sr$_{3}$Ti$_{2}$O$_{7}$ & 63704 & 3.26 & 2.40 & 3.67 & 3.21 & -- & 3.39 & 2.74 & -- \\
SrBiO$_{2}$Cl$^{a}$$^{b}$ & 84636 & 3.52 & 3.88 & 4.90 & 4.37 & 4.86 & 4.47 & 4.22 & -- \\
SrCl$_{2}$$^{b}$ & 196826 & 7.50 & 5.82 & 7.36 & 6.82 & 7.63 & 6.66 & 7.24 & 8.07 \\
SrCu$_{2}$O$_{2}$ & 25002 & 3.30 & 2.15 & 3.66 & 3.14 & 3.36 & 3.21 & 2.12 & -- \\
SrGe$_{2}$ & 152799 & 0.82 & 0.73 & 0.90 & 0.68 & -- & 0.79 & 0.63 & -- \\
SrMoO$_{4}$$^{a}$$^{b}$ & 239418 & 3.98 & 3.90 & 5.59 & 5.46 & -- & 5.62 & 4.97 & -- \\
SrO & 163625 & 6.10 & 3.79 & 5.99 & 5.54 & 6.16 & 4.75 & 4.93 & 5.81 \\
SrPbO$_{3}$ & 78682 & 1.78 & 1.00 & 2.08 & 1.83 & -- & 1.47 & 1.13 & -- \\
SrS & 249177 & 4.32 & 2.95 & 3.78 & 3.30 & 3.87 & 3.43 & 3.51 & 4.30 \\
SrSeO$_{4}$$^{a}$$^{b}$ & 40923 & 4.75 & 4.49 & 6.58 & 5.93 & -- & 6.05 & 6.23 & -- \\
SrSe & 52429 & 3.81 & 2.72 & 3.32 & 2.76 & 3.36 & 3.04 & 3.08 & 3.72 \\
SrSnO$_{3}$ & 238971 & 4.10 & 2.62 & 4.22 & 3.38 & -- & 3.68 & 3.59 & -- \\
SrTaO$_{2}$N & 411137 & 1.99 & 1.19 & 1.92 & 1.39 & 1.88 & 1.61 & 1.85 & -- \\
SrTiO$_{3}$ & 94573 & 3.25 & 2.22 & 3.45 & 3.01 & 3.61 & 3.28 & 2.65 & -- \\
SrZrS$_{3}$ & 154104 & 1.00 & 1.48 & 1.97 & 1.63 & -- & 1.95 & 1.72 & -- \\
TaON & 192423 & 2.40 & 2.33 & 3.08 & 2.62 & -- & 2.80 & 2.79 & -- \\
TeO$_{2}$$^{b}$ & 161691 & 3.50 & 3.56 & 4.45 & 3.97 & -- & 3.66 & 3.30 & -- \\
TeO$_{2}$$^{a}$ & 90733 & 3.41 & 3.62 & 4.62 & 4.14 & -- & 4.13 & 3.78 & -- \\
TiO$_{2}$ & 167953 & 3.03 & 2.16 & 3.21 & 2.83 & 3.36 & 3.66 & 2.96 & -- \\
TiO$_{2}$ & 92363 & 3.30 & 2.56 & 3.75 & 3.31 & 3.91 & 3.55 & 2.89 & -- \\
Tl$_{2}$AuPS$_{4}$ & 85680 & 2.20 & 1.56 & 2.18 & 1.85 & -- & 1.77 & 1.75 & -- \\
Tl$_{2}$Cu$_{2}$SnS$_{4}$ & 171221 & 1.40 & 0.60 & 1.13 & 0.63 & -- & 1.32 & 0.98 & -- \\
Tl$_{2}$TeBr$_{6}$ & 99127 & 2.06 & 2.40 & 2.95 & 2.64 & -- & 2.42 & 2.23 & -- \\
Tl$_{2}$TeI$_{6}$ & 99128 & 1.47 & 1.82 & 2.09 & 1.85 & -- & 2.12 & 1.88 & -- \\
Tl$_{2}$TeS$_{3}$ & 391285 & 1.60 & 1.48 & 1.93 & 1.69 & -- & 1.73 & 1.57 & -- \\
Tl$_{3}$AsSe$_{3}$ & 15148 & 1.30 & 1.16 & 1.45 & 1.14 & 1.39 & 0.00 & 0.00 & -- \\
Tl$_{4}$CdI$_{6}$ & 60756 & 2.80 & 2.50 & 2.90 & 2.45 & -- & 2.63 & 2.80 & -- \\
TlBr & 61532 & 2.66 & 2.26 & 2.25 & 1.76 & 2.14 & 2.54 & 2.83 & 2.41 \\
TlCl & 5253 & 3.23 & 2.54 & 2.91 & 2.48 & 2.88 & 2.97 & 3.36 & 3.27 \\
TlGaTe$_{2}$ & 16243 & 1.20 & 0.75 & 0.87 & 0.49 & 0.79 & 0.84 & 0.80 & -- \\
TlI & 258845 & 2.87 & 2.55 & 2.41 & 2.01 & 2.41 & 2.61 & 2.72 & -- \\
TlI & 60491 & 2.92 & 2.21 & 2.44 & 2.07 & 2.48 & 2.52 & 2.76 & 2.59 \\
TlI & 61520 & 2.10 & 2.22 & 1.93 & 1.56 & 1.95 & 2.24 & 2.44 & -- \\
TlInSe$_{2}$ & 180272 & 1.19 & 1.21 & 1.47 & 1.02 & 1.35 & 0.43 & 0.22 & -- \\
TlInTe$_{2}$ & 16245 & 0.96 & 0.92 & 1.04 & 0.65 & 0.95 & 0.93 & 0.85 & -- \\
TlS & 78161 & 1.36 & 1.08 & 1.53 & 1.25 & 1.56 & 1.34 & 1.14 & -- \\
TlScS$_{2}$ & 418474 & 1.51 & 1.14 & 1.55 & 1.29 & 1.42 & 1.40 & 1.58 & -- \\
TlScSe$_{2}$ & 418475 & 0.79 & 0.87 & 1.21 & 0.82 & 1.17 & 1.03 & 1.06 & -- \\
TlScTe$_{2}$ & 418476 & 0.85 & 0.08 & 0.11 & 0.81 & 0.16 & 0.43 & 0.29 & -- \\
TlSe & 104189 & 0.73 & 0.67 & 1.01 & 0.71 & 1.00 & 0.67 & 0.52 & -- \\
TlTaS$_{3}$ & 412385 & 0.78 & 0.76 & 1.20 & 0.92 & -- & 0.97 & 0.65 & -- \\
TlTiPS$_{5}$$^{b}$ & 171214 & 1.30 & 1.39 & 2.15 & 1.86 & -- & 1.83 & 1.26 & -- \\
WS$_{2}$ & 202366 & 1.35 & 1.36 & 1.77 & 1.64 & 1.83 & 1.53 & 1.19 & -- \\
WSe$_{2}$ & 40752 & 1.20 & 1.42 & 1.75 & 1.60 & 1.80 & 1.43 & 1.13 & -- \\
Xe$^{b}$ & 43428 & 9.29 & 6.77 & 8.01 & 7.41 & 8.29 & 7.38 & 8.15 & 8.75 \\
YF$_{3}$$^{b}$ & 6023 & 12.50 & 8.54 & 11.45 & 10.76 & 11.79 & 10.20 & 9.43 & -- \\
YNiSb & 105331 & 0.18 & 0.27 & 0.54 & 0.63 & 0.74 & 0.46 & 0.36 & 0.50 \\
YPtSb & 44970 & 0.16 & 0.66 & 0.53 & 0.21 & 0.50 & 0.92 & 0.61 & 0.58 \\
ZnAl$_{2}$S$_{4}$ & 35380 & 3.50 & 2.90 & 3.52 & 3.06 & 3.60 & 3.49 & 3.32 & -- \\
ZnAs$_{2}$ & 2021 & 0.97 & 0.49 & 0.75 & 0.48 & -- & 0.91 & 0.75 & -- \\
ZnGa$_{2}$Se$_{4}$ & 44887 & 2.18 & 2.23 & 2.71 & 1.97 & 2.62 & 2.68 & 2.74 & -- \\
ZnGa$_{2}$Te$_{4}$ & 290911 & 1.33 & 1.49 & 1.57 & 1.18 & 1.56 & 1.67 & 1.63 & -- \\
ZnGeAs$_{2}$ & 68324 & 1.15 & 0.72 & 0.91 & 0.34 & 0.84 & 1.15 & 1.18 & -- \\
ZnGeN$_{2}$ & 15144 & 3.36 & 2.50 & 3.45 & 2.76 & 3.30 & 3.12 & 2.91 & -- \\
ZnGeP$_{2}$ & 23706 & 2.02 & 1.45 & 1.77 & 1.52 & 1.81 & 2.07 & 1.91 & -- \\
ZnIn$_{2}$Te$_{4}$$^{a}$ & 25650 & 1.49 & 1.56 & 1.67 & 1.07 & 1.62 & 2.07 & 2.04 & -- \\
ZnP$_{2}$ & 250015 & 1.42 & 0.94 & 1.31 & 1.07 & -- & 1.48 & 1.27 & -- \\
ZnS & 291064 & 3.72 & 2.92 & 3.74 & 2.86 & 3.54 & 3.38 & 3.65 & 3.74 \\
ZnS & 67453 & 3.91 & 2.96 & 3.79 & 2.91 & 3.60 & 3.41 & 3.67 & 3.82 \\
ZnSb & 43265 & 0.50 & 0.66 & 0.76 & 0.41 & 0.73 & 0.66 & 0.61 & -- \\
ZnSiAs$_{2}$ & 68323 & 1.93 & 1.45 & 1.67 & 1.31 & 1.67 & 1.75 & 1.75 & -- \\
ZnSiP$_{2}$ & 23680 & 2.00 & 1.56 & 1.91 & 1.63 & 1.95 & 1.98 & 1.91 & -- \\
ZnSnAs$_{2}$ & 250389 & 0.75 & 0.56 & 0.70 & 0.01 & 0.57 & 0.93 & 0.99 & -- \\
ZnSnO$_{3}$ & 245943 & 3.90 & 2.28 & 3.85 & 2.75 & 3.61 & 3.36 & 3.22 & -- \\
ZnSnP$_{2}$ & 22179 & 1.68 & 1.44 & 1.76 & 1.13 & 1.64 & 1.87 & 1.85 & -- \\
ZnSnSb$_{2}$ & 651588 & 0.40 & 0.42 & 0.33 & 0.16 & 0.19 & 0.64 & 0.53 & -- \\
ZrS$_{2}$ & 604434 & 1.68 & 1.27 & 1.77 & 1.44 & 1.87 & 1.58 & 1.29 & 1.87 \\
ZrSe$_{2}$ & 109291 & 1.18 & 0.70 & 0.88 & 0.47 & 0.90 & 1.02 & 0.82 & 1.08 \\
\bottomrule
\end{longtable}
}

On the vacuum scale (main text, Fig.~3a), whether each surface straddles the redox levels of water is predicted in agreement with experiment for seven of the ten surfaces, compared with five for the base functional alone (Supplementary Table~\ref{tab:surf}).

\begin{table}[H]
\centering
\caption{\textbf{Surfaces.} IP and EA on the vacuum scale (eV): base alone (without spin--orbit coupling), with our correction (with spin--orbit coupling), GW (IP only)\cite{hinuma2014,chen2014} and experiment.}
\label{tab:surf}
\small
\begin{tabular}{lrrrrrrr}
\toprule
Surface & IP base & IP $+\ES$ & IP GW & IP exp. & EA base & EA $+\ES$ & EA exp. \\
\midrule
AlSb(110) & 4.92 & 4.84 & 5.36 & 5.21 & 3.53 & 3.44 & 3.52 \\
CdTe(110) & 5.57 & 5.56 & 5.93 & 5.78 & 4.12 & 3.91 & 4.30 \\
GaAs(110) & 5.26 & 5.32 & 5.63 & 5.52 & 4.11 & 3.96 & 4.00 \\
GaP(110) & 5.55 & 5.75 & 6.11 & 6.01 & 3.68 & 3.57 & 3.66 \\
InP(110) & 5.39 & 5.58 & 5.98 & 5.76 & 4.13 & 3.99 & 4.34 \\
MgO(100) & 5.49 & 6.94 & 6.84 & 7.15 & -0.10 & -0.59 & -0.52 \\
TiO$_{2}$(110) & 7.42 & 8.11 & -- & 8.10 & 5.24 & 4.88 & 5.07 \\
ZnO(10-10) & 6.90 & 8.12 & 8.15 & 7.82 & 5.27 & 4.90 & 4.38 \\
ZnS(110) & 6.54 & 7.10 & 7.38 & 7.49 & 3.66 & 3.39 & 3.77 \\
ZnSe(110) & 6.25 & 6.55 & 6.91 & 6.82 & 4.17 & 3.93 & 4.00 \\
ZnTe(110) & 5.47 & 5.44 & -- & -- & 3.45 & 3.25 & -- \\
\bottomrule
\end{tabular}
\end{table}

\paragraph{Heterojunctions.} Valence-band offsets of lattice-matched (110) heterojunctions follow from the same alignment as the surface IPs: $\Delta E_v = [\epsilon_\mathrm{VBM} - \langle V\rangle]_B - [\epsilon_\mathrm{VBM} - \langle V\rangle]_A + (\langle V\rangle_B - \langle V\rangle_A)_\mathrm{SL}$. The bulk terms come from LOCUS $+\ES$ of each material at the common lattice constant (the mean of the experimental ones), each with its own $g$ and with spin--orbit coupling. The potential step comes from a LOCUS calculation of a superlattice of eight (110) planes of each material (32 atoms, ideal zincblende positions, $4 \times 3 \times 1$ $k$-points), with $\langle V\rangle$ the macroscopic average of the electrostatic potential over the central half of each layer; our correction leaves the density, and hence the step, unchanged. Supplementary Table~\ref{tab:hj} lists AlAs/GaAs, ZnSe/GaAs and ZnSe/Ge, whose Kohn--Sham dielectric constants differ by factors of 1.5, 2.0 and 4.5. Our correction leaves the AlAs/GaAs offset close to experiment (0.57 against 0.53~eV) and raises the two offsets with ZnSe by 0.28 and 0.38~eV, because it lowers the VBM of ZnSe more than those of GaAs and Ge. These offsets, 1.33 and 1.97~eV, lie 0.08 and 0.12~eV above GW with vertex corrections (GW$\Gamma_1$@HSE, 1.25 and 1.85~eV\cite{hinuma2014,ghosh2022}) and above photoemission for these heterovalent interfaces, whose offsets depend on the interface structure (0.96--1.10 and 1.29--1.52~eV\cite{kowalczyk1982}). Applying $\ES$ to the superlattice itself, with the single $g$ of the cell, changes the offsets from the values obtained by aligning the bulk shifts by $-0.08$, $-0.21$ and $-0.34$~eV, increasing with the dielectric contrast: the edges of each layer are shifted with a screening that belongs to neither (Section~\ref{sec:localg}).

\begin{table}[H]
\centering
\caption{\textbf{Valence-band offsets of (110) heterojunctions} (eV, $E_v$ of the second material minus that of the first). $\varepsilon_\mathrm{KS} = 1/g$ of the two bulks; base functional with spin--orbit coupling and with our correction, both by alignment; experiment (photoemission). Last column: change of the offset when $\ES$ is applied to the superlattice itself with one $g$ for the whole cell (Section~\ref{sec:localg} replaces this single $g$ by a local screening).}
\label{tab:hj}
\small
\begin{tabular}{lcrrrr}
\toprule
Junction & $\varepsilon_\mathrm{KS}$ & Base & $+\ES$ & Exp. & One $g$ \\
\midrule
AlAs/GaAs & 7.9 / 11.5 & 0.50 & 0.57 & 0.53\cite{vurgaftman2001} & $-$0.08 \\
ZnSe/GaAs & 5.9 / 11.5 & 1.05 & 1.33 & 0.96\cite{kowalczyk1982} & $-$0.21 \\
ZnSe/Ge & 5.9 / 26.7 & 1.59 & 1.97 & 1.29--1.52\cite{kowalczyk1982} & $-$0.34 \\
\bottomrule
\end{tabular}
\end{table}

For the 12 solids of Supplementary Table~\ref{tab:dl} with both GW$_0$ and photoemission values, GW$_0$ and our correction place the d bands shallower than photoemission, by 0.6 and 0.5~eV on average\cite{fuchs2007}; this difference is common to GW and is usually attributed to vertex and final-state effects, so it is not specific to our correction.

\begin{table}[H]
\centering
\caption{\textbf{d bands.} Depth of the d-band centre below the VBM (eV). In the calculations the centre is the mean energy of the occupied states with a d character of at least 0.5, weighted with the $k$-point weights, of the element whose d band lies closest to the VBM. Columns: r$^2$SCAN + rVV10 (the base functional without the shell term), LOCUS alone, LOCUS with our correction, GW$_0$\cite{shishkin2007} and photoemission\cite{fuchs2007}.}
\label{tab:dl}
\small
\begin{tabular}{lrrrrr}
\toprule
Solid & r$^2$SCAN + rVV10 & LOCUS & $+\ES$ & GW$_0$ & Exp. \\
\midrule
AgBr & 2.45 & 2.89 & 3.46 & -- & 3.70 \\
AgCl & 1.75 & 2.25 & 2.83 & -- & 3.30 \\
AgI & 3.27 & 3.71 & 4.55 & -- & 4.40 \\
CdS & 7.52 & 8.40 & 9.23 & 9.1 & 9.40 \\
CdSe & 7.97 & 8.81 & 9.64 & 9.2 & 10.00 \\
CdTe & 8.29 & 9.23 & 10.18 & 9.7 & 10.50 \\
Cu$_{2}$O & 1.99 & 2.09 & 2.60 & -- & -- \\
CuBr & 1.21 & 1.21 & 1.89 & -- & -- \\
CuCl & 1.12 & 1.12 & 1.87 & -- & -- \\
CuI & 1.59 & 1.76 & 2.59 & -- & -- \\
GaAs & 15.16 & 17.14 & 17.95 & 18.1 & 18.90 \\
GaN & 13.51 & 15.68 & 16.96 & 16.8 & 17.00 \\
GaP & 14.92 & 16.91 & 17.81 & 17.9 & 18.60 \\
GaSb & 15.45 & 17.50 & 18.19 & 18.5 & -- \\
InAs & 14.77 & 16.09 & 16.53 & 16.6 & 16.80 \\
InP & 14.57 & 15.91 & 16.48 & 16.4 & 17.00 \\
InSb & 15.03 & 16.42 & 16.97 & 16.7 & 16.98 \\
ZnO & 4.80 & 5.67 & 6.65 & 6.7 & 7.50 \\
ZnS & 6.13 & 7.01 & 8.18 & 8.1 & 9.00 \\
ZnSe & 6.37 & 7.28 & 8.46 & 8.3 & 9.40 \\
\bottomrule
\end{tabular}
\end{table}

\begin{table}[H]
\centering
\caption{\textbf{Monolayers.} Gaps (eV) against G$_0$W$_0$ from C2DB\cite{haastrup2018}.}
\label{tab:2d}
\small
\begin{tabular}{lrrr}
\toprule
Monolayer & Base & $+\ES$ & G$_0$W$_0$ \\
\midrule
Te$_{2}$Ti$_{2}$ & 0.24 & 0.37 & 0.27 \\
Te$_{2}$Zr$_{2}$ & 0.32 & 0.44 & 0.45 \\
CrSTe & 0.29 & 0.58 & 0.70 \\
Au$_{2}$O$_{2}$ & 0.33 & 1.25 & 0.97 \\
MoTe$_{2}$ & 1.32 & 1.48 & 1.56 \\
Cu$_{2}$S$_{2}$ & 0.70 & 1.29 & 1.58 \\
S$_{2}$Tl$_{2}$ & 0.80 & 1.25 & 1.59 \\
As$_{4}$ & 1.12 & 1.40 & 1.80 \\
Ge$_{2}$Se$_{2}$ & 1.43 & 1.69 & 1.86 \\
P$_{4}$ & 1.06 & 1.39 & 2.03 \\
Pb$_{2}$S$_{2}$ & 2.01 & 1.88 & 2.23 \\
ISSb & 1.82 & 1.88 & 2.26 \\
S$_{2}$Si$_{2}$ & 1.65 & 2.15 & 2.88 \\
HfS$_{2}$ & 1.66 & 2.18 & 2.94 \\
PtS$_{2}$ & 2.01 & 2.42 & 2.95 \\
Cu$_{2}$Br$_{2}$ & 2.39 & 3.91 & 3.32 \\
Cu$_{2}$Cl$_{2}$ & 2.39 & 4.25 & 3.45 \\
CdBr$_{2}$ & 3.71 & 5.15 & 6.20 \\
BN & 5.10 & 6.20 & 7.12 \\
CaF$_{2}$ & 7.31 & 12.00 & 11.37 \\
\bottomrule
\end{tabular}
\end{table}

The bowing parameter of AlGaN, 0.57~eV, is below the reported 0.86--1.00~eV mainly because the calculated AlN gap lies 0.17~eV below experiment; at intermediate compositions the calculated gaps lie within 0.08~eV of the curve defined by the experimental end members and the reported bowing (Supplementary Table~\ref{tab:alloys}).

\begin{table}[H]
\centering
\caption{\textbf{Alloys.} Gaps (eV) averaged over special quasirandom structures at each composition $x$, with our correction on LOCUS and on r$^2$SCAN. Experimental values are given for the end members only; the intermediate compositions are compared through the bowing parameter $b$ fitted to the calculated series (spread of reported values: InGaN and AlGaN, refs~\citenum{wu2002,alam2020}; CIGS, ref.~\citenum{rincon2001}).}
\label{tab:alloys}
\small
\begin{tabular}{lrrrrr}
\toprule
Alloy & $x$ & Structures & LOCUS $+\ES$ & r$^2$SCAN $+\ES$ & Exp. \\
\midrule
AlGaN & 0.00 & 1 & 3.47 & 3.33 & 3.50 \\
AlGaN & 0.25 & 1 & 3.97 & 3.88 & -- \\
AlGaN & 0.50 & 2 & 4.57 & 4.53 & -- \\
AlGaN & 0.75 & 1 & 5.19 & 5.19 & -- \\
AlGaN & 1.00 & 1 & 5.94 & 6.04 & 6.11 \\
\multicolumn{3}{l}{AlGaN: bowing $b$} & 0.57 & & 0.86--1 \\ \midrule
CIGS & 0.00 & 1 & 0.78 & 0.02 & 1.04 \\
CIGS & 0.25 & 1 & 0.93 & 0.49 & -- \\
CIGS & 0.50 & 2 & 1.09 & 0.70 & -- \\
CIGS & 0.75 & 1 & 1.28 & 0.92 & -- \\
CIGS & 1.00 & 1 & 1.47 & 1.13 & 1.68 \\
\multicolumn{3}{l}{CIGS: bowing $b$} & 0.13 & & 0.198 \\ \midrule
InGaN & 0.00 & 1 & 3.47 & 3.33 & 3.50 \\
InGaN & 0.25 & 4 & 2.43 & 2.25 & -- \\
InGaN & 0.50 & 5 & 1.70 & 1.54 & -- \\
InGaN & 0.75 & 4 & 1.11 & 0.91 & -- \\
InGaN & 1.00 & 1 & 0.63 & 0.37 & 0.70 \\
\multicolumn{3}{l}{InGaN: bowing $b$} & 1.44 & & 1.4--3 \\ \midrule
\bottomrule
\end{tabular}
\end{table}

\begin{table}[H]
\centering
\caption{\textbf{Amorphous oxides.} Mobility gaps (eV): mean and range over cells, and experiment: a-SiO$_2$ ref.~\citenum{distefano1971}, a-Al$_2$O$_3$ and a-HfO$_2$ ref.~\citenum{ding2014} (ALD films), a-IGZO ref.~\citenum{fung2008} (Tauc gap).}
\label{tab:amo}
\small
\begin{tabular}{lrrr}
\toprule
System & Cells & $+\ES$ & Exp. \\
\midrule
SiO2 & 2 & 8.87 [8.68, 9.06] & 8.9 \\
Al2O3 & 2 & 6.39 [6.31, 6.47] & 6.71 \\
HfO2 & 2 & 6.07 [6.00, 6.14] & 5.77 \\
IGZO & 5 & 3.22 [3.13, 3.35] & 3.05 \\
\bottomrule
\end{tabular}
\end{table}

Supplementary Table~\ref{tab:amo2} adds cells prepared in other ways and further amorphous dielectrics and oxide semiconductors, computed with the same procedure (melt-quench with PFP in its r$^2$SCAN mode, LOCUS $+\ES$, the mobility criterion below). The six a-SiO$_2$ models, of 96 and 144 atoms in cubic and tetragonal cells and quenched in 13.5 or 300~ps or by stepwise cooling, give mobility gaps of 8.68--9.18~eV, and the mobility gaps of the tetragonal cells of a-Al$_2$O$_3$ and a-HfO$_2$ lie within 0.13~eV of those of the cubic ones. For a-Si$_3$N$_4$ at 3.0~g~cm$^{-3}$ the mobility gap is 4.86~eV; an optical gap of 5.43~eV has been obtained from the refractive-index dispersion of low-pressure chemical-vapour-deposited silicon nitride\cite{modreanu1998}, a definition that is not directly comparable. In a-IGZO the gap rises with the Ga content and falls with the In content, from 2.95~eV for In:Ga:Zn = 2:1:1 to 3.40~eV for 1:2:1, following the order of the corrected gaps of In$_2$O$_3$, ZnO and Ga$_2$O$_3$ (Supplementary Table~\ref{tab:gaps}); replacing a quarter of the In of the 1:1:1 composition by Sn gives 3.43~eV. On the r$^2$SCAN + rVV10 base, the corrected mobility gap of the five a-IGZO cells of the main text (Fig.~5b) is 3.01~eV, against 3.22~eV on LOCUS and 3.05~eV in experiment.

\begin{table}[H]
\centering
\caption{\textbf{Further amorphous cells.} Mobility gaps of the base alone (Base) and with our correction (mean and range over cells, eV); LOCUS base unless stated. a-IGZO compositions are In:Ga:Zn ratios; a-IGZTO is In$_{12}$Ga$_{16}$Zn$_{16}$Sn$_4$O$_{66}$.}
\label{tab:amo2}
\small
\begin{tabular}{>{\raggedright\arraybackslash}p{0.27\textwidth}>{\raggedright\arraybackslash}p{0.34\textwidth}rrr}
\toprule
System & Preparation & $n$ & Base & $+\ES$ \\
\midrule
a-SiO$_2$ & 96 atoms, cubic cell (Fig.~5b) & 2 & 6.18 & 8.87 [8.68, 9.06] \\
a-SiO$_2$ & 96 atoms, tetragonal cell & 2 & 6.08 & 8.78 [8.77, 8.78] \\
a-SiO$_2$ & 144 atoms, quench in 300~ps & 1 & 6.45 & 9.18 \\
a-SiO$_2$ & 144 atoms, stepwise cooling & 1 & 6.24 & 8.94 \\
a-Al$_2$O$_3$ & 100 atoms, cubic cell (Fig.~5b) & 2 & 4.40 & 6.39 [6.31, 6.47] \\
a-Al$_2$O$_3$ & 100 atoms, tetragonal cell & 1 & 4.49 & 6.52 \\
a-HfO$_2$ & 96 atoms, cubic cell (Fig.~5b) & 2 & 4.63 & 6.07 [6.00, 6.14] \\
a-HfO$_2$ & 96 atoms, tetragonal cell & 1 & 4.60 & 6.03 \\
a-Si$_3$N$_4$ & 112 atoms, stepwise cooling & 3 & 3.76 & 4.86 [4.72, 4.97] \\
a-Si$_5$N$_6$ & 110 atoms, N-deficient & 2 & 3.84 & 4.80 [4.51, 5.09] \\
a-IGZO 2:1:1 & 114 atoms & 2 & 1.54 & 2.95 [2.95, 2.95] \\
a-IGZO 1:1:2 & 108 atoms & 2 & 1.59 & 3.05 [3.01, 3.10] \\
a-IGZO 1:1:1 & 112 atoms (Fig.~5b) & 5 & 1.74 & 3.22 [3.13, 3.35] \\
a-IGZO 1:2:1 & 114 atoms & 2 & 1.89 & 3.40 [3.33, 3.47] \\
a-IGZTO & 114 atoms, In$_{12}$Ga$_{16}$Zn$_{16}$Sn$_4$O$_{66}$ & 2 & 1.92 & 3.43 [3.39, 3.46] \\
a-IGZO 1:1:1, r$^2$SCAN + rVV10 base & 112 atoms (Fig.~5b cells) & 5 & 1.47 & 3.01 [2.85, 3.12] \\
\bottomrule
\end{tabular}
\end{table}

Mobility gaps are the gaps between extended states, those whose edge-uniformity index $\kappa \geq 0.4$. The index is $\kappa = \mathrm{PR}/\mathrm{PR}_\mathrm{uniform}$, where $\mathrm{PR} = (\sum_a w_a)^2/\sum_a w_a^2$ is the participation ratio of the atomic weights $w_a = \sum_i |P_{ai}|^2$ of the state (PAW projections) and $\mathrm{PR}_\mathrm{uniform}$ the value for equal weights on chemically equivalent atoms; $\kappa = 1$ for a state spread uniformly over equivalent atoms and $\kappa \ll 1$ for a state localized on a few. Supplementary Table~\ref{tab:mobsens} gives the sensitivity of the mobility gaps to the threshold: between 0.3 and 0.4 they change by about 0.1~eV, while at 0.5 the gap of SiO$_2$ rises by 0.55~eV.

\begin{table}[H]
\centering
\caption{\textbf{Mobility gaps against the uniformity threshold.} Mean over melt-quenched cells (eV) for thresholds 0.3, 0.4 (used) and 0.5, and experiment.}
\label{tab:mobsens}
\small
\begin{tabular}{lrrrrr}
\toprule
System & Cells & 0.3 & 0.4 & 0.5 & Exp. \\
\midrule
SiO$_2$ & 2 & 8.74 & 8.87 & 9.42 & 8.9 \\
Al$_2$O$_3$ & 2 & 6.28 & 6.39 & 6.67 & 6.71 \\
HfO$_2$ & 2 & 5.95 & 6.07 & 6.19 & 5.77 \\
InGaZnO$_4$ & 5 & 3.11 & 3.22 & 3.31 & 3.05 \\
\bottomrule
\end{tabular}
\end{table}

For the transport gaps of the four aromatic hydrocarbon crystals and solid C$_{60}$ the MAE is 0.33~eV; the corrected gaps of naphthalene, anthracene and pentacene lie within the experimental ranges, and that of benzene is underestimated by 1.1~eV (Supplementary Table~\ref{tab:molcr}).

\begin{table}[H]
\centering
\caption{\textbf{Molecular crystals.} Transport gaps (eV) of the base alone and with our correction, G$_0$W$_0$@PBE and evGW from ref.~\citenum{rangel2016}, and experiment (acenes, ref.~\citenum{rangel2016}; C$_{60}$, ref.~\citenum{lof1992}).}
\label{tab:molcr}
\small
\begin{tabular}{lrrrrr}
\toprule
Crystal & Base & $+\ES$ & G$_0$W$_0$ & evGW & Exp. \\
\midrule
benzene & 4.76 & 6.67 & 7.3 & 8.2 & 7.6--8.0 \\
naphthalene & 3.32 & 5.04 & 5.5 & 6.1 & 5.0--5.5 \\
anthracene & 2.26 & 3.93 & 4.0 & 4.5 & 3.9--4.2 \\
pentacene & 0.80 & 2.22 & 2.2 & 2.8 & 2.2--2.4 \\
C$_{60}$ & 1.06 & 2.18 & -- & -- & 2.3 \\
\bottomrule
\end{tabular}
\end{table}

\section{Local screening at interfaces}
\label{sec:localg}

The screening $g$ of our correction is one number for a calculation. In a cell that contains materials of different dielectric constants, one $g$ shifts the edges of every layer with a screening that belongs to none of the layers (Section~\ref{sec:all}, Heterojunctions). A local screening $g(\mathbf{r})$, which varies in space, removes this restriction without a new parameter. Each material $M$ keeps the $g_M$ of its bulk, and the screening at $\mathbf{r}$ is
\begin{equation}
  g(\mathbf{r}) = \sum_M g_M\, h_M(\mathbf{r}), \qquad h_M(\mathbf{r}) = \frac{\sum_{a \in M} \rho_a(\mathbf{r})}{\sum_a \rho_a(\mathbf{r})},
\end{equation}
with Hirshfeld-type weights from atomic densities $\rho_a$ (here Gaussians of width 1.2~\AA) and, at the one-centre points of an atom, the $g_M$ of its material. The level shifts follow from the operators $W_X = s\,g(\mathbf{r})\,u_X(\mathbf{r})$ ($X = $ V, C), with the bare shifts $u_X$ of the main-text Methods evaluated on the integration points of our correction. To first order, $\Delta\epsilon_n = \langle\psi_n|\,{-W_\mathrm{V}}\,|\psi_n\rangle$ for occupied and $\langle\psi_n|W_\mathrm{C}|\psi_n\rangle$ for empty states; with a constant $g$ this reproduces the shifts of our correction to $10^{-6}$~eV. Where the edge states of one layer are resonant with the bands of the other, the first-order shifts of the mixed states average the two screenings; the operator $V_S = (1 - P)\,W_\mathrm{C}\,(1 - P) - P\,W_\mathrm{V}\,P$, with $P$ the projector on the occupied states, is then added to the Kohn--Sham Hamiltonian in a block of occupied and empty bands around the gap and diagonalized once. The occupied--empty block of $V_S$ vanishes, so that the density and the total energy of a calculation with integer occupations remain those of the base functional. Supplementary Table~\ref{tab:localg} applies this to the superlattices of Supplementary Table~\ref{tab:hj}. With one $g$ for the cell, the difference of the valence-edge shifts of the two layers deviates from the value obtained by aligning the bulk shifts by $-0.08$, $-0.21$ and $-0.34$~eV. With $g(\mathbf{r})$ the deviation is at most 0.07~eV to first order and at most 0.03~eV after the diagonalization; the diagonalization matters most for ZnSe/Ge. The conduction states of the ZnSe layers are resonant with the conduction bands of GaAs and Ge and have no edge localized on ZnSe, so that only AlAs/GaAs has conduction-band entries. The procedure is part of the released code.

\begin{table}[H]
\centering
\caption{\textbf{Local screening on (110) superlattices.} Change of the difference of the shifts of the two layers' edges (eV; layer edges, the highest occupied and lowest empty states with a weight above 0.7 on the layer) relative to the value obtained by aligning the bulk shifts, for $\ES$ applied to the superlattice with one $g$ for the cell (first order), with $g(\mathbf{r})$ (first order) and with $g(\mathbf{r})$ and the diagonalization of $H + V_S$ (24 occupied and 16 empty bands per $k$-point). $\varepsilon_\mathrm{KS} = 1/g$ of the two bulks. --, edge not defined: the selected state has a weight on the layer below 0.75 before or after the diagonalization (the state is resonant with a band of the other material).}
\label{tab:localg}
\small
\begin{tabular}{lcrrrrrr}
\toprule
 & & \multicolumn{3}{c}{Valence band} & \multicolumn{3}{c}{Conduction band} \\
Junction & $\varepsilon_\mathrm{KS}$ & One $g$ & $g(\mathbf{r})$ & $+$diag. & One $g$ & $g(\mathbf{r})$ & $+$diag. \\
\midrule
AlAs/GaAs & 7.9 / 11.5 & $-$0.08 & $-$0.01 & $-$0.01 & +0.03 & +0.01 & +0.01 \\
ZnSe/GaAs & 5.9 / 11.5 & $-$0.21 & $-$0.03 & +0.02 & -- & -- & -- \\
ZnSe/Ge & 5.9 / 26.7 & $-$0.34 & $-$0.07 & $-$0.02 & -- & -- & -- \\
\bottomrule
\end{tabular}
\end{table}

\section{Wide-gap systems}
\label{sec:wide}

The 35 held-out crystals with experimental gaps above 6~eV are listed in Supplementary Table~\ref{tab:wide}. Our correction underestimates the gaps of most of them (mean error $-0.36$~eV), with the largest errors in the rare-gas solids (Ne $-2.0$~eV) and the alkali halides. On PBE states, the exact-exchange correction of the CBM of these systems is 1.42 times the CBM bare shift (median, against 1.06 over all held-out solids); the bare shift therefore underestimates the exchange correction of the diffuse CBM. The same deficit places the lowest unoccupied levels of molecules below their reference values, and the electron affinities of bound anions are consequently overestimated by 1.0~eV on average (Section~\ref{sec:mol}). The per-edge decomposition of G$_0$W$_0$ (Supplementary Table~\ref{tab:gw}) shows that the underestimated gaps originate at the CBM: for the rare-gas solids Ne, Ar and Kr and the alkali halides LiF, NaF, LiCl, NaCl and KCl, the G$_0$W$_0$ correction of the CBM is $+1.28$~eV on average, against $+0.69$~eV from $\ES$ on the PBE base, whereas the corrections of the VBM agree ($-2.75$ and $-2.62$~eV). The largest positive errors (CaCO$_3$, BaF$_2$, AlPO$_4$) are in solids whose experimental references carry flags (the exciton flag in all three, the low-reference flag also for CaCO$_3$) and may partly reflect the references.

{\small
\begin{longtable}{lrrrrrrr}
\caption{\textbf{Crystals with experimental gaps above 6 eV} (eV), sorted by the error of our correction.}\label{tab:wide}\\
\toprule
Solid & Exp. & $+\ES$ & Error & HSE06 & mBJ & G$_0$W$_0$@PBE & QSG$\hat{W}$ \\
\midrule
\endfirsthead
\toprule
Solid & Exp. & $+\ES$ & Error & HSE06 & mBJ & G$_0$W$_0$@PBE & QSG$\hat{W}$ \\
\midrule
\endhead
Ne & 21.48 & 19.45 & -2.03 & 14.30 & 18.85 & 19.59 & 22.70 \\
Ar & 14.15 & 12.66 & -1.49 & 10.34 & 12.55 & 13.41 & 13.64 \\
Xe & 9.29 & 8.01 & -1.28 & 7.38 & 8.15 & 10.30 & 8.75 \\
Kr & 11.59 & 10.37 & -1.22 & 8.67 & 10.64 & 11.03 & 10.92 \\
KBr & 7.80 & 6.66 & -1.14 & 5.51 & 6.46 & 6.93 & 7.41 \\
LaF$_{3}$ & 10.10 & 9.01 & -1.09 & 9.68 & 6.25 & -- & -- \\
NaCl & 8.75 & 7.66 & -1.09 & 6.46 & 7.71 & 8.07 & 8.44 \\
YF$_{3}$ & 12.50 & 11.45 & -1.05 & 10.20 & 9.43 & -- & -- \\
NaBr & 7.10 & 6.22 & -0.88 & 5.32 & 6.21 & 6.59 & 6.97 \\
RbI & 6.37 & 5.50 & -0.87 & 4.89 & 5.51 & 6.91 & 5.98 \\
CaF$_{2}$ & 12.10 & 11.29 & -0.81 & 9.34 & 10.45 & 10.61 & 12.13 \\
KCl & 8.69 & 7.92 & -0.77 & 6.52 & 7.78 & 8.16 & 8.77 \\
LiBr & 7.60 & 6.87 & -0.73 & 6.17 & 6.49 & 7.14 & 7.44 \\
LiCl & 9.40 & 8.67 & -0.73 & 7.74 & 8.27 & 9.01 & 9.33 \\
MgF$_{2}$ & 12.40 & 11.73 & -0.67 & 9.47 & 10.53 & 11.54 & -- \\
CaO & 6.88 & 6.22 & -0.66 & 5.31 & 5.38 & 5.26 & 6.95 \\
LiIO$_{3}$ & 6.10 & 5.50 & -0.60 & 1.31 & 1.05 & 0.23 & -- \\
RbCl & 8.30 & 7.71 & -0.59 & 6.36 & 7.58 & 7.88 & 8.37 \\
RbBr & 7.20 & 6.61 & -0.59 & 5.51 & 6.44 & 6.84 & 7.15 \\
BN & 6.36 & 5.95 & -0.41 & 5.79 & 5.77 & 6.71 & 6.50 \\
NaF & 11.70 & 11.35 & -0.35 & 8.41 & 10.25 & 10.77 & 11.92 \\
RbF & 10.40 & 10.10 & -0.30 & 7.64 & 8.78 & 9.00 & 10.01 \\
BeO & 10.59 & 10.35 & -0.24 & 9.23 & 9.17 & 10.04 & -- \\
MgO & 7.67 & 7.49 & -0.18 & 6.44 & 6.76 & 7.10 & 7.91 \\
BaCl$_{2}$ & 7.00 & 6.84 & -0.16 & 6.33 & 6.33 & 7.43 & -- \\
AlN & 6.11 & 5.96 & -0.15 & 5.50 & 5.30 & 5.53 & 6.25 \\
SrCl$_{2}$ & 7.50 & 7.36 & -0.14 & 6.66 & 7.24 & 7.68 & 8.07 \\
SrO & 6.10 & 5.99 & -0.11 & 4.75 & 4.93 & 4.82 & 5.81 \\
LiF & 13.60 & 13.76 & 0.16 & 11.39 & 12.08 & 13.63 & 14.52 \\
Al$_{2}$O$_{3}$ & 8.80 & 8.97 & 0.17 & 8.05 & 7.86 & 8.65 & -- \\
SiO$_{2}$ & 8.90 & 9.09 & 0.19 & 7.87 & 8.23 & 9.07 & -- \\
CaCl$_{2}$ & 6.90 & 7.75 & 0.85 & 6.97 & 6.72 & 9.09 & -- \\
AlPO$_{4}$ & 7.89 & 9.61 & 1.72 & 7.80 & 8.26 & -- & -- \\
BaF$_{2}$ & 9.10 & 10.87 & 1.77 & 8.73 & 9.48 & 9.51 & -- \\
CaCO$_{3}$ & 6.00 & 8.78 & 2.78 & 7.14 & 7.14 & 7.69 & -- \\
\bottomrule
\end{longtable}
}

\section{Dependence on the base functional}
\label{sec:base}

Supplementary Table~\ref{tab:notopened} lists the solids, among the 307 computed with three bases, whose corrected gap stays below half of the experimental one on at least one base. On PBE these are mostly copper and silver d$^{10}$ compounds, such as chalcopyrites, kesterites and Ag$_2$O, and narrow-gap pnictides; r$^2$SCAN opens the gaps of most of them, and the shell term opens those of the chalcopyrites. The bare-shift coefficients are the same for all bases; only $s$ is refitted. On these 307 solids the corrected gaps have MAEs of 0.61, 0.40 and 0.38~eV with PBE, r$^2$SCAN and LOCUS, so that between r$^2$SCAN and LOCUS the average errors differ little. On a PBE base, which costs less than half as much as r$^2$SCAN, our correction approaches the accuracy of HSE06 and mBJ wherever the PBE gap opens (MAE 0.53~eV on the 384 of the 437 benchmark solids computed with PBE that are not such failures).

{\small
\begin{longtable}{lrrrr}
\caption{\textbf{Solids whose corrected gap stays below half of the experimental one on at least one base} (eV). $^*$, corrected gap below half of the experimental gap.}\label{tab:notopened}\\
\toprule
Solid & Exp. & PBE $+\ES$ & r$^2$SCAN $+\ES$ & LOCUS $+\ES$ \\
\midrule
\endfirsthead
\toprule
Solid & Exp. & PBE $+\ES$ & r$^2$SCAN $+\ES$ & LOCUS $+\ES$ \\
\midrule
\endhead
Ag$_{2}$O & 1.30 & 0.40$^*$ & 1.01 & 1.24 \\
AgF & 2.80 & 0.13$^*$ & 1.27$^*$ & 2.64 \\
AgGaSe$_{2}$ & 1.81 & 0.83$^*$ & 1.46 & 1.78 \\
AgGaTe$_{2}$ & 1.32 & 0.53$^*$ & 1.10 & 1.35 \\
AgInS$_{2}$ & 2.04 & 0.98$^*$ & 1.58 & 1.92 \\
AgInSe$_{2}$ & 1.24 & 0.04$^*$ & 0.94 & 1.26 \\
As$_{2}$Te$_{3}$ & 0.90 & 0.44$^*$ & 0.61 & 0.56 \\
BiNSr$_{3}$ & 0.89 & 0.39$^*$ & 0.76 & 0.67 \\
Ca$_{2}$Si & 1.90 & 0.44$^*$ & 0.74$^*$ & 0.70$^*$ \\
Ca$_{2}$Sn & 0.90 & 0.20$^*$ & 0.51 & 0.44$^*$ \\
Cu$_{2}$CdGeSe$_{4}$ & 1.20 & 0.39$^*$ & 0.41$^*$ & 0.79 \\
Cu$_{2}$GeS$_{3}$ & 1.53 & 0.49$^*$ & 0.84 & 1.12 \\
Cu$_{2}$SnS$_{3}$ & 0.86 & 0.32$^*$ & 0.21$^*$ & 0.48 \\
Cu$_{2}$ZnGeSe$_{4}$ & 1.29 & 0.25$^*$ & 0.04$^*$ & 0.59$^*$ \\
Cu$_{2}$ZnSnS$_{4}$ & 1.51 & 0.09$^*$ & 0.96 & 1.33 \\
Cu$_{3}$AsS$_{4}$ & 1.24 & 0.05$^*$ & 0.47$^*$ & 0.82 \\
Cu$_{3}$PSe$_{4}$ & 1.40 & 0.26$^*$ & 0.73 & 1.01 \\
CuBr & 3.07 & 1.36$^*$ & 2.12 & 2.77 \\
CuCl & 3.40 & 1.69$^*$ & 2.35 & 3.04 \\
CuGaSe$_{2}$ & 1.70 & 0.63$^*$ & 1.22 & 1.57 \\
CuGaTe$_{2}$ & 1.23 & 0.45$^*$ & 1.00 & 1.26 \\
CuInS$_{2}$ & 1.53 & 0.02$^*$ & 0.92 & 1.40 \\
CuInSe$_{2}$ & 1.04 & 0.05$^*$ & 0.02$^*$ & 0.82 \\
CuInTe$_{2}$ & 1.06 & 0.25$^*$ & 0.70 & 0.96 \\
GaSb & 0.82 & 0.07$^*$ & 0.50 & 0.67 \\
InTe & 1.16 & 0.28$^*$ & 0.37$^*$ & 0.30$^*$ \\
LiZnAs & 1.61 & 0.75$^*$ & 1.37 & 1.54 \\
Mg$_{2}$Ge & 0.69 & 0.30$^*$ & 0.66 & 0.63 \\
Mg$_{3}$Sb$_{2}$ & 0.80 & 0.30$^*$ & 0.82 & 0.74 \\
Na$_{3}$Sb & 1.10 & 0.44$^*$ & 1.11 & 1.04 \\
SbNSr$_{3}$ & 1.15 & 0.53$^*$ & 1.01 & 0.95 \\
ScN & 1.10 & 0.20$^*$ & 0.94 & 0.92 \\
TlGaTe$_{2}$ & 1.20 & 0.49$^*$ & 0.79 & 0.87 \\
TlScTe$_{2}$ & 0.85 & 0.81 & 0.16$^*$ & 0.11$^*$ \\
ZnGeAs$_{2}$ & 1.15 & 0.34$^*$ & 0.84 & 0.91 \\
ZnSnAs$_{2}$ & 0.75 & 0.01$^*$ & 0.57 & 0.70 \\
ZrSe$_{2}$ & 1.18 & 0.47$^*$ & 0.90 & 0.88 \\
\bottomrule
\end{longtable}
}

\section{Halide perovskites}
\label{sec:pvk}

The snapshots come from Langevin molecular dynamics with PFP\cite{takamoto2022} version 9.0.0 on Matlantis\cite{matlantis} in its r$^2$SCAN calculation mode\cite{shinagawa2026} (main text, Methods; 0.5~fs with the standard hydrogen mass for three of the FAPbI$_3$ snapshots). The cells are fixed, so the dynamics samples local distortions and tilts within the cell but not a change of the cell shape. Supplementary Table~\ref{tab:pvk} lists every group of snapshots. Entries marked poly.\ are polymorphous structures, relaxed at fixed cell from the last snapshot at 300~K (residual forces below 0.02~eV~\AA$^{-1}$); the high-symmetry crystal structures, at the experimental lattice, are part of Supplementary Table~\ref{tab:gaps}. The MAE over 11 compositions uses, for each, the 300~K snapshots in its largest cell of up to 96 atoms: the cubic cell with the experimental lattice constant of the cubic phase for the caesium and formamidinium compounds and the low-temperature orthorhombic cell for MAPbI$_3$; the $\gamma$-CsPbI$_3$ cell and a 320-atom cell of CsPbI$_3$ (below) are additional checks outside this MAE. MAPbI$_3$ is compared with the gap of its orthorhombic phase\cite{dar2016}, not with that of the room-temperature tetragonal phase. Spin--orbit coupling is included in second variation in a window of as many bands as the cell has atoms on each side of the gap (main text, Methods); a window is needed because in the second-variation basis the spin--orbit shifts of lead halides do not converge with the number of empty bands. Doubling the window changes the corrected gaps of the snapshots by at most 0.015~eV. With all computed bands the gaps of the lead compounds are 0.06--0.10~eV lower, an uncertainty of about 0.1~eV in their spin--orbit shifts. Spin--orbit calculations without a band window for the orthorhombic phases, with PBE and DSH\cite{wang2022}, give shifts of $-0.99$ to $-1.08$~eV for CsPbI$_3$ and CsPbBr$_3$, 0.2--0.3~eV smaller in magnitude than the spin--orbit shifts computed here within the window for the orthorhombic structures ($-1.29$ and $-1.23$~eV), and for $\gamma$-CsSnI$_3$ shifts 0.1~eV larger ($-0.39$ to $-0.41$ against $-0.29$~eV); the structures differ (here, Materials Project structures scaled to the experimental volume). If the window-free calculations of ref.~\citenum{wang2022} are taken as the reference, the corrected gaps of the lead compounds would be 0.2--0.3~eV higher. Applying our correction before the spin--orbit step, instead of adding the spin--orbit shifts, changes the gap of CsPbI$_3$ by less than 0.01~eV.

Five independent trajectories of CsPbI$_3$ in the 40-atom cell at 300~K (24 snapshots) give a corrected gap of 1.68~eV with a standard error of 0.016~eV, within 0.01~eV of the six-snapshot average of Supplementary Table~\ref{tab:pvk}; the means of the trajectories scatter by 0.03~eV, as expected for independent snapshots, so the sampling uncertainty is below the differences between cells of 40 and 80 atoms.

On the five caesium lead and tin halides computed in ref.~\citenum{wang2022}, the MAE of our correction is 0.07~eV, compared with 0.13~eV for the dielectric-dependent hybrid scheme of that work, whose structures are the experimental orthorhombic ones; the difference comes mainly from CsPbCl$_3$. Like the high-symmetry structures (main text), the polymorphous structures give gaps below the measured room-temperature values, by up to 0.5~eV for the tin compounds (Supplementary Table~\ref{tab:pvk}).

The other uncertainties of the protocol are the following. Gaps from cells of 40 and 80 atoms differ by 0.05--0.11~eV at each temperature (main text, Fig.~4b). The cells keep the shape and volume of the cubic phase (for MAPbI$_3$, of the low-temperature orthorhombic phase), whereas several of these compounds are orthorhombic at room temperature; for $\gamma$-CsPbI$_3$ in its own 80-atom cell the average gap is 0.06~eV below that of the cubic cell of the same size. The classical dynamics at fixed cell includes neither zero-point motion nor thermal expansion, and the structures come from a machine-learned potential at the r$^2$SCAN level rather than from LOCUS itself.

Three snapshots of CsPbI$_3$ at 300~K in a $4\times4\times4$ cubic cell (320 atoms, the same molecular-dynamics protocol), computed with the final form of LOCUS at the $\Gamma$ point alone (Section~\ref{sec:cost}), have corrected gaps of 1.50--1.51~eV, below all six snapshots of the 80-atom cell (1.56--1.74~eV) and 0.14~eV below their average. Of this difference, 0.08~eV is already in the base gap (2.04 against 2.12~eV on average), about three times the standard error of the difference between the two averages (0.03~eV); this part is therefore attributable to the larger cell rather than to the choice of snapshots. The remaining 0.06~eV corresponds to the $\Gamma$-point sampling, which lowers $g$: the 40-atom cell at $2\times2\times2$ $k$-points, the same sampling, gives $g = 0.195$ instead of 0.208 and a corrected gap 0.05~eV lower than at $3\times3\times3$; the 320-atom snapshots have $g$ of 0.194--0.195.

{\small
\begin{longtable}{lrrrrrr}
\caption{\textbf{Halide perovskite snapshots.} Temperature (K; poly., polymorphous structure), atoms per cell, number of snapshots, mean base gap (without spin--orbit coupling), mean and sample standard deviation of the corrected gap (with spin--orbit coupling; none for single structures), and experiment (eV). The standard errors of the means (standard deviation divided by the square root of the number of snapshots) are 0.01--0.06~eV, assuming independent snapshots (see text). Experiment: refs~\citenum{wang2022,sutton2016,lau2016,eperon2014,koh2015,dar2016}.}\label{tab:pvk}\\
\toprule
Composition & $T$ & Atoms & Snapshots & Base & $+\ES$ & Exp. \\
\midrule
\endfirsthead
\toprule
Composition & $T$ & Atoms & Snapshots & Base & $+\ES$ & Exp. \\
\midrule
\endhead
CsPbBr$_{3}$ & 300 & 40 & 6 & 2.43 & 2.28 $\pm$ 0.09 & 2.31 \\
CsPbBr$_{3}$ & 300 & 80 & 6 & 2.50 & 2.35 $\pm$ 0.06 & 2.31 \\
CsPbBr$_{3}$ & 450 & 40 & 6 & 2.54 & 2.47 $\pm$ 0.12 & 2.31 \\
CsPbBr$_{3}$ & 450 & 80 & 3 & 2.49 & 2.35 $\pm$ 0.03 & 2.31 \\
CsPbBr$_{3}$ & poly. & 40 & 1 & 2.45 & 2.23 & 2.31 \\
CsPbCl$_{3}$ & 300 & 40 & 6 & 2.81 & 2.87 $\pm$ 0.04 & 2.99 \\
CsPbCl$_{3}$ & poly. & 40 & 1 & 2.77 & 2.77 & 2.99 \\
CsPbI$_{3}$ & 300 & 40 & 6 & 2.14 & 1.69 $\pm$ 0.09 & 1.72 \\
CsPbI$_{3}$ & 300 & 80 & 6 & 2.12 & 1.64 $\pm$ 0.07 & 1.72 \\
CsPbI$_{3}$ & 450 & 40 & 6 & 2.15 & 1.72 $\pm$ 0.13 & 1.72 \\
CsPbI$_{3}$ & 600 & 40 & 6 & 2.14 & 1.74 $\pm$ 0.10 & 1.72 \\
CsPbI$_{3}$ & 600 & 80 & 3 & 2.08 & 1.65 $\pm$ 0.03 & 1.72 \\
CsPbI$_{3}$ & poly. & 40 & 1 & 2.02 & 1.41 & 1.72 \\
CsPbI$_{3}$ Br$_{16}$of$_{24}$ & 300 & 40 & 6 & 2.30 & 2.02 $\pm$ 0.10 & 2.05 \\
CsPbI$_{3}$ Br$_{8}$of$_{24}$ & 300 & 40 & 6 & 2.20 & 1.86 $\pm$ 0.08 & 1.92 \\
CsPbI$_{3}$ Sn$_{2}$of$_{8}$ & 300 & 40 & 9 & 1.69 & 1.36 $\pm$ 0.14 & -- \\
CsPbI$_{3}$ Sn$_{4}$of$_{8}$ & 300 & 40 & 9 & 1.45 & 1.26 $\pm$ 0.14 & -- \\
CsPbI$_{3}$ Sn$_{6}$of$_{8}$ & 300 & 40 & 9 & 1.11 & 1.11 $\pm$ 0.16 & -- \\
CsPbI$_{3}$ ($\gamma$) & 300 & 80 & 3 & 2.06 & 1.59 $\pm$ 0.01 & 1.72 \\
CsPbI$_{3}$ ($\gamma$) & poly. & 80 & 1 & 2.04 & 1.51 & 1.72 \\
CsSnBr$_{3}$ & 300 & 40 & 6 & 1.29 & 1.78 $\pm$ 0.04 & 1.81 \\
CsSnBr$_{3}$ & poly. & 40 & 1 & 0.96 & 1.29 & 1.81 \\
CsSnI$_{3}$ & 300 & 40 & 9 & 0.95 & 1.15 $\pm$ 0.13 & 1.25 \\
CsSnI$_{3}$ & poly. & 40 & 1 & 0.82 & 0.88 & 1.25 \\
FAPbI$_{3}$ & 300 & 96 & 9 & 2.02 & 1.51 $\pm$ 0.12 & 1.48 \\
FAPbI$_{3}$ & poly. & 96 & 1 & 1.95 & 1.35 & 1.48 \\
FAPbI$_{3}$ Br$_{24}$of$_{24}$ & 300 & 96 & 6 & 2.43 & 2.26 $\pm$ 0.09 & 2.23 \\
FAPbI$_{3}$ Sn$_{2}$of$_{8}$ & 300 & 96 & 6 & 1.69 & 1.36 $\pm$ 0.15 & -- \\
FAPbI$_{3}$ Sn$_{4}$of$_{8}$ & 300 & 96 & 6 & 1.43 & 1.21 $\pm$ 0.14 & -- \\
FAPbI$_{3}$ Sn$_{6}$of$_{8}$ & 300 & 96 & 6 & 1.32 & 1.28 $\pm$ 0.13 & -- \\
FAPbI$_{3}$ Sn$_{8}$of$_{8}$ & 300 & 96 & 6 & 1.09 & 1.35 $\pm$ 0.10 & 1.41 \\
MAPbI$_{3}$ (orth.) & 300 & 48 & 6 & 1.91 & 1.65 $\pm$ 0.05 & 1.65 \\
MAPbI$_{3}$ (orth.) & poly. & 48 & 1 & 1.85 & 1.51 & 1.65 \\
\bottomrule
\end{longtable}
}

\section{Where our correction should be used with care}
\label{sec:limits}

Our correction is less accurate in the following cases, but even there it is on average more accurate than the semilocal functionals alone; where it cannot act, it leaves the result of the base functional unchanged. Above 6~eV it underestimates the gaps (Section~\ref{sec:wide}), with an MAE of 0.80~eV against 2.75~eV for LOCUS alone, 3.37~eV for PBE and 1.88~eV for HSE06. Molecular electron affinities remain overestimated, although their MAE falls from 2.31 to 1.04~eV (Section~\ref{sec:mol}). The gaps of the d$^0$ oxides are overestimated, through their VBM shift, by 0.46~eV on average for the 13 whose experimental reference does not carry the low-reference flag (main text, Discussion; Section~\ref{sec:sz} for the origin at the VBM), whereas LOCUS and PBE alone underestimate their gaps by 0.74 and 1.18~eV; on the PBE base, the corrected gaps of the same 13 oxides have an MAE of 0.22~eV. Open-shell transition-metal oxides are only partly described (Supplementary Table~\ref{tab:os}): the corrected gaps of Fe$_2$O$_3$ and Cr$_2$O$_3$ are close to experiment (2.31 and 3.46~eV, compared with 2.2 and 3.2~eV), and the gaps of NiO and MnO shift by about 1.2~eV towards experiment but remain underestimated by 0.7--1.1~eV, which points to the on-site character of their upper Hubbard bands that a global screening does not describe. Our correction cannot open a gap that the base closes (Section~\ref{sec:base}): CoO and FeO, with partly filled t$_{2g}$ minority shells, did not converge to an insulating state of the base functional in the AFM-II order (600 iterations from a high-spin start), and without orbital symmetry breaking the base is metallic. Defect formation energies and transition levels, which depend on total energies, are those of the base functional.

\begin{table}[H]
\centering
\caption{\textbf{Open-shell transition-metal oxides} in high-spin antiferromagnetic order (eV). Experiment: NiO ref.~\citenum{sawatzky1984}; MnO, CoO refs~\citenum{vanelp1991mno,vanelp1991coo}; FeO ref.~\citenum{bowen1975}; Fe$_2$O$_3$ ref.~\citenum{parkinson2016} (indirect gap); Cr$_2$O$_3$ ref.~\citenum{kasikov2019} (direct gap of films; reported values 3.0--3.8~eV).}
\label{tab:os}
\small
\begin{tabular}{llrrr}
\toprule
Oxide & Order & Base & $+\ES$ & Exp. \\
\midrule
NiO & AFM-II & 2.05 & 3.21 & 4.30 \\
Cr$_{2}$O$_{3}$ & AFM & 2.51 & 3.46 & 3.20 \\
Fe$_{2}$O$_{3}$ & AFM & 1.46 & 2.31 & 2.20 \\
MnO & AFM-II & 1.89 & 3.16 & 3.90 \\
CoO & AFM-II & \multicolumn{2}{l}{no gap (base SCF metallic, not converged)} & 2.50 \\
FeO & AFM-II & \multicolumn{2}{l}{no gap (base SCF metallic, not converged)} & 2.40 \\
\bottomrule
\end{tabular}
\end{table}

\section{Cost}
\label{sec:cost}

Twenty-one held-out solids of up to 16 atoms were chosen to span sp semiconductors, d$^{10}$ compounds and oxides; Supplementary Table~\ref{tab:node} gives the wall times of the 19 whose steps all ran in sequence on the same node (16 cores of an Intel Xeon Gold 6254, 16 MPI processes). mBJ (maximum 300 iterations, default mixer) did not converge for AlN, CuCl, NaCl, SrTiO$_3$ and TiO$_2$, whose mBJ times are lower bounds ($^*$). The steps are the self-consistent calculation of each base, including the PBE starting point of the meta-GGAs, the $\ES$ step (empty bands up to twice the occupied ones), and mBJ (GPAW's implementation on the PBE setups, also from a PBE start). About 7~s of the $\ES$ step (start-up and reading the calculation) does not grow with the system and dominates for the smallest cells. Supplementary Table~\ref{tab:cost} gives the time of the $\ES$ step relative to the self-consistent calculation over the benchmark calculations, whose cells are larger than those of the same-node comparison and whose steps may have run on different nodes; with the PBE base the median is 59\%, because the PBE self-consistent calculation is faster than those of the meta-GGA bases.

\paragraph{Comparison with HSE06.} For a molecular-dynamics snapshot of CsPbI$_3$ (40 atoms, $3\times3\times3$ $k$-points, 600~eV) on a node of the same type with 16 MPI processes, the LOCUS ground state took 2,631~s and the $\ES$ step 915~s, with a peak memory of 3.5 and 5.4~GiB per rank. HSE06 in GPAW, with the adaptively compressed exchange operator, started from the converged PBE state and with only the occupied bands and eight empty ones, had not completed its first exchange-operator build after 3,798~s, at 6.2~GiB per rank; the run was stopped there. A converged HSE06 calculation needs several such builds (8 for the five-atom cell of CsPbI$_3$, with $3\times3\times3$ $k$-points and 500~eV), so the HSE06 ground state of the 40-atom cell would take several hours.

\paragraph{320 atoms.} For one of the 320-atom snapshots of CsPbI$_3$ (Section~\ref{sec:pvk}; $\Gamma$ point, 600~eV), the ground state and the $\ES$ step with spin--orbit coupling ran on one node with 16 MPI processes and 192~GiB of memory. The band step of $\ES$ (empty bands up to twice the occupied ones) needed 7.3~GiB per rank; its evaluation step, which is parallelized over $k$-points and therefore, with the $\Gamma$ point alone, ran on one process, needed the largest memory of the calculation, 124~GiB. HSE06 from the converged PBE state of another of these snapshots, with the occupied bands and 32 empty ones, used 97~GiB in its first iteration, before any exchange-operator build, and exceeded the memory of the node in the first build; with 704 empty bands the run exceeded the node memory at the start. A systematic study of time and memory against the cell size remains to be done.

\begin{table}[H]
\centering
\caption{\textbf{Same-node wall times} (s) of the self-consistent calculation and the $\ES$ step for each base, and of mBJ. $^*$, mBJ not converged within 300 iterations (lower bound).}
\label{tab:node}
\small
\begin{tabular}{lrrrrrrrr}
\toprule
Solid & Atoms & PBE & $+\ES$ & r$^2$SCAN + rVV10 & $+\ES$ & LOCUS & $+\ES$ & mBJ \\
\midrule
AgCl & 2 & 15 & 11 & 71 & 12 & 33 & 16 & 139 \\
AlP & 2 & 16 & 10 & 25 & 10 & 34 & 14 & 49 \\
C & 2 & 10 & 7 & 13 & 7 & 16 & 8 & 32 \\
CuCl & 2 & 18 & 12 & 29 & 15 & 41 & 17 & 195$^*$ \\
LiF & 2 & 14 & 9 & 19 & 9 & 21 & 11 & 105 \\
MgO & 2 & 12 & 8 & 16 & 8 & 22 & 11 & 78 \\
NaCl & 2 & 15 & 9 & 33 & 9 & 34 & 12 & 157$^*$ \\
ZnS & 2 & 18 & 12 & 27 & 12 & 36 & 17 & 97 \\
AlN & 4 & 22 & 11 & 36 & 12 & 40 & 15 & 170$^*$ \\
BeO & 4 & 12 & 8 & 17 & 8 & 23 & 10 & 46 \\
CdS & 4 & 19 & 13 & 35 & 15 & 72 & 19 & 216 \\
SrTiO$_{3}$ & 5 & 21 & 13 & 36 & 15 & 49 & 19 & 220$^*$ \\
Cu$_{2}$O & 6 & 22 & 12 & 53 & 14 & 49 & 20 & 173 \\
SnO$_{2}$ & 6 & 35 & 24 & 74 & 31 & 90 & 35 & 244 \\
TiO$_{2}$ & 6 & 21 & 13 & 38 & 12 & 45 & 13 & 175$^*$ \\
CuGaS$_{2}$ & 8 & 42 & 23 & 115 & 29 & 170 & 34 & 115 \\
Al$_{2}$O$_{3}$ & 10 & 39 & 17 & 86 & 21 & 95 & 24 & 230 \\
HfO$_{2}$ & 12 & 110 & 49 & 206 & 66 & 211 & 68 & 605 \\
CdIn$_{2}$S$_{4}$ & 14 & 64 & 30 & 260 & 39 & 317 & 46 & 308 \\
\bottomrule
\end{tabular}
\end{table}

\begin{table}[H]
\centering
\caption{\textbf{Time of the $\ES$ step} (\% of the self-consistent calculation; median and interquartile range), split into the extra empty bands, the evaluation and input/output.}
\label{tab:cost}
\small
\begin{tabular}{lrrrrr}
\toprule
Base & Calculations & Total & Empty bands & Evaluation & I/O \\
\midrule
PBE & 447 & 59 [50, 69] & 16 & 20 & 8 \\
r2SCAN + rVV10 & 485 & 32 [24, 42] & 11 & 7 & 3 \\
LOCUS & 789 & 21 [9, 30] & 12 & 3 & 1 \\
\bottomrule
\end{tabular}
\end{table}

\section{Independent validation set}
\label{sec:indep}

As an independent test, 50 compounds were drawn at random from the compilation of experimental band gaps of Zhuo et al.\cite{zhuo2018}, among those not used elsewhere in this work, with integer stoichiometry, an experimental gap of at least 0.1~eV and a Materials Project\cite{jain2013} ground-state structure that is experimentally sourced, on the convex hull, non-magnetic, non-metallic and of at most 20 atoms per cell. They were computed with the protocol of the benchmark, on meshes that contain the band edges (Supplementary Table~\ref{tab:indep}); TcS$_2$ was not computed, because the protocol has no PAW setup for Tc.

References more than 0.1~eV below the gap of LOCUS with spin--orbit coupling were not used. Semilocal functionals, the meta-GGA base included, place the gap below the fundamental gap of the computed structure, so that a measured value well below the base gap most likely refers to another polymorph or composition, to absorption by defects or to an error of the compilation. This criterion excludes 15 of the 49 compounds (31\%), against 36 of the 438 solids of the curated Borlido benchmark\cite{borlido2019} (8\%); 13 of these 36 carry the low-reference flag. For 8 of the 15 excluded compounds the reference lies more than 0.1~eV below even the semilocal gap listed by the Materials Project for the same structure, against 1 of the 34 retained compounds; the references of the retained compounds lie on average 0.96~eV above the Materials Project gap. For the remaining 34 compounds the corrected gaps have an MAE of 0.28~eV (ME $+0.10$~eV; MAE 0.22~eV without the 4 with the exciton flag, marked x), against 0.68~eV for LOCUS alone and 0.72~eV for LOCUS with spin--orbit coupling.

{\small
\begin{longtable}{llrrrrl}
\caption{\textbf{Independent validation set.} Materials Project identifier of the structure, experimental gap\cite{zhuo2018}, LOCUS without and with spin--orbit coupling, and the corrected gap (eV). e, reference not used (more than 0.1~eV below LOCUS with spin--orbit coupling); z, mesh enlarged to contain the band edges; x, exciton flag: exciton binding energy of at least 0.5~eV or Bohr radius of at most 4~\AA\ (hydrogenic estimate with the screening of our correction; main text, Methods).}\label{tab:indep}\\
\toprule
Compound & Structure & Exp. & LOCUS & $+$SOC & $+\ES$ & \\
\midrule
\endfirsthead
\toprule
Compound & Structure & Exp. & LOCUS & $+$SOC & $+\ES$ & \\
\midrule
\endhead
PtSe$_{2}$ & mp-aaaaabqx & 0.10 & 0.64 & 0.63 & 0.93 & e \\
TlBiTe$_{2}$ & mp-aaaabopi & 0.12 & 0.63 & 0.22 & 0.42 & e \\
Bi$_{2}$Te$_{2}$Se & mp-aaaabrxa & 0.30 & 0.69 & 0.18 & 0.36 &  \\
P$_{4}$Os & mp-aaackuxm & 0.50 & 1.49 & 1.48 & 1.77 & e \\
Li$_{3}$Bi & mp-aaaabije & 0.70 & 0.74 & $-$0.04 & 0.17 &  \\
LiMgAs & mp-aaaaaspa & 0.70 & 1.74 & 1.64 & 2.04 & e,z \\
CsSb & mp-aaabgqkg & 0.70 & 0.76 & 0.72 & 1.08 &  \\
KSb & mp-aaaaachc & 0.90 & 0.94 & 0.89 & 1.24 &  \\
TiS$_{3}$ & mp-aaaaaoro & 0.90 & 0.49 & 0.49 & 0.96 &  \\
SbRuSe & mp-aaackstv & 0.90 & 0.86 & 0.85 & 1.10 &  \\
Ca$_{2}$CdAs$_{2}$ & mp-aaackici & 0.95 & 1.41 & 1.34 & 1.70 & e \\
GeP & mp-aaackifz & 0.95 & 1.09 & 1.09 & 1.39 & e,z \\
As$_{2}$Ir & mp-aaaaaxdx & 1.00 & 0.88 & 0.81 & 1.06 & z \\
TcS$_{2}$ & mp-aaaaaoar & 1.00 & \multicolumn{4}{l}{not computed (no PAW setup)} \\
RbBiS$_{2}$ & mp-aaaabsll & 1.26 & 1.65 & 1.07 & 1.56 &  \\
POsS & mp-aaacksze & 1.40 & 1.03 & 0.93 & 1.25 & z \\
SiAs & mp-aaaaactr & 1.48 & 1.47 & 1.46 & 1.81 & z \\
GeI$_{2}$ & mp-aaaabphy & 1.50 & 2.27 & 2.15 & 2.64 & e \\
Tl$_{2}$Hg$_{3}$Se$_{4}$ & mp-aaackxqq & 1.57 & 1.48 & 1.47 & 1.91 &  \\
CsSbSe$_{2}$ & mp-aaaaaekf & 1.70 & 1.57 & 1.55 & 2.10 &  \\
K$_{2}$Te & mp-aaaaacpf & 1.90 & 2.49 & 2.27 & 3.21 & e \\
AgSbS$_{2}$ & mp-aaaaafuw & 1.90 & 1.70 & 1.68 & 2.18 &  \\
Sc$_{2}$S$_{3}$ & mp-aaaaaapl & 2.00 & 1.40 & 1.37 & 2.00 &  \\
ZnCu$_{2}$GeS$_{4}$ & mp-aaaaajmm & 2.10 & 1.00 & 0.99 & 1.77 &  \\
K$_{2}$S & mp-aaaaabni & 2.10 & 2.65 & 2.63 & 4.09 & e \\
P$_{2}$Pd$_{3}$S$_{8}$ & mp-aaaaaelq & 2.12 & 1.66 & 1.60 & 2.39 & z \\
ZrS$_{3}$ & mp-aaaaaorp & 2.23 & 1.48 & 1.47 & 2.04 &  \\
Na$_{2}$S & mp-aaaaaayy & 2.40 & 3.02 & 3.00 & 4.12 & e \\
NbCu$_{3}$Se$_{4}$ & mp-aaaaafzn & 2.45 & 1.50 & 1.47 & 2.39 & z \\
Na$_{4}$MgGe$_{2}$Se$_{6}$ & mp-aaackute & 2.53 & 2.16 & 2.12 & 2.80 &  \\
HgPS$_{3}$ & mp-aaaabofi & 2.60 & 2.04 & 2.03 & 2.69 &  \\
Cs$_{2}$Hg$_{3}$S$_{4}$ & mp-aaackxgd & 2.80 & 2.22 & 2.21 & 3.02 &  \\
K$_{2}$NbCuS$_{4}$ & mp-aaaaaoln & 2.82 & 2.22 & 2.22 & 3.54 & x \\
TaCu$_{3}$S$_{4}$ & mp-aaaaapxk & 3.00 & 2.15 & 2.09 & 3.20 & z \\
Al$_{2}$Se$_{3}$ & mp-aaaaarha & 3.12 & 2.49 & 2.47 & 3.15 &  \\
Li$_{2}$CdSnS$_{4}$ & mp-aaacpqom & 3.26 & 2.42 & 2.40 & 3.27 &  \\
ZnSiAg$_{2}$S$_{4}$ & mp-aaacpsfy & 3.28 & 2.56 & 2.55 & 3.37 &  \\
Cs$_{2}$Te & mp-aaabgqtv & 3.50 & 2.17 & 2.01 & 2.95 &  \\
Sr$_{2}$ScSbO$_{6}$ & mp-aaackykw & 3.57 & 4.13 & 4.13 & 5.81 & e \\
HgBr$_{2}$ & mp-aaaabilw & 3.59 & 2.86 & 2.77 & 4.00 &  \\
Ca$_{2}$ScSbO$_{6}$ & mp-aaacprrc & 3.68 & 4.27 & 4.27 & 5.92 & e,x \\
NaSbF$_{6}$ & mp-aaaaaivb & 3.76 & 6.26 & 6.25 & 11.11 & e,x \\
BHPb$_{2}$O$_{4}$ & mp-aaackwzb & 3.76 & 3.35 & 3.34 & 4.70 & x \\
As$_{2}$O$_{3}$ & mp-aaaaaciw & 4.00 & 4.29 & 4.29 & 5.61 & e,x \\
MgSiN$_{2}$ & mp-aaaaafll & 4.05 & 4.75 & 4.75 & 5.89 & e \\
BeS & mp-aaaaaaqg & 4.17 & 3.54 & 3.51 & 4.24 & z \\
HgCl$_{2}$ & mp-aaaabhvb & 4.45 & 3.27 & 3.25 & 4.70 &  \\
CdCl$_{2}$ & mp-aaaabhwb & 5.75 & 4.33 & 4.33 & 5.87 &  \\
CsI & mp-aaabizep & 6.37 & 4.33 & 3.98 & 5.51 & x \\
CsF & mp-aaaaacqq & 10.00 & 6.02 & 5.99 & 9.62 & x \\
\bottomrule
\end{longtable}
}

\section{Generalization}
\label{sec:general}

Three analyses test how far our correction generalizes beyond its calibration.

The first concerns the number of calibration solids; the calibration enters the gaps only through the ratio $k_C/k_V$ (Section~\ref{sec:calib}). Supplementary Table~\ref{tab:calsize} gives, for random subsets of the exact-exchange corrections of 448 solids of the benchmark, computed on the PBE states on which the coefficients were fixed, the spread of the ratio relative to the fit on all solids and the resulting MAE. With ten solids the 5--95\% range of the ratio is 0.37 to 1.54, but the MAE stays within 0.43--0.48~eV; three solids give 0.43--0.52~eV. The wide range of the ratio comes from 24 solids in which the lowest empty state at the PBE gap has valence character and therefore a negative exact-exchange correction. Of these, 22 have PBE gaps below 0.2~eV, half of them copper chalcogenides such as CuInSe$_2$, in which the gap is inverted or nearly closed; none of the ten calibration solids has such an edge. On the 405 solids whose PBE gap is at least 0.2~eV, the 5--95\% range of the ratio for ten solids is 0.75 to 1.27 and the MAE stays within 0.43--0.45~eV.

The second is the error by chemical class (Supplementary Table~\ref{tab:classes}). Our correction is most accurate for phosphides, arsenides, antimonides, nitrides and tellurides (0.2--0.3~eV) and least accurate for fluorides and the elemental solids, whose errors come mainly from the rare-gas solids, consistent with Section~\ref{sec:wide}.

The third is the transfer of $s$ from molecules to solids (Section~\ref{sec:sz}) and of the bare-shift coefficients between base functionals (main text, Fig.~2c).

\begin{table}[H]
\centering
\caption{\textbf{Number of calibration solids.} Median and 5--95\% range over 300 random subsets of the exact-exchange corrections on PBE states.}
\label{tab:calsize}
\small
\begin{tabular}{rll}
\toprule
Solids & $(k_C/k_V)$ / all solids & MAE (eV) \\
\midrule
3 & 1.192 [-0.309, 1.807] & 0.451 [0.427, 0.519] \\
5 & 1.129 [0.048, 1.643] & 0.446 [0.426, 0.491] \\
10 & 1.110 [0.373, 1.543] & 0.443 [0.426, 0.480] \\
20 & 1.017 [0.562, 1.377] & 0.438 [0.426, 0.463] \\
50 & 1.004 [0.752, 1.258] & 0.437 [0.427, 0.453] \\
100 & 1.010 [0.846, 1.167] & 0.437 [0.430, 0.447] \\
448 & 1.000 [1.000, 1.000] & 0.437 [0.437, 0.437] \\
\bottomrule
\end{tabular}
\end{table}

\begin{table}[H]
\centering
\caption{\textbf{Error by chemical class} (most electronegative anion), held-out crystals (eV).}
\label{tab:classes}
\small
\begin{tabular}{lrrrrrr}
\toprule
Class & $n$ & MAE & ME & $n$ (HSE06) & HSE06 & $+\ES$ on HSE06 set \\
\midrule
sulfides & 96 & 0.40 & 0.22 & 96 & 0.37 & 0.40 \\
oxides & 79 & 0.59 & 0.42 & 73 & 0.65 & 0.61 \\
selenides & 62 & 0.41 & 0.05 & 62 & 0.40 & 0.41 \\
iodides & 36 & 0.40 & -0.03 & 31 & 0.55 & 0.36 \\
bromides & 31 & 0.52 & 0.07 & 28 & 0.61 & 0.51 \\
tellurides & 28 & 0.25 & -0.05 & 28 & 0.32 & 0.25 \\
chlorides & 20 & 0.62 & 0.00 & 19 & 0.83 & 0.60 \\
antimonides & 18 & 0.28 & 0.20 & 18 & 0.28 & 0.28 \\
phosphides & 16 & 0.20 & -0.07 & 16 & 0.27 & 0.20 \\
arsenides & 15 & 0.18 & 0.03 & 15 & 0.13 & 0.18 \\
nitrides & 14 & 0.28 & 0.05 & 14 & 0.32 & 0.28 \\
fluorides & 13 & 0.84 & 0.03 & 12 & 1.81 & 0.86 \\
elements & 8 & 0.90 & -0.76 & 8 & 2.08 & 0.90 \\
other & 8 & 0.42 & -0.18 & 8 & 0.42 & 0.42 \\
intermetallic and other & 4 & 0.24 & -0.24 & 4 & 0.49 & 0.24 \\
carbides & 4 & 0.07 & -0.04 & 4 & 0.08 & 0.07 \\
hydrides & 2 & 0.19 & 0.19 & 2 & 0.66 & 0.19 \\
\bottomrule
\end{tabular}
\end{table}

\FloatBarrier